\documentclass[aps,prd,twocolumn,superscriptaddress,nofootinbib,eqsecnumm,showpacs]{revtex4}
\usepackage[pass,letterpaper]{geometry}
\pdfoutput=1  
\usepackage{graphicx,color}
\usepackage{latexsym,amsmath,amsfonts,amssymb,booktabs}
\usepackage{hyperref}
\usepackage{slashed,upgreek,amscd,cancel,tensor,color}
\usepackage{url}
\usepackage{placeins}
\usepackage{doi}
\usepackage[normalem]{ulem}

\definecolor{MyBlue}{rgb}{0.15,0.15,0.70}
\hypersetup{
colorlinks=true,
citecolor=MyBlue,
linkcolor=MyBlue,
urlcolor=MyBlue
}

\newcommand{\be}{\begin{equation}}
\newcommand{\ee}{\end{equation}}
\newcommand{\beq}{\begin{equation}}
\newcommand{\eeq}{\end{equation}}
\newcommand{\bea}{\begin{eqnarray}}
\newcommand{\eea}{\end{eqnarray}}

\newcommand{\bee}{{\bf{e}}}
\newcommand{\obs}{_{\rm O}}

\newcommand{\Gal}{_{\rm G}}

\newcommand{\dd}{\text{d}}
\newcommand{\bx}{{\bf{x}}}

\newcommand{\bk}{{\mathbf k}}

\newcommand{\HH}{{\cal H}}

\newcommand{\m}{{\rm m}}

\newcommand{\nn}{{\nonumber}}

\usepackage[stable]{footmisc}

\newcommand\ees{\end{eqnarray}}
\newcommand\bees{\begin{eqnarray}}

\begin{document}
\title{Properties of the stochastic astrophysical gravitational wave background: astrophysical sources dependencies}

\author{Giulia Cusin}
\email{giulia.cusin@physics.ox.ac.uk}
\affiliation{Astrophysics Department, University of Oxford, DWB, Keble Road, Oxford OX1 3RH, UK}
\author{Irina Dvorkin}
\email{irina.dvorkin@aei.mpg.de}
\affiliation{Max Planck Institute for Gravitational Physics (Albert Einstein Institute), Am M\"{u}hlenberg 1, Potsdam-Golm, 14476, Germany}
\author{Cyril Pitrou}
\email{ pitrou@iap.fr}
\affiliation{Institut d'Astrophysique de Paris, CNRS UMR 7095, \\
98 bis, Bd Arago, 75014 Paris, France}
\author{Jean-Philippe Uzan}
\email{uzan@iap.fr}
\affiliation{Institut d'Astrophysique de Paris, CNRS UMR 7095, \\
98 bis, Bd Arago, 75014 Paris, France}
\date{\today}

\begin{abstract}
This article explores the properties  (amplitude and shape) of the angular power spectrum of the anisotropies of the astrophysical gravitational wave background (AGWB)  focusing on the signatures of the astrophysical models describing sub-galactic physics. It demonstrates that while some parameters have negligible impact others, and in particular the stellar evolution models, the metallicity and the merger time delay distribution can result in relative differences of order 40\% in the angular power spectrum of anisotropies in both the LIGO/Virgo and LISA frequency bands. It is also shown that the monopole and the anisotropic components of the AGWB are complementary and sensitive to different astrophysical parameters. It follows that AGWB anisotropies are a new observable with the potential to provide new astrophysical information that can not be accessed otherwise.
\end{abstract}
\maketitle

\tableofcontents
\section{Introduction}

Diffuse stochastic backgrounds arise  from the  incoherent superposition  of signals from resolved and unresolved sources. Many such backgrounds for different kinds of radiation have  been observed in astronomy. Electromagnetic backgrounds of radiation include the cosmic microwave background (CMB) with its black body spectrum~\cite{Penzias:1965wn}, the cosmic infrared background (CIB) from stellar dust~\cite{2001ARA&A..39..249H} and the extragalactic background made up of all the electromagnetic radiation emitted by stars, galaxies, galaxy clusters etc. since their formation~\cite{1967ApJ...148..377P,1991Natur.353..315S}. Similarly, there should exist a neutrino background~\cite{2006ARNPS..56..137H} and a background of gravitational waves (GW).  

The GW background can be split into a stochastic background of gravitational radiation of cosmological origin, e.g. produced during inflation,  and one of astrophysical origin (AGWB). The latter results from the superposition of a large number of resolved and unresolved sources from the onset of stellar activity until today. The nature of the AGWB is expected to be significantly different from its cosmological counterpart, which is expected to be, at least for inflation, stationary, unpolarized, almost statistically Gaussian and isotropic, by analogy with the cosmic microwave background. Many different astrophysical sources contribute to the AGWB, including merging stellar-mass black hole (BH) and neutron star (NS) binaries~\cite{TheLIGOScientific:2016wyq, Regimbau:2016ike, Mandic:2016lcn, Dvorkin:2016okx, Nakazato:2016nkj, Dvorkin:2016wac, Evangelista:2014oba}, merging supermassive black hole binaries~\cite{Kelley:2017lek}, rotating neutron stars~\cite{Surace:2015ppq, Talukder:2014eba, Lasky:2013jfa}, stellar core collapse~\cite{Crocker:2017agi, Crocker:2015taa} and population III binaries~\cite{Kowalska:2012ba}.\\

The observational landscape is growing and covers a large range of frequencies; see e.g. Ref.~\cite{Moore:2014lga} for a review.  At extremely low frequencies $\sim 10^{-16}$ Hz observational bounds come mainly from the analysis of CMB B-modes while frequencies in the range $10^{-10}-10^{-6}$~Hz are covered by pulsar timing arrays: the Parkes Pulsar Timing Array\footnote{{\tt http://www.atnf.csiro.au/research/pulsar/ppta/}} (PPTA), the Large European Array for Pulsar Timing\footnote{{\tt http://www.leap.eu.org}} (LEPTA), and the North American Nanohertz Observatory for Gravitational Waves (NANOGrav), all of which form the International Pulsar Timing Array Consortium\footnote{{\tt http://www.ipta4gw.org}} (IPTA). Frequencies in the range $10^{-4}-10^{-1}$~Hz will be probed with the space-based Laser Interferometer Space Antenna\footnote{{\tt www.lisamission.org}} (LISA) scheduled to be launched in 2034. Higher frequencies ($1-10^{3}$~Hz) are accessible with ground-based interferometers, including Advanced LIGO (aLIGO) \cite{2015CQGra..32g4001L} and Advanced Virgo (aVirgo) \cite{2015CQGra..32b4001A} which already conducted two observational runs (O1 and O2) during 2014-2017 and are currently entering into the third observational run,  KAGRA interferometer which is expected to become operational in 2018-2019 and  LIGO India which is currently under construction. A third generation of ground-based interferometers, the Einstein Telescope\footnote{{\tt http://www.et-gw.eu}} (ET) and the Cosmic Explorer (CE) \cite{Evans:2016mbw} are in their design stages.

The latest upper bounds obtained in Ref.~\cite{2019arXiv190302886T} using the first and second aLIGO observing runs are $\Omega_{\rm GW}(f=25 \text{Hz}) < 4.8\times10^{-8}$, assuming a population of compact binary sources, and  $\Omega_{\rm GW}(f=25 \text{Hz}) < 6\times 10^{-8}$ for a frequency-independent background for the frequency ranges $20-92$ Hz and $20-80$ Hz, respectively, where $\Omega_{\rm GW}=\dd\rho_{\rm GW}/\dd\ln f/\rho_c$ is the energy density in GW per logarithmic frequency interval in units of the critical density of the Universe. This improves bounds on the stochastic background obtained from the analysis of big-bang nucleosynthesis~\cite{Maggiore:1999vm, Allen:1996vm},  and of the cosmic microwave background~\cite{Smith:2006nka,Henrot-Versille:2014jua} at 100~Hz. LIGO-Virgo upper bounds are well above current theoretical predictions, for example the population model derived from the O1+O2 source catalog \cite{2018arXiv181112907T,2019arXiv190302886T} predicts an amplitude of the total background (binary black holes and binary neutron stars) of $\Omega_{\rm GW}(f=25 \text{Hz})=8.9_{-5.6}^{+12.6}\times 10^{-10}$ and $\Omega_{\rm GW}(f=25 \text{Hz})=5.3_{-2.5}^{+4.2}\times 10^{-10}$ from binary black holes alone, where the uncertainties are due to $90\%$ confidence limits on the merger rates.  Assuming the most probable rate for compact binary mergers,  Ref.~\cite{Abbott:2017xzg} concludes that the total background may be detectable with a signal-to-noise-ratio  of  3  after  40  months  of  total  observation  time.  At low frequencies, Pulsar Timing Arrays give a bound  $\Omega_{\rm GW}<1.3\times10^{-9}$ for $f=2.8 \times 10^{-9}$~Hz~\cite{Shannon:2013wma}. The possibility of measuring and mapping the gravitational wave background is discussed in Refs.~\cite{Allen:1996gp, Cornish:2001hg, Mitra:2007mc, Thrane:2009fp, Romano:2015uma, Romano:2016dpx, Renzini:2018vkx} while different methods employed by LIGO and LISA to reconstruct an angular resolved map of the sky are presented in Ref.~\cite{TheLIGOScientific:2016xzw}. An analogous discussion for  Pulsar Timing Arrays  can be found in Refs.~\cite{Mingarelli:2013dsa,Taylor:2013esa, Gair:2014rwa}. 

The latest observational constraints from  the first and second aLIGO runs~\cite{2019arXiv190308844T} are derived for multipoles up to $\ell=4$ with upper limits on the amplitude in the range $\Omega_{\rm GW}(f=25 \text{Hz},\Theta)<0.64-2.47\times 10^{-8}$ sr$^{-1}$ for a population of binary compact objects and assuming that the angular and frequency dependencies factorize, an assumption we shall investigate below for realistic astrophysical models.\\

From a theoretical perspective, as any background of radiation, the AGWB is fully characterized in terms of Stokes parameters, intensity and polarization, as a function of direction of frequency; see Refs.~\cite{Romano:2016dpx, Cusin:2018avf} for a definition of Stokes parameters for a background of spin-2 radiation. The first prediction of the AGWB angular power spectrum was presented in our analysis~\cite{Cusin:2018rsq} following our seminal formalism\footnote{Note that a first attempt to describe anisotropies of the AGWB with a Boltzmann approach was proposed by Ref.~\cite{Contaldi:2016koz} while Ref.~\cite{Cusin:2018avf} refines it by introducing an emissivity function that realistically describes GW emission at the galactic scale.}  introduced in Refs.~\cite{Cusin:2017mjm, Cusin:2017fwz}. This formalism is very flexible and splits the cosmological, large-scale structure and sub-galactic scales so that it can be applied to any source contributions and any frequency band.  It significantly differs from the simpler computation of the monopole~\cite{Regimbau:2011rp,Dvorkin:2016okx} in which the sources were assumed to be homogeneously and isotropically distributed. The anisotropies were further studied in Refs.~\cite{Jenkins:2018uac, Jenkins:2018kxc} using a different set of astrophysical models (see Ref.~\cite{Cusin:2018ump} for a critical analysis of these works). The first study of the generation of polarization induced by the diffusion by massive structures is presented in Ref.~\cite{Cusin:2018avf}.\\

The goal of this article is to extend our previous analysis \cite{Cusin:2018avf} for the contribution of BH mergers in the LIGO/Virgo frequency band, in the following directions: 
\begin{enumerate}
\item describe in details the properties of the angular power spectrum of AGWB;
\item explore the astrophysical dependencies of the angular power spectrum; 
\item study the contribution of binary NS mergers;
\item extend the analysis to lower frequencies (LISA band and lower). 
\end{enumerate}

In the LIGO band where the background is dominated by mergers of compact objects, we  explore different stellar models for the evolution of BH stellar progenitors and we study the dependence of the result on the distribution function of  orbital parameters of binary objects, mass distribution and initial mass function. We also analyze how different BH populations contribute to the background.  We explicitly show that anisotropies are very sensitive to changes in the astrophysical model used to describe the sub galactic process of formation and evolution of GW sources and the sub-galactic process of GW emission. 
We present a detailed discussion of which are the astrophysical parameters and functions the angular power spectrum is most sensitive to and hence that we will be able to constrain the first, in both the LIGO and the LISA frequency bands. 

The article is organized as follows. Section~\ref{initial} summarizes the main results of our formalism~\cite{Cusin:2017mjm, Cusin:2017fwz}  describing the computation of the anisotropies of the AGWB (power spectrum and cross-correlations with other cosmological probes). We stress that three main building blocks are present: choice of a cosmological framework, description of large scale structures and astrophysical modeling of the GW sources. While the first two are standard lore in cosmology, we focus on the sub-galactic astrophysics which is less constrained today. Section~\ref{general} describes the general properties of the angular power spectrum of anisotropies. Using the Limber approximation, we provide an analytic approximation of the angular power spectrum of anisotropies,  useful to derive order of magnitude estimates and to understand  our numerical results. We then propose a general derivation of shot-noise and we explain that cross-correlation with galaxy number counts can help to extract a map of AGWB anisotropies, even for shot-noise dominated background maps.  We then discuss  the frequency-direction factorization hypothesis used in current directional searches (e.g. by LIGO-Virgo). We demonstrate that this approximation fails in capturing the physics in the upper part of the LIGO-Virgo frequency band. Section~\ref{sub} defines the building blocks of the astrophysical modeling. It describes the normalisation of the model, which is necessary to have a meaningful model comparison. Section~\ref{exploration} explores the signature of these models on the monopole, the angular power spectrum and the various cross-correlations.  In Section~\ref{NS} this analysis is extended to include the contribution of NS mergers and in Section~\ref{LISAsec} to the LISA frequency band.

\section{Anisotropies of AGWB}\label{initial}

The dimensionless energy density of the GW background per unit of solid angle, $\dd^2\bee$, and logarithmic frequency, $\dd f/f$, can be split into an homogenous and isotropic component and a directional dependent one as 
\begin{align}\label{background}
\Omega_{\rm GW}(\bee, f)&=\frac{f}{\rho_c}\frac{\dd^3\rho_{\rm GW}}{\dd^2\bee\, \dd f}(\bee, f)\nn\\
&=\frac{\bar{\Omega}_{\rm GW}(f)}{4\pi}+\delta\Omega_{\rm GW}(\bee, f)\,, 
\end{align}
where $\rho_c$ is the critical energy density of the universe and $\rho_{\rm GW}$ is the energy density of GW.  Using the standard expression for the energy density in terms of the wave amplitude, see e.g. Ref.~\cite{Maggiore:1900zz},  and recalling that the definition of energy requires an average over several periods of the wave,  we find 
\be\label{preintensity}
\Omega_{\rm GW}(\bee, f)=\frac{c^2}{4G\rho_c}\frac{1}{T_{\obs}}f^3 \sum_{A=+\,, \times} |\tilde{h}_A(f, \bee)|^2\,,
\ee
where $T_{\obs}$ comes from the time average and represents the period of observation of the detector.\footnote{The dependence on $T_{\obs}$ may appear strange at first, but note that for a \emph {continuous} background signal centered around a frequency $f$ with width $\Delta f$ one has for $T_{\obs} \gg 1/f$, $T_{\obs}\propto\int_0^{T_{\obs}} |h_A(t)|^2\dd t \simeq \int |h_A(f')|^2\dd f' \simeq \Delta f |h_A(f)|^2$ and hence $ |h_A(f)|^2 \propto T_{\obs}$. }

In this section we describe the generic properties of monopole and anisotropies. To that purpose, we rely on the  reference model, fully defined in \S~\ref{reference} below. We first recall the computation of the isotropic contribution (\S~\ref{sub2.1}), of the power spectrum of anisotropies (\S~\ref{sub2.2}) and of the cross-correlations with other cosmological probes (\S~\ref{sub2.3}). We conclude in \S~\ref{sub2.4} with a schematic illustration of the general computation strategy and numerical implementation, focusing on the different scales that enter in the discussion.

\subsection{Isotropic contribution}\label{sub2.1}

The background component in Eq.~(\ref{background}) is given by the line of sight integration on our past lightcone
 \begin{align}\label{back}
\bar{\Omega}_{\rm GW}(f)=&\int_{\eta_*}^{\eta_{\obs}} \dd\eta\,\partial_{\eta}\bar{\Omega}_{\rm GW}(f, \eta)\,,
\end{align}
where $\eta_*$ stands for a maximal distance (or, equivalently, maximal redshift) above which there are no astrophysical sources. For future convenience, we introduce the redefinition 
\be\label{link}
\partial_{\eta}\bar{\Omega}_{\rm GW}=\frac{f}{\rho_c} \mathcal{A}(f, \eta)\,, 
\ee
with 
\be\label{AA}
\mathcal{A}(\eta, f)\equiv a^4 \int \dd\theta_{\Gal}\bar{n}_{\Gal}(\eta, \theta_{\Gal})\mathcal{L}_{\Gal}(\eta, f_{\Gal}, \theta_{\Gal})\,, 
\ee
where $\mathcal{L}_{\Gal}$ is the effective GW luminosity of a galaxy per unit of emitted frequency, $f_{\Gal}$, characterized by the set of parameters $\theta_{\Gal}$ (mass, metallicity...). This effective luminosity has been introduced in Ref.~\cite{Cusin:2017fwz}. In the galaxy rest-frame, it represents the sum of the luminosity of the GW emitted by all astrophysical sources contained in that galaxy, averaged on the distribution function of their peculiar velocity. As shown in Ref.~\cite{Cusin:2017fwz}, at linear order, the effects of the peculiar motion of a source in its host galaxy can be neglected  on  average. The relation between the effective luminosity and the emitted strain is established in Eqs.~(79) and~(80) of Ref.~\cite{Cusin:2017mjm}. In Eq.\,(\ref{AA}), $n_{\Gal}$ is the comoving number density of galaxies, $a$ is the scale factor of the Friedman-Lema\^{\i}tre spacetime normalized to 1 today and $\eta$ its comoving time. The frequencies at emission and observation are related by
\be\label{fgf}
f_{\Gal}=(1+z_{\Gal})f
\ee
where $z_{\Gal}$ is the redshift.

\begin{figure}[!htb]
\includegraphics[width=\columnwidth]{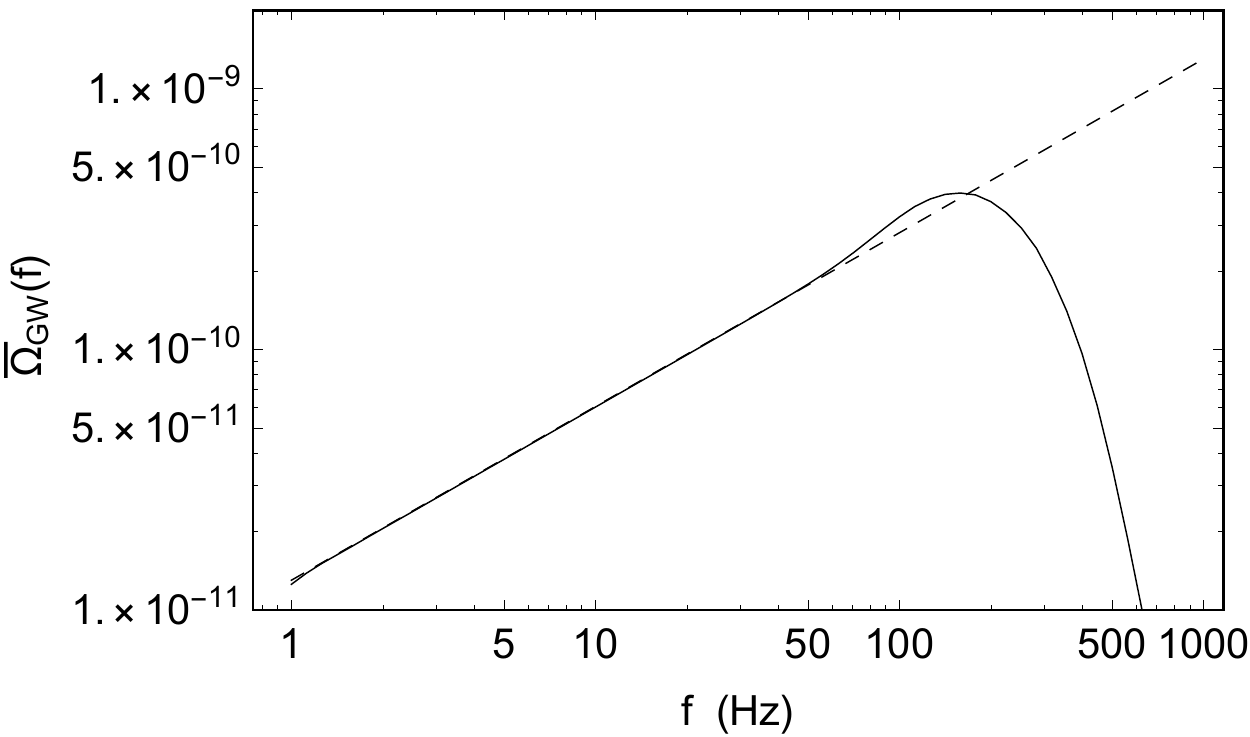}
\caption{\label{back} Background energy density of the stochastic astrophysical GW as a function of frequency compared with the fit with $\propto f^{2/3}$. The reference astrophysical model is defined in \S~\ref{reference}.}\label{BackgroundGW}
\end{figure}

\begin{figure*}[!htb]
\includegraphics[width=0.47\linewidth]{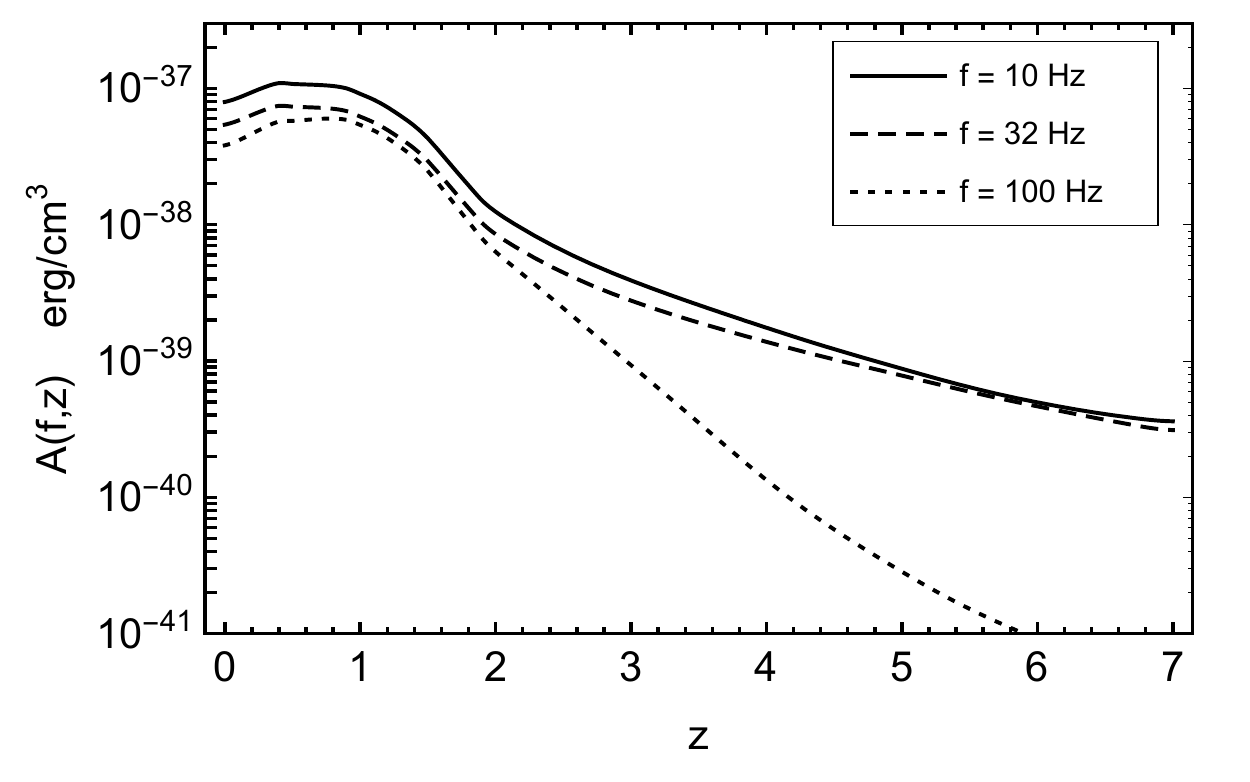}
\includegraphics[width=0.47\linewidth]{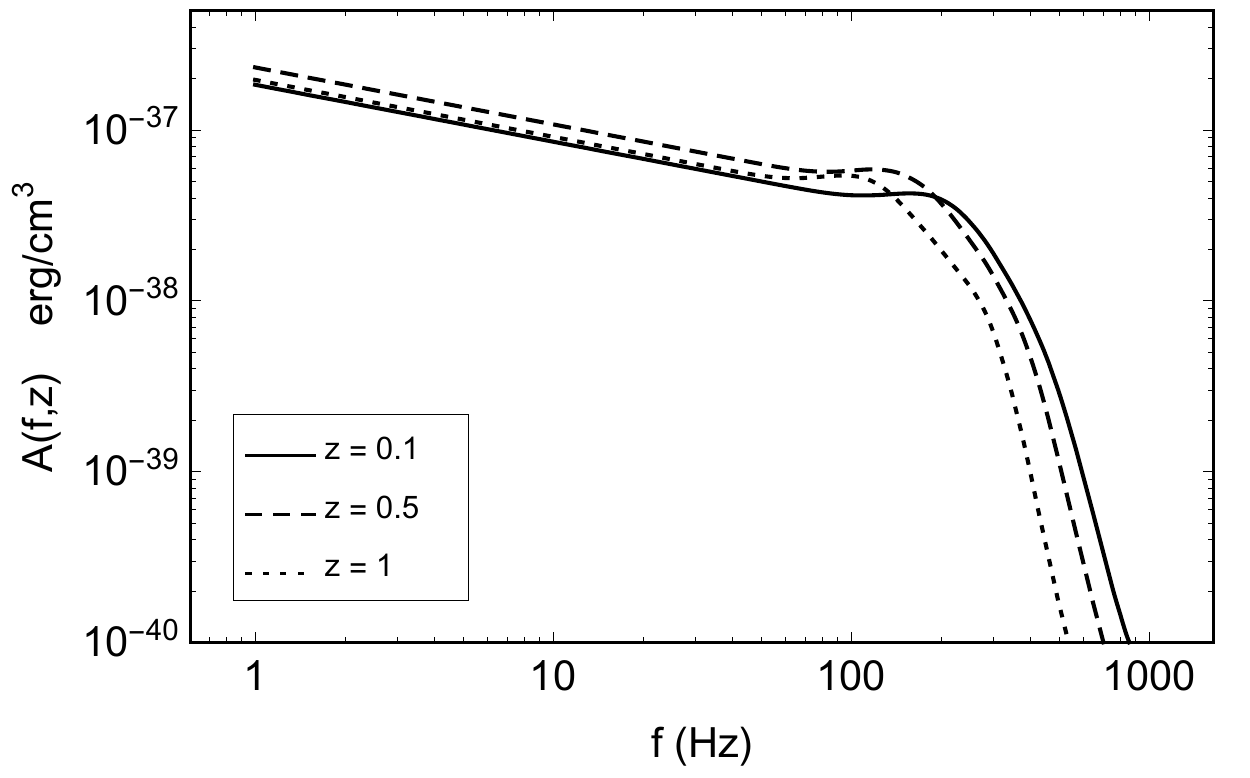}
\caption{The astrophysical kernel $\mathcal{A}(z, f)$ defined in Eq.~(\ref{link}) as a function of redshift (Left) and as a function of frequency (Right). The reference astrophysical model is defined in \S~\ref{reference}.}\label{ANuZ}
\end{figure*}

Figure~\ref{back} depicts the background contribution for the reference astrophysical model. It shows that the low frequency part is well-fitted by a power law $\propto f^{2/3}$. Figure~\ref{ANuZ} presents the astrophysical kernel  $\mathcal{A}(z, f)$ defined in Eq.~(\ref{link}) as a function of redshift (left panel) and frequency (right panel). 

\subsection{Anisotropies}\label{sub2.2}

The anisotropic component of the background energy density of Eq. (\ref{background}) is given by
 \begin{align}\label{ssss}
&\delta\Omega_{\rm GW}(\bee,  f)=\nn\\
&=\frac{f}{4\pi\rho_c}\int_{\eta_*}^{\eta_{\obs}} \dd\eta\,\mathcal{A}\left(\eta, f\right) \left[\delta_{G}+4\Psi-2\bee\cdot \nabla v+6\int_{\eta}^{\eta_{\obs}}\dd\eta' \dot{\Psi}\right] \nn\\
&+\frac{f}{4\pi\rho_c}\int_{\eta_*}^{\eta_{\obs}} \dd\eta\,\mathcal{B}(\eta, f) \left[\,\bee\cdot \nabla v-\Psi-2\int_{\eta}^{\eta_{\obs}} d\eta' \dot{\Psi}\right]\,,
\end{align}
where the astrophysical kernel $\mathcal{A}$ is defined in Eq.~(\ref{AA}) while 
\begin{align}
\mathcal{B}(\eta, f)\equiv f\, a^3\bar{n}_{\Gal}(\eta)\int d\theta_{\Gal}\frac{\partial \mathcal{L}_{\Gal}}{\partial f_{\Gal}}\Big|_{{\bar{f}}_{\Gal}}(\eta,  f_{\Gal}, \theta_{\Gal})\,,\label{BB}
\end{align}
in which the relation between the frequencies at emission and observation needs only to be evaluated at lowest order so that Eq.~(\ref{fgf}) reduces to $f_{\Gal}=f/a$. 

In Eq.~(\ref{ssss}), $n_{\Gal}$ stands for the comoving number density of galaxies, ${\Psi} $ for the gravitational potential, $v$ for the comoving velocity field and $\delta_{\rm G}$ is the galaxy over-density. This latter is related to the dark matter over-density, $\delta_{\rm m}$, by the bias $b$, defined in comoving gauge, such that
\be
\delta_{\rm G}^{\rm c} = b \delta_{\rm m}^{\rm c} \quad \Rightarrow \quad \delta_{\rm G} + 3\HH v= b \left(\delta_{\rm m} + 3 \HH v\right)\,,
\ee
where $\HH\equiv \dd\ln a/\dd\eta$ is the comoving Hubble parameter.

The statistical properties of $\delta \Omega_{\rm GW}(\bee,f)$ are first encoded in its angular correlation function or, equivalently, in its angular power spectrum 
\be\label{Cell}
C_{\ell}(f)=\frac{2}{\pi}\int \dd k\,k^2 |\delta\Omega_{\ell}(k, f)|^2\,, 
\ee
where ${\delta\Omega}_{\ell}(k,f)$, derived in Ref.~\cite{Cusin:2017fwz}, is given by
\begin{align}\label{Rkk}
&{\delta\Omega}_{\ell}(k,f)=\frac{f}{4\pi\rho_c}\Bigg\{\int_{\eta_*}^{\eta_{\obs}} \dd\eta\,
  \mathcal{A}(\eta, f)\times\\
&\quad\left[\left(4{\Phi}_k(\eta)+b\delta_{\rm m, k}(\eta)+(b-1) 3
  \mathcal{H} v_k(\eta)\right)
  j_{\ell}(k\Delta\eta)\right.\nn\\
&\qquad\left.-2k {v}_k(\eta)j'_{\ell}(k\Delta\eta)\right]\nn\\
&+\int_{\eta_*}^{\eta_{\obs}} \dd\eta\,  \mathcal{B}(\eta, f) \left[-{\Phi}_k(\eta)j_{\ell}(k\Delta\eta)+k {v}_k(\eta)j'_{\ell}(k\Delta\eta)\right]\nn\\
&+\int_{\eta_*}^{\eta_{\obs}}
   \dd\eta\left[6\mathcal{A}(\eta,
   f)-2\mathcal{B}(\eta, f)\right]\int_{\eta}^{\eta_{\obs}}\dd\tilde\eta\, {\Phi}'_k(\tilde\eta) j_{\ell}(k\Delta\tilde\eta)\Bigg\}\,, \nn
\end{align}
where $\Delta\eta=\eta_0-\eta$,  $j_\ell$ stands for the spherical Bessel function, $k$ is the wavenumber and $X_k(\eta)$ stands for the Fourier modes of $X$ and it has been assumed that the bias has no scale dependence.

To conclude, we introduce the reduced angular power spectrum by normalizing over the monopole  as 
\be
C_{\ell}^{\text{rel}}\equiv C_{\ell}\frac{(4\pi)^2}{ \bar{\Omega}^2_{\rm GW}}\,.
\ee
From now on we will refer to this dimensionless quantity as the angular power spectrum of \emph{relative anisotropies}.

\subsection{Cross-correlation}\label{sub2.3}

Since the AGWB anisotropy depends on cosmological perturbations, see Eq.~(\ref{ssss}),  it correlates with any other cosmological probe, such as galaxy number counts and weak lensing convergence. The cross-correlation power spectra have been presented  in Ref.~\cite{Cusin:2017fwz}, 
\be\label{Bchi}
B_\ell^X(f) \equiv \frac{2}{\pi}\int \dd k\,k^2 \frac{4\pi}{\bar \Omega_{\rm GW}(f)} \delta\Omega^*_{\ell}(k,f)\, X_{\ell}(k)\,.
\ee
For weak lensing, $X_\ell$ is the cosmic convergence 
\be
\kappa_{\ell}=-\frac{\ell(\ell+1)}{2}\int_0^{\chi_H}d\chi\, g(\chi)\, \hat{\Psi}_k(\chi) j_{\ell}(k\chi)\,,
\ee
so that $B_\ell\equiv B_\ell^\kappa$ is given by
\begin{align}\label{Bell}
B_{\ell}(f)&=\frac{2}{\pi}\int \dd k\,k^2  \frac{4\pi}{\bar \Omega_{\rm GW}(f)}\delta\Omega^*_{\ell}(k,f)\,\kappa_{\ell}(k)\,.
\end{align}
In these expressions, $\chi_H$ corresponds to the maximal depth of a given survey and 
\be
g(\chi)\equiv\frac{1}{\chi} \int_{\chi}^{\chi_H}d\chi' p_{\chi}(\chi') \frac{(\chi'-\chi)}{\chi'}\,,
\ee
where the function $p_{\chi}$ is the sources distribution function. For the cross-correlation with galaxy,  $X_\ell$ is number counts $\Delta_{\ell}(k, z)$, defined e.g. in Eq.  (44) of  \cite{Bonvin:2011bg} so that  $D_\ell\equiv B_\ell^\Delta$ 
\begin{align}
D_{\ell}(f, z)&=\frac{2}{\pi}\int \dd k\,k^2 \frac{4\pi}{\bar \Omega_{\rm GW}(f)} \delta\Omega^*_{\ell}(k,f)\,\Delta_{\ell}(k, z)\,.
\end{align}

\begin{figure*}[!htb]
\includegraphics[width=\columnwidth]{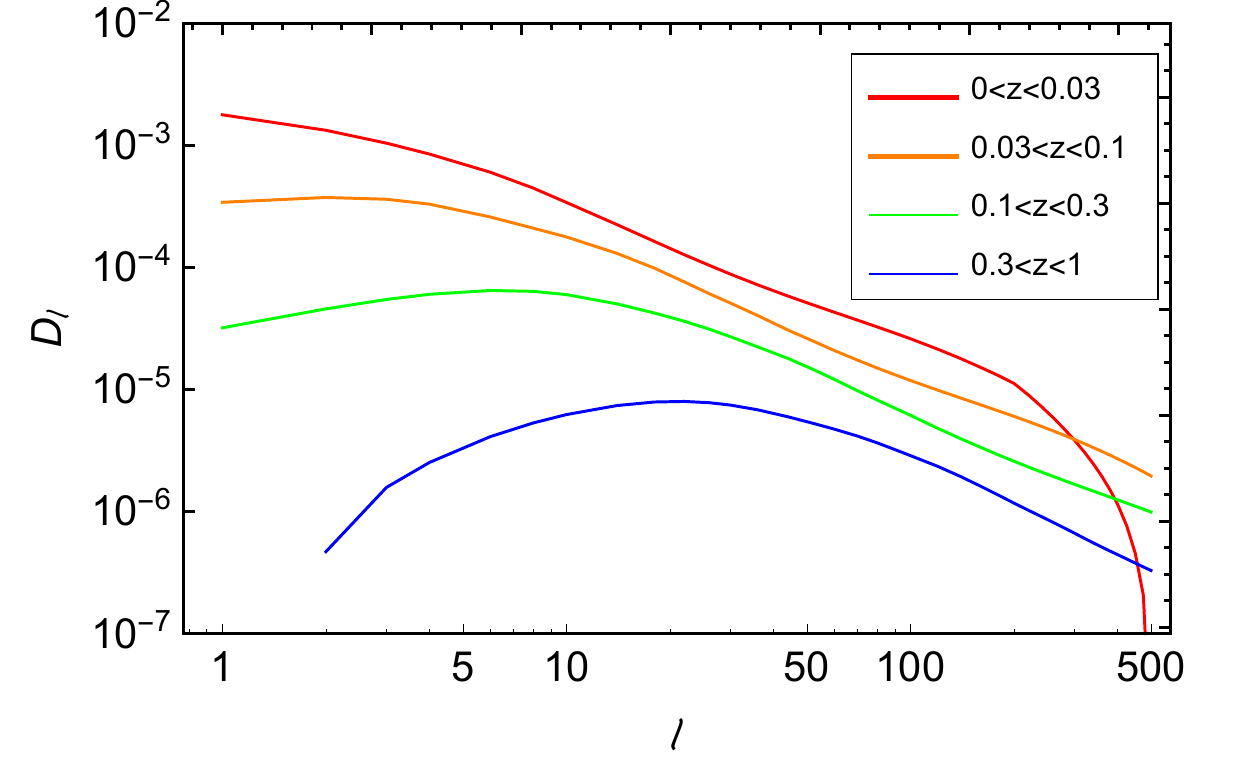}
\includegraphics[width=\columnwidth]{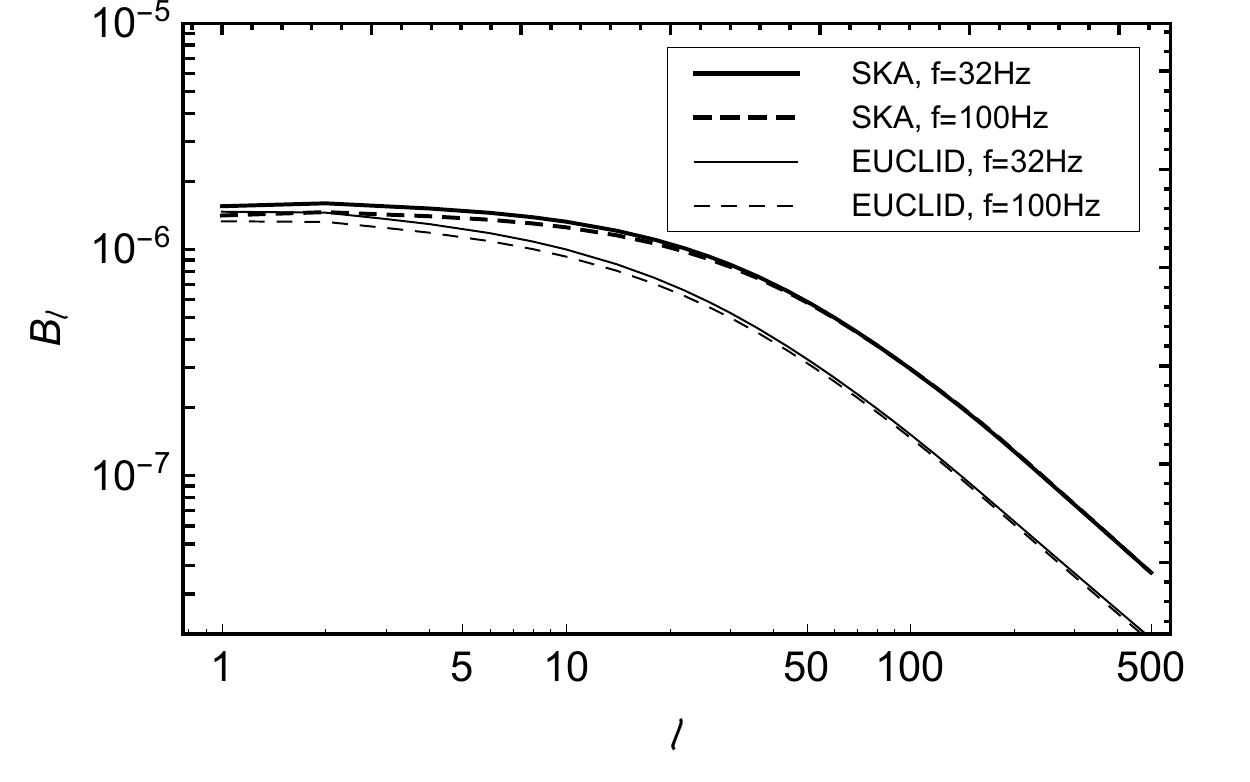}
\caption{\emph{Left:} Cross-correlation with galaxy number counts in different redshift bins and integrated over frequency in the range 10 Hz$<$ f$<$100 Hz. \emph{Right:} Cross-correlation with weak lensing convergence using the SKA~\cite{Andrianomena:2014sya} and Euclid~\cite{2011arXiv1110.3193L} redshift distributions. The reference astrophysical model used for this plot is defined in \S~\ref{reference}.}\label{cross}
\end{figure*} 

For our reference astrophysical model, the cross-correlations with weak lensing convergence (for both SKA and Euclid source distributions)\footnote{For SKA, we use the source distribution of Ref.\,\cite{Andrianomena:2014sya}, and we verified that the results for the cross-correlation are quantitatively similar to  what one would obtained using the SKA2 source distribution Ref.\,\cite{Harrison:2016stv}.} and with galaxy number counts are presented in Fig.~\ref{cross}.  Note that cross-correlating with galaxy number counts at different redshifts is equivalent to filtering the astrophysical kernel $\mathcal{A}$ with a window function that selects different redshift bins. The shift of the peak of $D_\ell$ to higher multipoles as we consider higher redshift bins can be understood from the Limber relation between redshift and multipoles, as detailed in \S~\ref{general}.  Interestingly, the cross-correlation with galaxy number counts can help to reconstruct the astrophysical kernel $\mathcal{A}$ as a function of redshifts. Furthermore, the study of the cross-correlation with galaxy number counts is useful to distinguish in observations a AGWB from cosmological backgrounds, which are not expected to be correlated with the galaxy distribution.


\subsection{Coarse graining approach}\label{sub2.4}

As can be seen for the previous sections, three main ingredients enter the computation of the angular power spectrum. They are related to the three main building blocks describing the astrophysics of GW sources (${\cal A}$ and ${\cal B}$), the large scale distribution of the sources (through the cosmological variables) and the properties of the cosmological model. We can thus distinguish three scales in the problem:
\footnote{We stress that we refer here to different physical processes that take place on different scales. These processes impact the shape of the power spectrum on all angular scales.}
\begin{enumerate}
\item {\bf{cosmological scale}}: the large scale structure can be effectively described by cosmological scalar perturbations in the metric and the matter distribution; structures are then assumed to move with the cosmic flow;
\item {\bf{galactic scale}}: each galaxy is characterized by a set of parameters $\theta_{\Gal}$ (mass, metallicity...) and an effective GW luminosity resulting from the contributions of the various sources it contains. The galaxy number density is computed from the halo mass function; 
\item {\bf{sub-galactic scale}}: different classes of GW sources (binary compact objects, rotating neutron stars, etc.) are  characterized by parameters (masses, orbital parameters...). Each source emits an energy spectrum which depends on these parameters, in a typical range of frequencies.
\end{enumerate}

\begin{figure*}[!htb]
\begin{center}
\includegraphics[width=.8\textwidth]{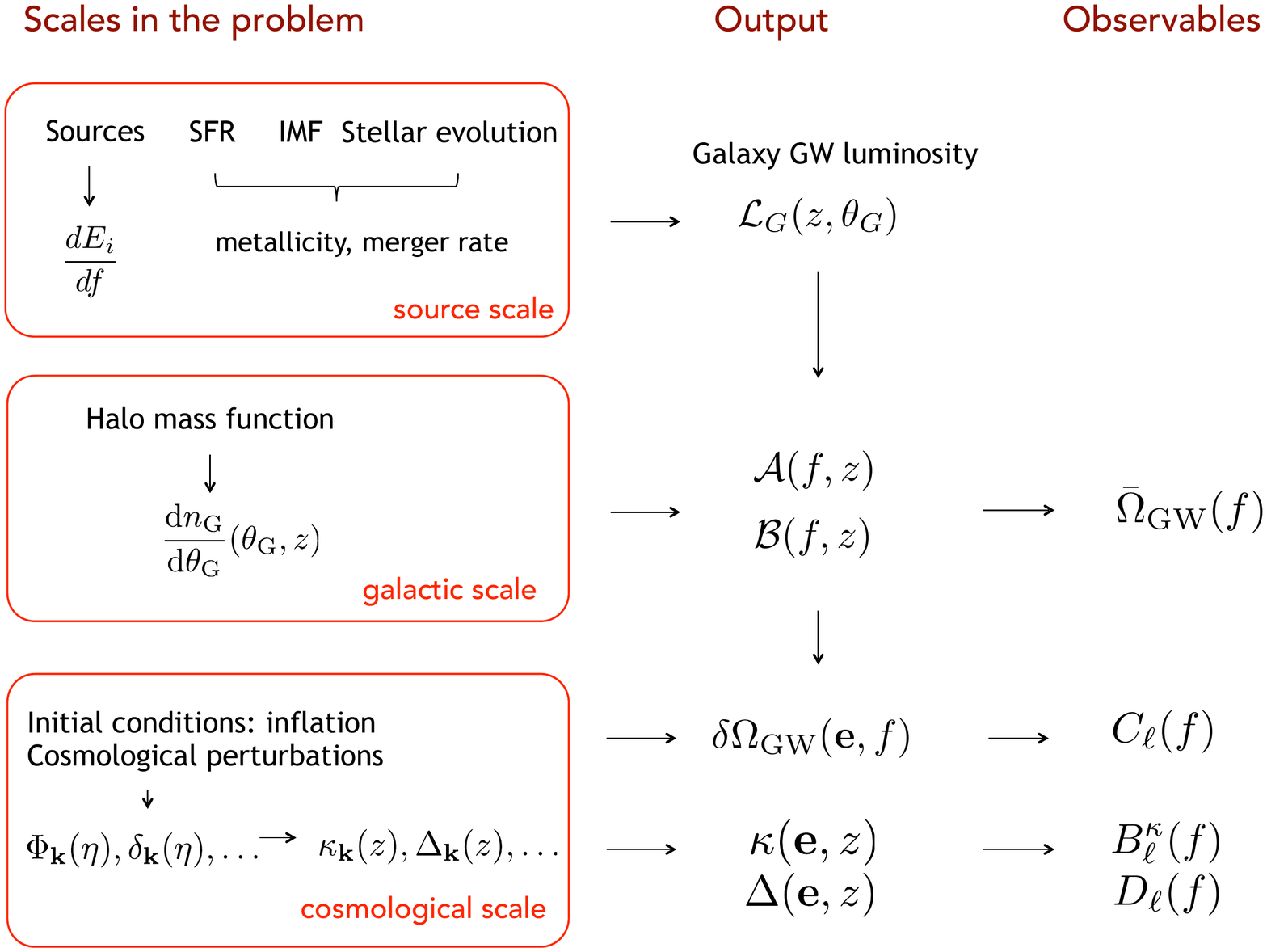}
\end{center}
\caption{The general structure of the computation.}\label{fig_strategy}
\end{figure*}

This approach is schematically summarized in Fig. \ref{fig_strategy}. It illustrates how the effective GW luminosity associated with each galaxy is first obtained from the properties of the galaxy by integrating over all its GW sources and then integrated over the halo mass  distribution to give the astrophysical functions $\mathcal{A}$ and $\mathcal{B}$. These two functions then enter the equation for the background energy density (\ref{background}) and the master equation (\ref{ssss}) for the anisotropies. To finish, anisotropies depend also on the cosmological perturbation variables which need to be evolved during the cosmic history, hence depending on both the initial primordial power spectrum and a set of transfer functions. This last step is nowaday part of the standard lore of cosmology. Putting all these ingredients together, the angular power spectrum can be computed along with the cross-correlation with the various cosmological probes.\\

In the present work, we use the standard $\Lambda$CDM cosmological model in which the universe is described by a Friedmann-Lema\^{\i}tre spacetime with perturbations that describe the large scale structure.  In the theory of cosmological perturbations, any variable, $X(\eta, x^i)$ say, is a stochastic field. It can be decomposed in Fourier modes, $ X(\eta,\bk)$, which can be expressed as the product of a transfer function and of the initial metric perturbation: $ X(\eta,\bk)= X_k(\eta)\Phi^P(\bk)$.  The power spectrum of $\Phi^P(\bk)$ is predicted e.g. from inflation and constrained from CMB analysis. We use Planck satellite~\cite{Ade:2015xua}  cosmological parameters. Linear transfer functions are obtained from {\tt CMBquick}~\cite{CMBquick} and we use Halofit \cite{Smith:2002dz} to account for the non-linearities in the matter power spectrum. 

Figure~\ref{xir} presents the galaxy  correlation function used in this work, at a redshift $z=0.6$. It assumes that galaxies follow the evolution of the underlying dark matter field so that they are related by a bias function. We assume a scale-independent bias model scaling as $\propto \sqrt{1+z}$~\cite{Marin:2013bbb, Rassat:2008ja}. Explicitly,
\be
b(z)=b_0 \sqrt{1+z}\qquad b_0 = 1.5\,.
\ee
At  $z=0.6$ the bias is $1.8$, and one can check that our correlation function is consistent with the one of SDSS VIPER, see Figs.  3 and 4 of Ref.~\cite{2012MNRAS.427.3435A}. Since very small scales do not contribute to the final power spectrum, as we will demonstrate in \S~\ref{general}, the use of a more refined scale-dependent model for the bias would not significantly affect our results.  As already mentioned, the non-linear evolution of the density growth has been taken into account by using the Halofit approximation~\cite{Smith:2002dz}. 

A detailed description of how this work treats the sub-galactic physics is postponed to \S~\ref{sub}.

\begin{figure}
\includegraphics[width=1.\columnwidth]{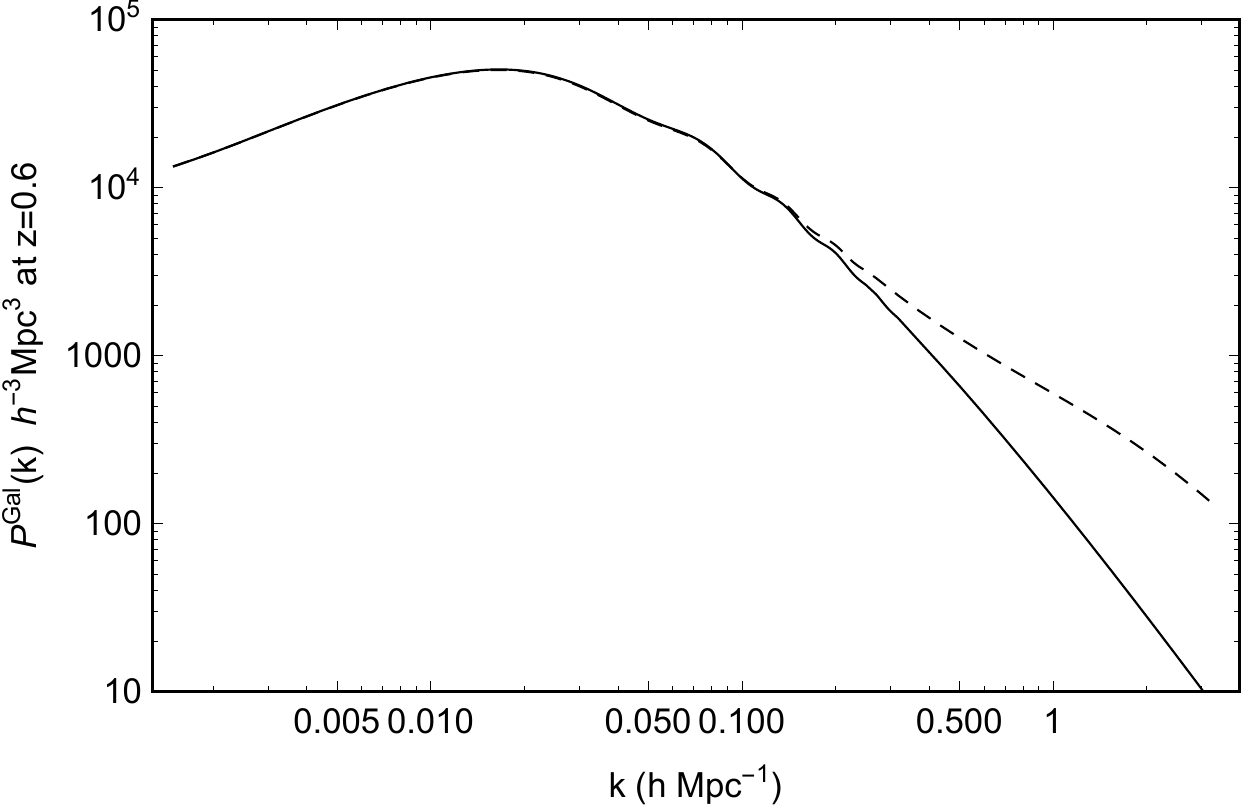}
\caption{Matter power spectrum, linear (solid line) and including non-linearities with Halofit (dashed line).}\label{kPk}
\end{figure}

\begin{figure}[!htb]
\includegraphics[width=\columnwidth]{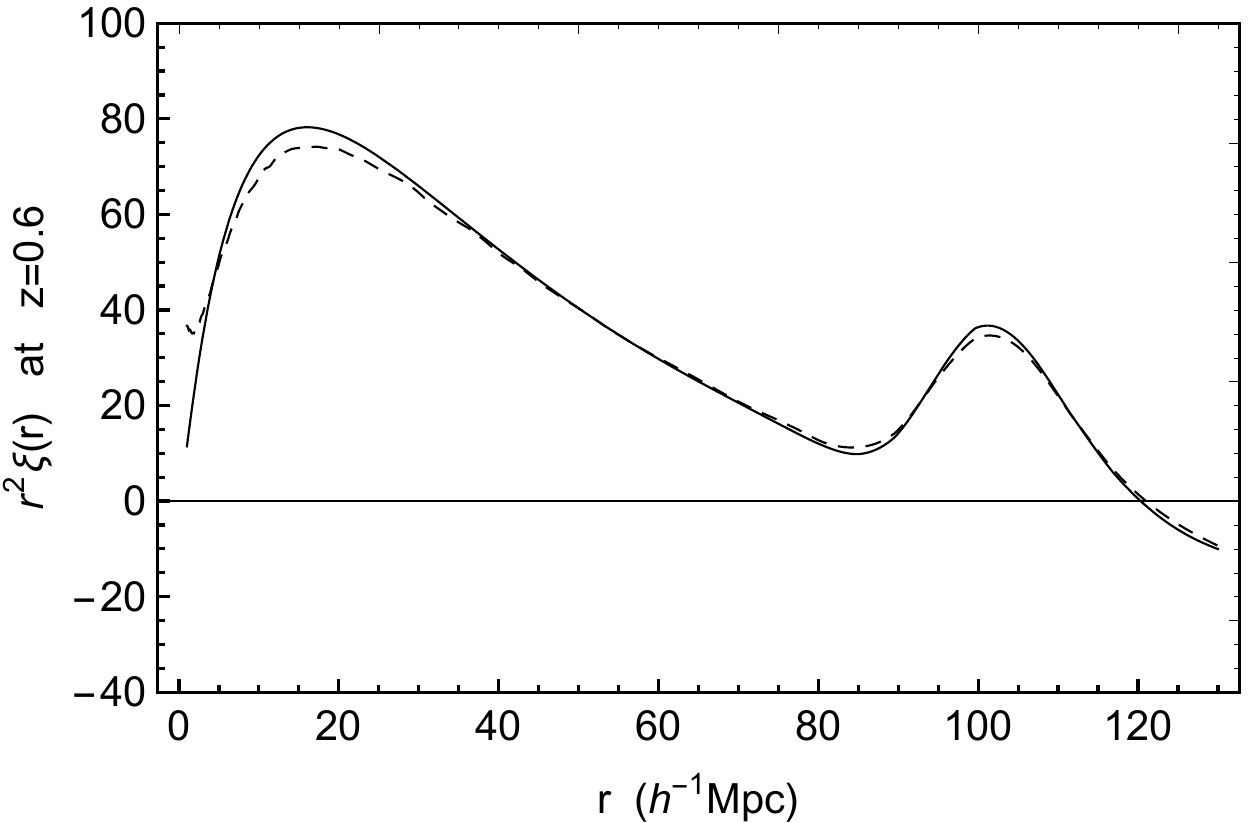}
\caption{Galaxy correlation function at $z=0.6$. The dotted lines include non-linearities described with Halofit.}\label{xir}
\end{figure}

\section{General properties of the angular power spectrum}\label{general}

Before turning to the definition and investigation of the imprint of various astrophysical models, we take some time to describe some general features of the angular power spectrum of  the AGWB anisotropies. Again, in order to illustrate our purpose, we use our reference astrophysical  model  described below in \S~\ref{reference}. The goal of this section is three-fold. First in \S~\ref{sub3.1} we describe an analytic approximation that will allow us to understand our numerics, then in \S~\ref{sub3.2} we discuss the shot noise contribution that will inevitably limit our predictions. To conclude,  in \S~\ref{fac} we investigate to which extent the direction-frequency factorization is a good hypothesis.

\subsection{Analytic approximation of the angular power spectrum}\label{sub3.1}

\begin{figure*}[!htb]
\includegraphics[width=0.9\columnwidth]{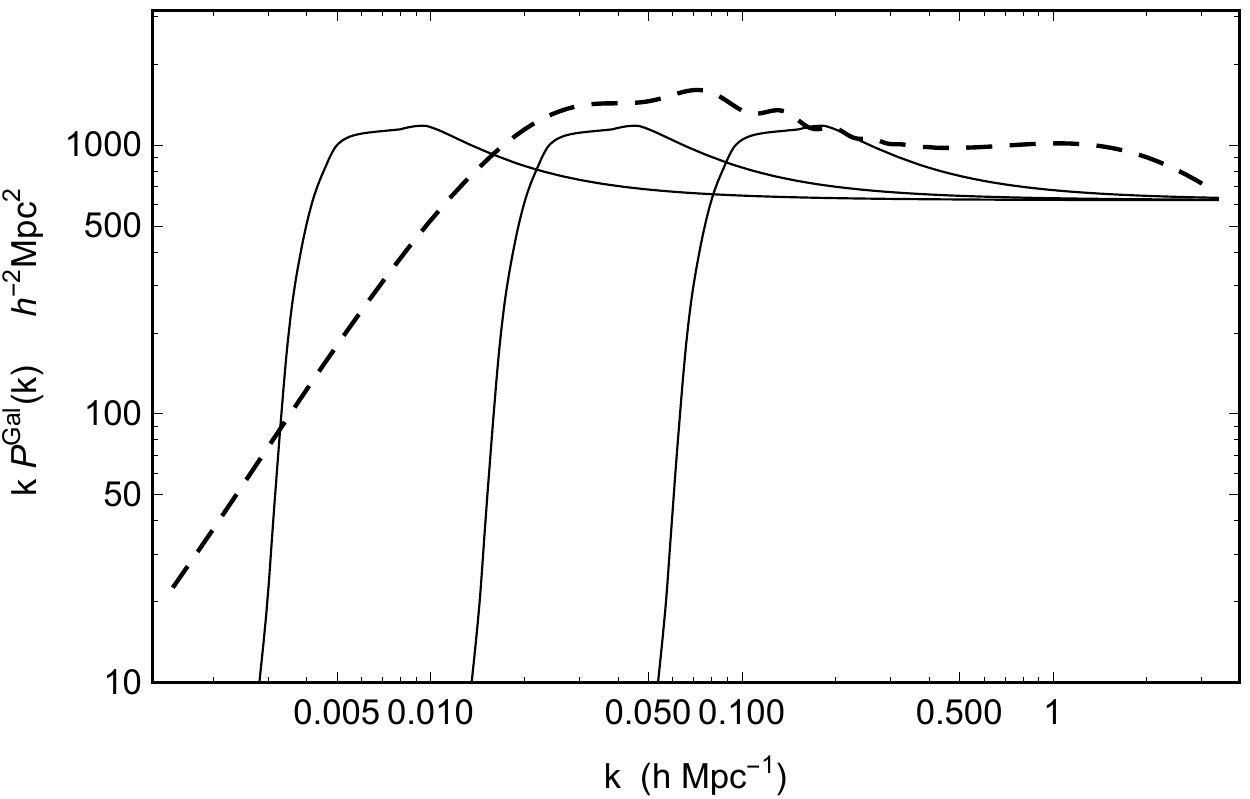}\qquad
\includegraphics[width=0.93\columnwidth]{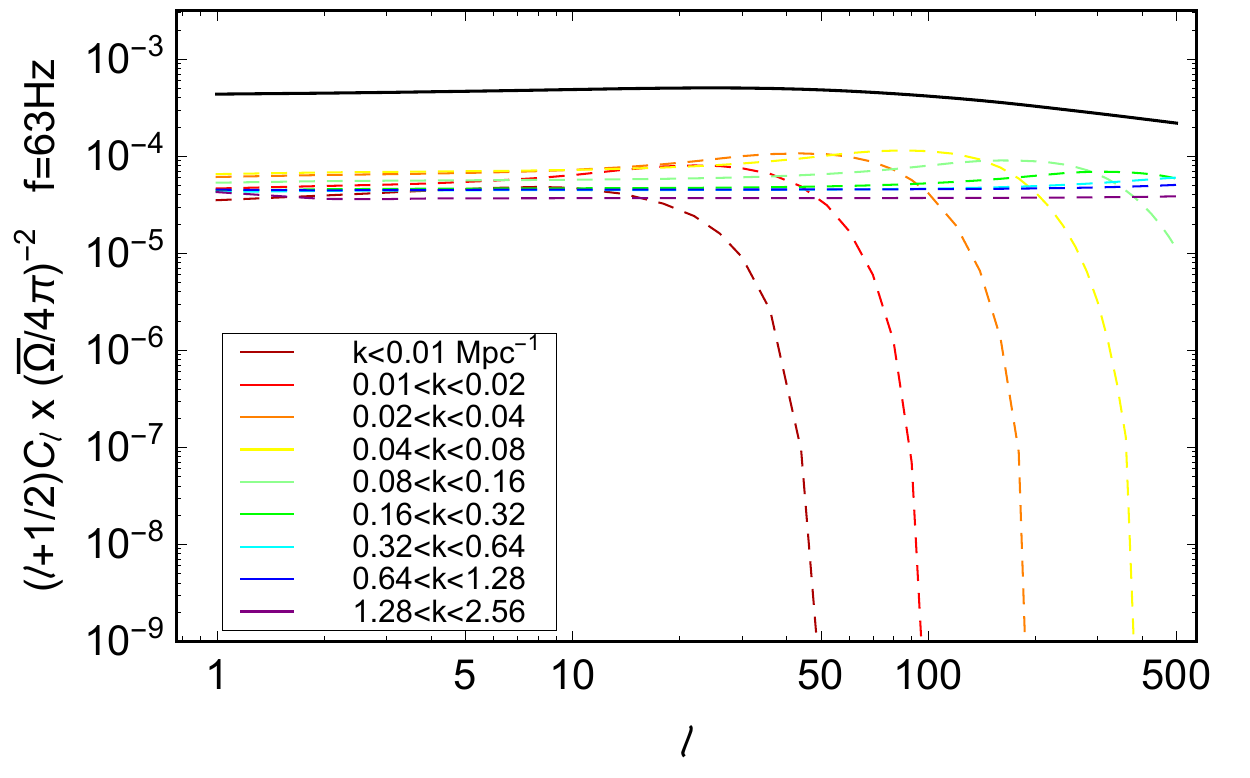}
\caption{{\it Left}: Galaxy power spectrum today (multiplied by $k$) in dashed line and the window function $\sim |\partial_{\eta}\bar{\Omega}_{\rm GW}|^2$ (solid lines) evaluated at time satisfying the constraint (\ref{LimberConstraint}) for $\ell=10$, $\ell=50$ and $\ell=200$ (from left to right respectively). {\it Right}: Angular power spectrum for different bins of $k$ and the
total sum (black line).  The reference astrophysical model used for this plot is defined in \S~\ref{reference}.}
\label{Pschiaccio}
\end{figure*}

Keeping only the dominant contribution in Eq. (\ref{ssss}) and using the expression~(\ref{link}) of the astrophyical kernel, we get 
\be\label{sss4}
\delta\Omega_{\rm GW}(\bee,  f)=\int_{\eta_*}^{\eta_{\obs}} \dd\eta\,\partial_{\eta}\left(\frac{\bar{\Omega}_{\rm GW}}{4\pi}\right)\delta_{G}({\bf{e}}, \eta)\,.
\ee
Hence the angular power spectrum simplifies to 
\be\label{ClStandardMethod}
C_{\ell}\simeq \frac{2}{\pi} \int k^2 \dd k P_{\rm Gal}(k) \left|\int \dd\eta
  \partial_{\eta} \left(\frac{\bar \Omega_{\rm GW}}{4\pi}\right)j_\ell(k \Delta\eta)\right|^2\,, 
\ee
where $P_{\rm Gal}(k)$ is the galaxy power spectrum and $\Delta \eta=\eta_0-\eta$. The Limber approximation~\cite{LoVerde:2008re,PitrouFlat} can then be used to derive the slope of the angular power spectrum. One method consists in noticing that for any test function $f(x)$
\be\label{Limberjl}
\int \dd x j_\ell(x) f(x) \simeq \sqrt{\frac{\pi}{2\ell+1}}f(\ell+1/2)\,.
\ee 
It follows that
\begin{eqnarray}
&C^{\rm Limber}_\ell  \simeq&  \left(\ell+\tfrac{1}{2}\right)^{-1} \times \nn \\
&&\, \int \dd \log k \, k P_{\rm Gal}(k) \left|\partial_{\eta} \left(\frac{\bar  \Omega_{\rm GW}}{4\pi}\right)\right|^2\,,\label{Limberk}
\end{eqnarray}
where $k$ and $\Delta\eta$ must satisfy the Limber constraint
 \be\label{LimberConstraint}
k\, \Delta\eta = \ell +\frac{1}{2}\,\,.
\ee
Note that the integrand function can be thought as the product of $k P_{\rm Gal}$ and of the window function $\left|\partial_{\eta} \bar  \Omega_{\rm GW} (\eta)\right|^2$  evaluated at the value of $\eta$ satisfying the constraint (\ref{LimberConstraint}). In Fig.~\ref{Pschiaccio} we plot the galaxy power spectrum today together with the window functions at different values of $\ell$. It clearly shows that the window function selects different areas below the power spectrum. For sufficiently small $\ell$, the dominant contribution to the integral comes from the peak of the function $k P_{\rm Gal}$, around $k\sim 0.05$ Mpc$^{-1}$ corresponding to length scale of $\sim 120$ Mpc. For larger $\ell$, the peak of the power spectrum is cut out and the integral is dominated by large modes.  

Formally, this is similar to the computation of CMB angular power spectrum. However for the CMB the visibility function is sharply peaked around the recombination, hence selecting (for different multipoles $\ell$) a very narrow region of the power spectrum. For the GW background, the \emph{visibility function} extends typically from $z\simeq 4$ (corresponding to a comoving distance of order $7000$~Mpc) down to $z=0$ as depicted in Fig.~\ref{ANuZ}.  It follows that there is no direct relation between a wavemode $k$ and an angular mode $\ell$. For a fixed observed frequency, a given wavemode contributes to all multipoles such that $\ell \lesssim \ell_{\rm max} (k)$, where $\ell_{\rm max} (k)$ is related to $z_{\rm max}$ (the typical maximum redshift for sources observed at that frequency) by
\be
\ell_{\rm max} \equiv k [\eta_0-\eta(z_{\rm max})]\,.
\ee
For a given multipole $\ell$ and a given wavemode $k$, if the corresponding distance is too large, the number of GW sources is suppressed and so is the $C_\ell$. Fig.~\ref{Pschiaccio} illustrates the contributions of various bins of $k$ to the total signal. It is clear that for each bin in $k$ the contribution scales as $1/\ell$ on large scales, and drops beyond $\ell_{\rm max}$. 

It is useful to introduce a further assumption, namely that the emission depends mildly on redshift. More precisely, using Eq.~(\ref{Limberk}) and assuming that we can ignore the time variation of $\partial_\eta \bar \Omega_{\rm GW}$, we find that the multipoles are approximated by
\begin{align}\label{ClProp}
&C^{\text{Limber+static}}_\ell \propto \frac{1}{\ell+\tfrac{1}{2}}\left|\partial_{\eta} \left(\frac{\bar  \Omega_{\rm GW}}{4\pi}\right)\right|^2_{\eta=\eta_0} \times \nn\\
& \qquad\qquad\qquad\qquad \int_{k_{\rm min}(\ell)} P_{\rm Gal}(k) \dd k\,,
\end{align}
where $k_{\rm min}(\ell)$ is set by the fact that there is a maximum distance $r_{\rm max}$ at which we can find GW sources and thus a minimum Fourier mode set by the Limber constraint~(\ref{LimberConstraint}).  We note that with our galaxy power spectrum which describes correctly the large scales, and thus the small Fourier modes, the proportionality relation
(\ref{ClProp}) is insensitive to $k_{\rm min}$ and one can replace it by $0$. Indeed for $k<k_{\rm eq}$ (with $k_{\rm eq} \simeq 0.01 \,{\rm Mpc}^{-1}$ the Fourier mode entering the horizon at matter-radiation equivalence), $P_{\rm Gal}(k) \propto k^\alpha $ with $\alpha\simeq 1$. In particular we find that  $C_\ell \propto \ell^{-1}$. 

\begin{figure*}[!htb]
\includegraphics[width=0.9\columnwidth]{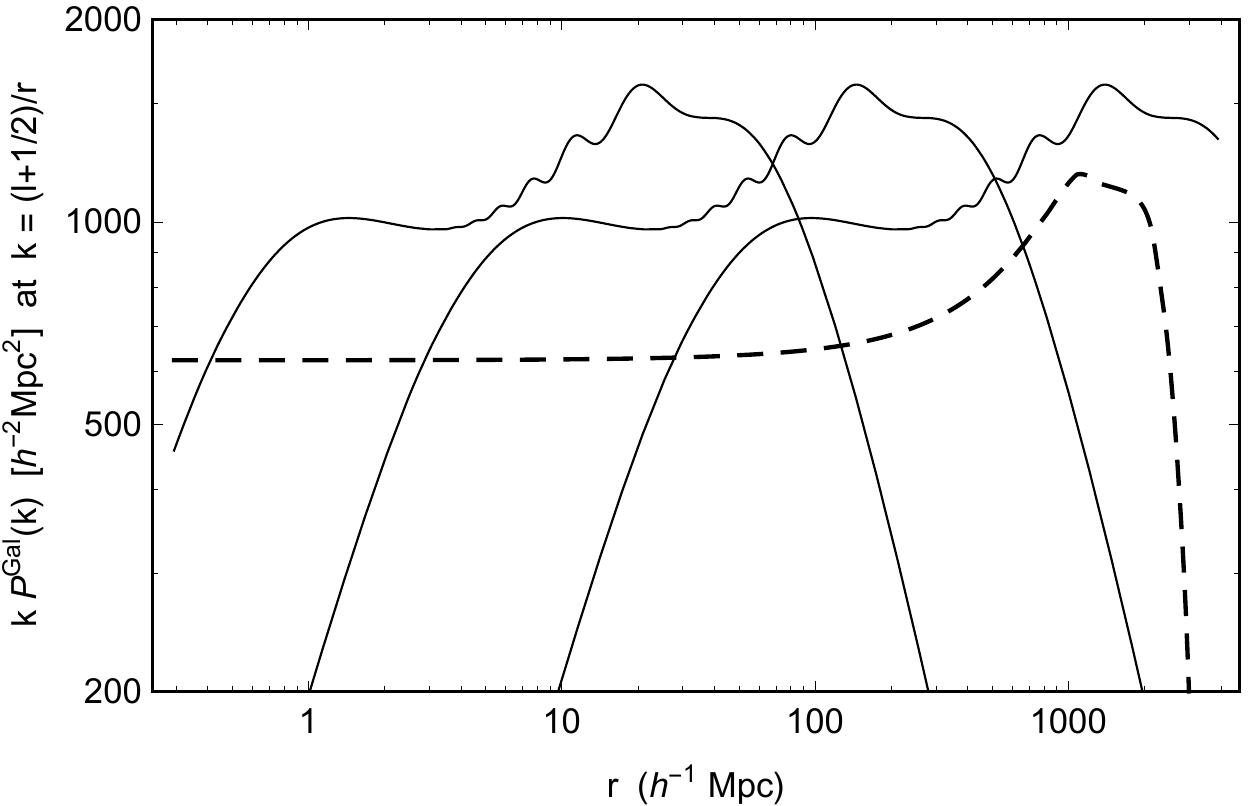}\qquad
\includegraphics[width=0.94\columnwidth]{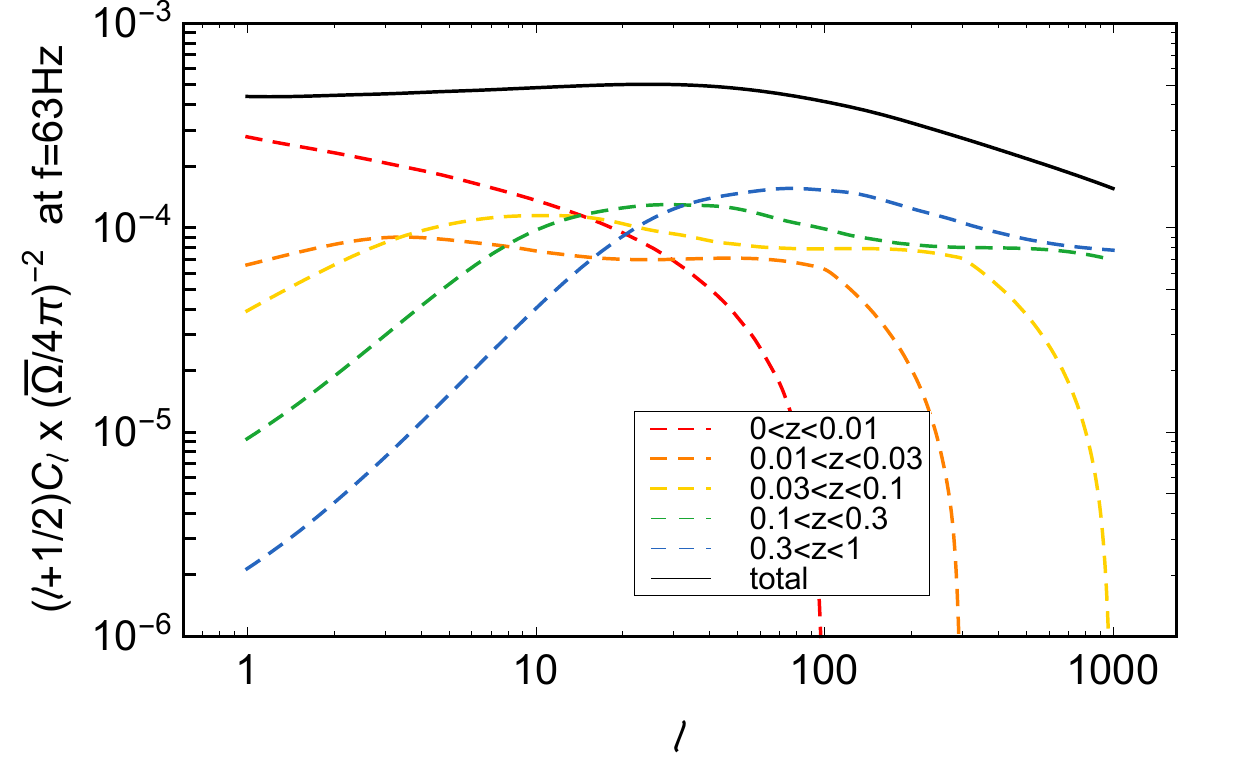}
\caption{{\it Left}: Window function $\sim|\partial_{\eta}\bar{\Omega}_{\rm GW}|^2$ in dashed line and the galaxy power spectrum today (multiplied by $k$) in solid line evaluated at a wavemode $k$ satisfying the constrain $k=(\ell+1/2)/r$ for $\ell=1$, $\ell=10$ and $\ell=100$ (from left to right respectively). {\it Right}: Angular power spectrum for different bins of redshift and their total sum (black line). The reference astrophysical model used for this plot is defined in \S~\ref{reference}.}
\label{Pschiaccio2}
\end{figure*}

An equivalent formulation of the Limber approximation is given by 
\begin{align}
C^{\rm Limber}_\ell \simeq \left(\ell+\tfrac{1}{2}\right)^{-1} \int \dd \log r\, k P_{\rm Gal}(k) \left|\partial_r \left(\frac{\bar  \Omega_{\rm GW} (r)}{4\pi}\right)\right|^2\,,\label{Limberr}
\end{align}
where we have used the fact that the comoving distance and conformal time are related by $r=\Delta \eta$, hence the Limber constraint can be rewritten as  $k\, r = \ell +1/2$.  In this case, the function $\left|\partial_r \bar  \Omega_{\rm GW} (r)\right|^2$ is fixed while the power spectrum has to be evaluated at $k$ satisfying the Limber constraint, for different $\ell$.  The left panel of  Fig.~\ref{Pschiaccio2} shows this fixed window function and the power spectrum evaluated at the Limber constraint for $\ell=1, 10, 100$. 

Figure~\ref{Pschiaccio2}  shows that a given bin in  $r$ contributes more efficiently to multipoles such that  $\ell\sim 1/(r k_{\rm peak})$. This can also be understood by separating the contributions coming from various bins of distance, or equivalently various bins of redshift, as in the right panel of  Fig.~\ref{Pschiaccio2}.  As expected, the lowest redshifts, corresponding to the shortest comoving distances, contribute the most on large scales (small $\ell$).

\subsection{General treatment of shot noise}\label{shot noise}\label{sub3.2}

When working with galaxy data (e.g. with a galaxy catalogue) it is important to keep in mind that the background angular power spectrum suffers a shot noise component which adds to the theoretical predictions,
\be
C^{\text{exp}}_{\ell}=C_{\ell}+S_n\,
\ee
where ``exp'' indicates the angular power spectrum computed from data, $C_{\ell}$ is the theoretical power spectrum (i.e. what we compute in this work) and $S_n$ denotes the shot noise contribution (see e.g. Ref.~\cite{Jenkins:2019uzp}).  This contribution is flat in $\ell$ space and gives a constant offset to the angular power spectrum
\be\label{prediction}
S_n=\frac{1}{(4\pi)^2}\int \dd r \Big|\frac{\partial \bar \Omega_{\rm GW}}{\partial r}\Big|^2 \frac{1}{r^2} \frac{1}{\bar{n}_{\Gal}(r)}\,.
\ee
The derivation of this result can be found in appendix \ref{derivation}. There are two important points to keep in mind: (1) this is an offset. Since the angular power spectrum decreases with $\ell$, it will affect more large $\ell$ and (2)  the prediction (\ref{prediction}) diverges for $r=0$, hence the contribution of the Poisson noise depends on the cut-off used to regularize this integral. From an observational point of view, the physical quantity on which the cut-off has to be set is the observed flux: sources with a flux bigger than a given threshold can be resolved and are therefore filtered out.

Using the fact that the flux $\Phi$ received per unit of frequency from a  source at redshift $z$ is related to the luminosity per unit of emitted frequency by $\Phi(f)=1/(4\pi)\mathcal{L}_{\Gal}/(1+z)$, we see that an upper bound on $\Phi$ defines the region of integration in the plane $(z, \mathcal{L}_{\Gal})$. In other terms, we introduce a selection function which is 1 for $ \Phi<\Phi_{\text{cut}}$, and 0 otherwise, i.e. we multiply the integrand function in Eq.~(\ref{prediction}) by  the selection function 
\be
W(z, \mathcal{L}_{\Gal})=\left\{
\begin{array}{ccc}
1 &\text{for}& \mathcal{L}_{\Gal}<4\pi\Phi_{\text{cut}}(1+z)\,,\\
0&\text{for}&\mathcal{L}_{\Gal}>4\pi\Phi_{\text{cut}}(1+z)\,.
\end{array}
\right.
\ee
Of course, if one can assume that all galaxies have the same luminosity, then the cut-off in the flux translates directly into a lower cut-off in redshift (or analogously in $r$). The same selection function has to be applied to the integral defining the theoretical curve of the angular power spectrum. A detailed derivation of the shot noise component and comments on the regularization procedure can be found in appendix~\ref{derivation}.

We conclude this section with a few remarks. First of all, we observe that when one derives a prediction for the angular power spectrum using a galaxy catalogue, this predictions contains both the "theoretical part" and the shot noise components. Since the angular power spectrum decreases with $\ell$ and the shot noise component is an offset, the high-$\ell$ part of the angular power spectrum will be dominated by the shot-noise part of the result. This explains the shape of the curve of Ref.~\cite{Jenkins:2018uac}, where a simulated galaxy catalogue is used: at high $\ell$, $C_\ell\sim \text{cnst}$ indicating the fact that shot noise dominates over the signal for those multipoles.  Second, the part of the angular power spectrum dominated by Poisson noise also contains astrophysical information so that an adequate understanding and modeling of both the "theoretical" and shot noise components is necessary to extract astrophysical quantities out of future observations. 

\begin{figure}[!htb]
\includegraphics[width=\columnwidth]{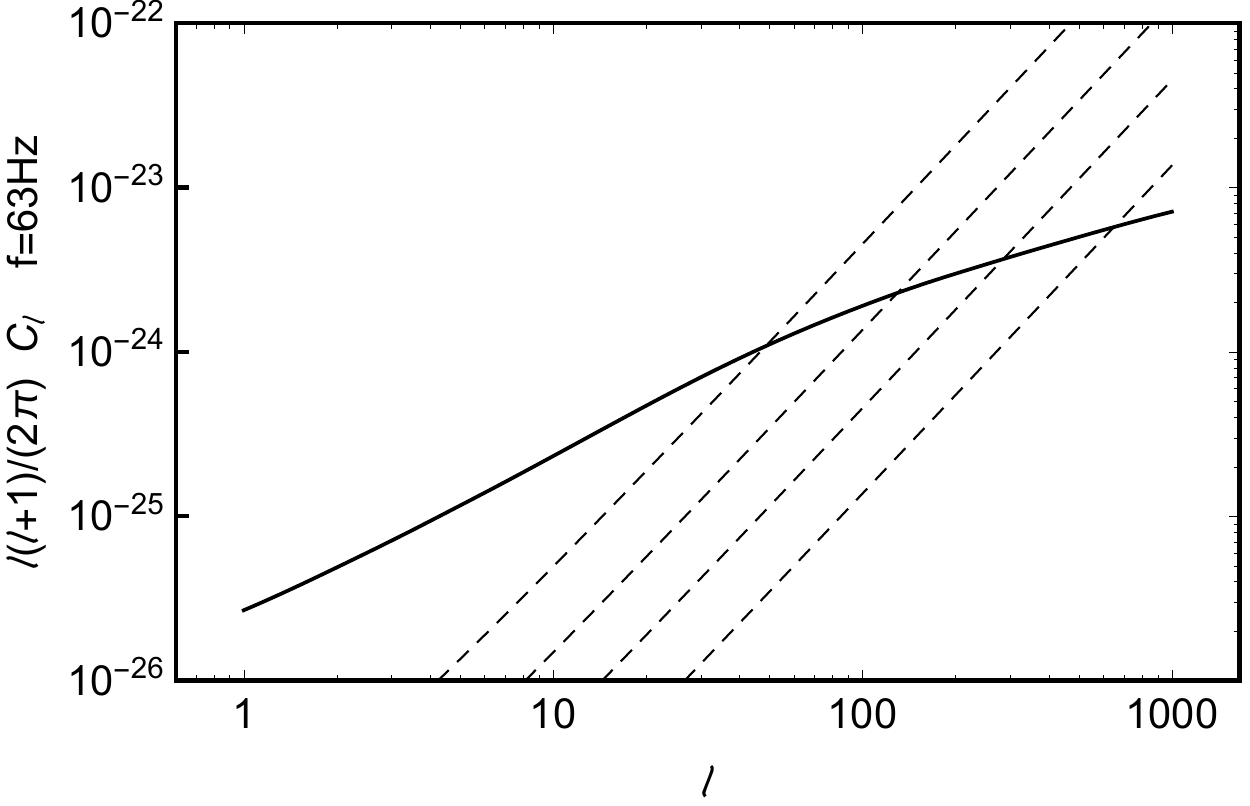}
\includegraphics[width=\columnwidth]{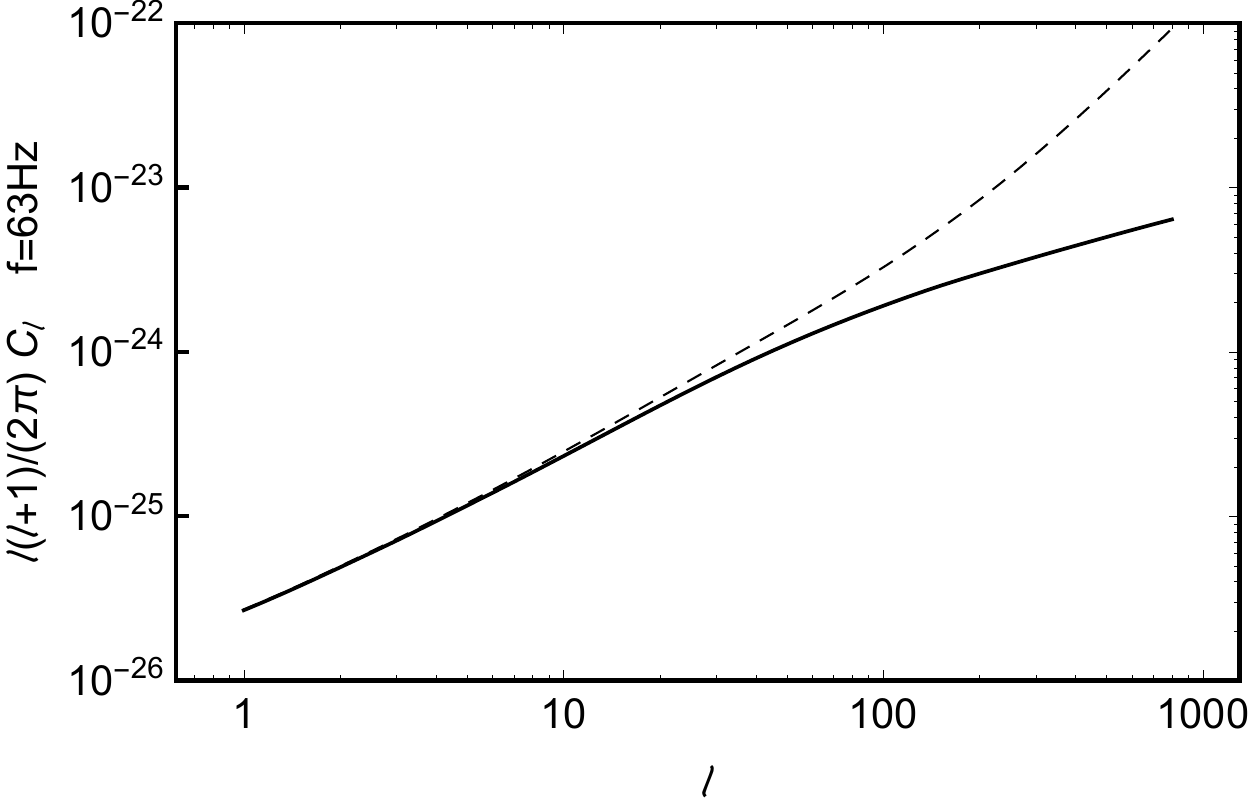}
\caption{ \emph{Top}: Theoretical prediction of the angular power spectrum (continuous line) and Poisson noise for various cut-offs (0.03,0.1,0.3,1) Mpc. \emph{Bottom}: We draw the attention onto the fact that with a cut-off at 0.1 Mpc, we get a curve similar in shape to the one in Jenkins et al.\cite{Jenkins:2018uac}. 
The reference astrophysical model used for this plot is defined in \S~\ref{reference}.}\label{PoissonNoise}
\end{figure}

We emphasize that the shot noise we explore in this section is due to the discreteness of the spatial distribution of galaxies. This shot noise component (that we will refer to as \emph{spatial shot noise}) is always present, and has to be added to the theoretical predictions for the angular power spectrum in any frequency band. It is due to the fact that when deriving theoretical predictions, we treat the galaxy number density as a continuous field. The same type of shot noise is present for the case of the CIB, see e.g.  Ref.~\cite{Ade:2013zsi}.

When the signal is dominated by popcorn-like events (i.e. events with a short duration with respect to the time of integration, and no time overlap), there is a second contribution to shot noise coming from the ``discreetness" of events in time, which has to be added to the spatial shot noise. This second component of the shot noise is present in the LIGO frequency band where BH and NS mergers give the dominant contribution. It is absent in the LISA frequency band, where the background comes from radiation emitted by binary systems of compact objects in the inspiraling phase, and can be treated as a continuous (almost) stationary background.  To take into account in an effective way this time-like shot noise component, one can multiply the factor $n_G$ at the denominator in Eq.~(\ref{prediction}) by the fraction of galaxies which contain a merger in the observation time. 


\subsection{Shot-noise and cross-correlation}

As observed in Ref.~\cite{Jenkins:2019uzp}, in the LIGO-Virgo frequency band the contribution of temporal shot noise to the angular power spectrum of anisotropies is dominating the signal, for observation times of the order of years. We suggest here that using cross-correlation with galaxy number counts can help in overcoming the problem. 
By considering the cross-correlation with galaxy number counts it may be possible to extract anisotropies of the AGWB even if the map of anisotropies is shot-noise dominated. We sketch here the derivation. A future work will be dedicated to a detailed study of the properties of the cross-correlation map. 

We schematically write galaxy number counts and AGWB anisotropies as 
\begin{align}\label{short}
\Delta_N&=\frac{N_G-\bar{N}_G}{\bar{N}_G}\,,\\
\delta \Omega_{GW}&=L (n_G-\bar{n}_G)\,,\label{short2}
\end{align}
where we denote as $N_G$ the number of galaxies in a given position in the sky and as $\bar{N}_G$ its spatial average. $n_G$ is the number of galaxies which contain a merger in the time $T_o$, and $\bar{n}_G=f \bar{N}_G$ where the fraction $f\leq 1$. The quantity  $L$ is a typical galaxy luminosity. We assume that the variable number of galaxies $N_G$ follows a Poisson distribution with average $\bar{N}_G$ and variance $\bar{N}^2_G$. Using basic properties of Poisson distributions it is easy to verify that
\begin{align}\label{Q}
\langle \delta\Omega_{GW} \delta\Omega_{GW}\rangle_P &=L^2 \bar{N}_G f\,,\\
\langle \delta\Omega_{GW} \Delta_{N}\rangle_P &= L f\,\\
\langle \Delta_{N} \Delta_{N}\rangle_P &=1/\bar{N}_G\,,
\end{align}
where $P$ stays for Poisson. \footnote{To derive the expression for the cross-correlation we used the fact that $N_G=n_G+m_G$ where $m_G$ is the number of galaxies which do not contain a merger over the time of observation. The variables $n_G$ and $m_G$ follow independent Poisson distributions with average $\bar{n}_G=f \bar{N}_G$ and  $\bar{m}_G=(1-f) \bar{N}_G$. Then $\langle (n_G-\bar{n}_G)(N_G-\bar{N}_G)\rangle=\langle n_G N_G\rangle-f \bar{N}_G^2=\langle n_G n_G\rangle+\langle n_G\rangle\langle m_G\rangle-f \bar{N}_G^2=Var(n_G)=f \bar{N}_G$.}

  We can compute now the signal part of these cross-correlations. We simply use the fact that we can rewrite Eqs. (\ref{short}) and (\ref{short2}) as $\Delta_N=\delta_G$ and $\delta\Omega_{GW}=L f \bar{N}_G\delta_G$ and we use that  the galaxy over density  $\delta_G$ is a stochastic variable. Then we have
\begin{align}\label{P}
\langle \delta\Omega_{GW} \delta\Omega_{GW}\rangle_S &=L^2 f^2 \bar{N}^2_G \langle \delta_G\delta_G\rangle\,,\\
\langle \delta\Omega_{GW} \Delta_{N}\rangle_S &= L f  \bar{N}_G \langle \delta_G\delta_G\rangle\,\\
\langle \Delta_{N} \Delta_{N}\rangle_S &=\langle \delta_G\delta_G\rangle\,,
\end{align}
where $S$ stays for signal. Then comparing Eqs. (\ref{Q}) and (\ref{P}) we can have a rough estimate of the ratio signal over Poisson-noise for the autocorrelation $\langle \delta\Omega_{GW} \delta\Omega_{GW}\rangle$ and cross correlation $\langle \delta\Omega_{GW} \Delta_N\rangle$
We have
\begin{align}
&\left(\frac{S}{N}\right)_{\text{auto}}\sim f\,,\\
& \left(\frac{S}{N}\right)_{\text{cross}}\sim \text{independent of}\, f\,,
\end{align}
in other words the signal to noise of cross-correlation is boosted with respect to the one of the auto-correlation of a factor $1/f\gg 1$. We stress that this is a simplistic derivation to illustrate the idea of using cross-correlations to overcome the shot-noise problem. A full and realistic study of cross-correlation map will be presented in a  future work.

 \subsection{Frequency-direction factorization}\label{fac}
  
Searches for anisotropies (e.g. at LIGO-Virgo) usually rely on the assumption that the frequency and the direction dependencies of the background energy density can be factorized~\cite{2017PhRvL.118l1102A,2019arXiv190308844T}, i.e. that
  \be\label{factorization}
  \Omega^{\text{fac}}(\bee, f)=\frac{2\pi^2 f^3}{3 H_0^2} H(f) P(\bee)\,,
  \ee
  where
  \be
  H(f)=\left(\frac{f}{f_{\text{ref}}}\right)^{\alpha-3}\,,
  \ee
  with $f_{\text{ref}}$ some reference frequency  and $\alpha=2/3$ for a background from merging compact binaries. The factorization assumption used in the LIGO-Virgo analysis \cite{2019arXiv190308844T} relies on the models in Ref.\cite{Jenkins:2018uac, Jenkins:2018kxc}.
Note that since $H(f)$ is dimensionless in natural units, the quantity $P(\bee)$ in Eq.~(\ref{factorization}) has dimensions of a time. The angular power spectrum of the direction-dependent factor in Eq.~(\ref{factorization}) is then defined using the decomposition
    \be
  P(\bee)=\sum_{\ell m} Y_{\ell m}(\bee) a_{\ell m}^{\text{fac}}\,,
  \ee
as 
  \be
  C_{\ell}^{\text{fac}}=\frac{1}{2\ell+1}\sum_m\langle a_{\ell m}^{\text{fac}}a_{\ell m}^{*\text{fac}}\rangle\,. 
  \ee
  Observational constraints are then set on the dimensionless spectrum
  \be
  C_{\ell}^{\Omega}=\left(\frac{2 \pi^2}{3 H_0^2}\right)^2 f_{\text{ref}}^6 C_{\ell}^{\text{fac}}\,.
  \ee
 It is easy to verify that the relation between this spectrum and the angular power spectrum defined above in Eq.~(\ref{Cell}) is
  \be\label{above}
  C_{\ell}(f)=\left(\frac{f}{f_{\text{ref}}}\right)^{2\alpha} C_{\ell}^{\Omega}\,.
  \ee
  
\begin{figure}
\includegraphics[width=\columnwidth]{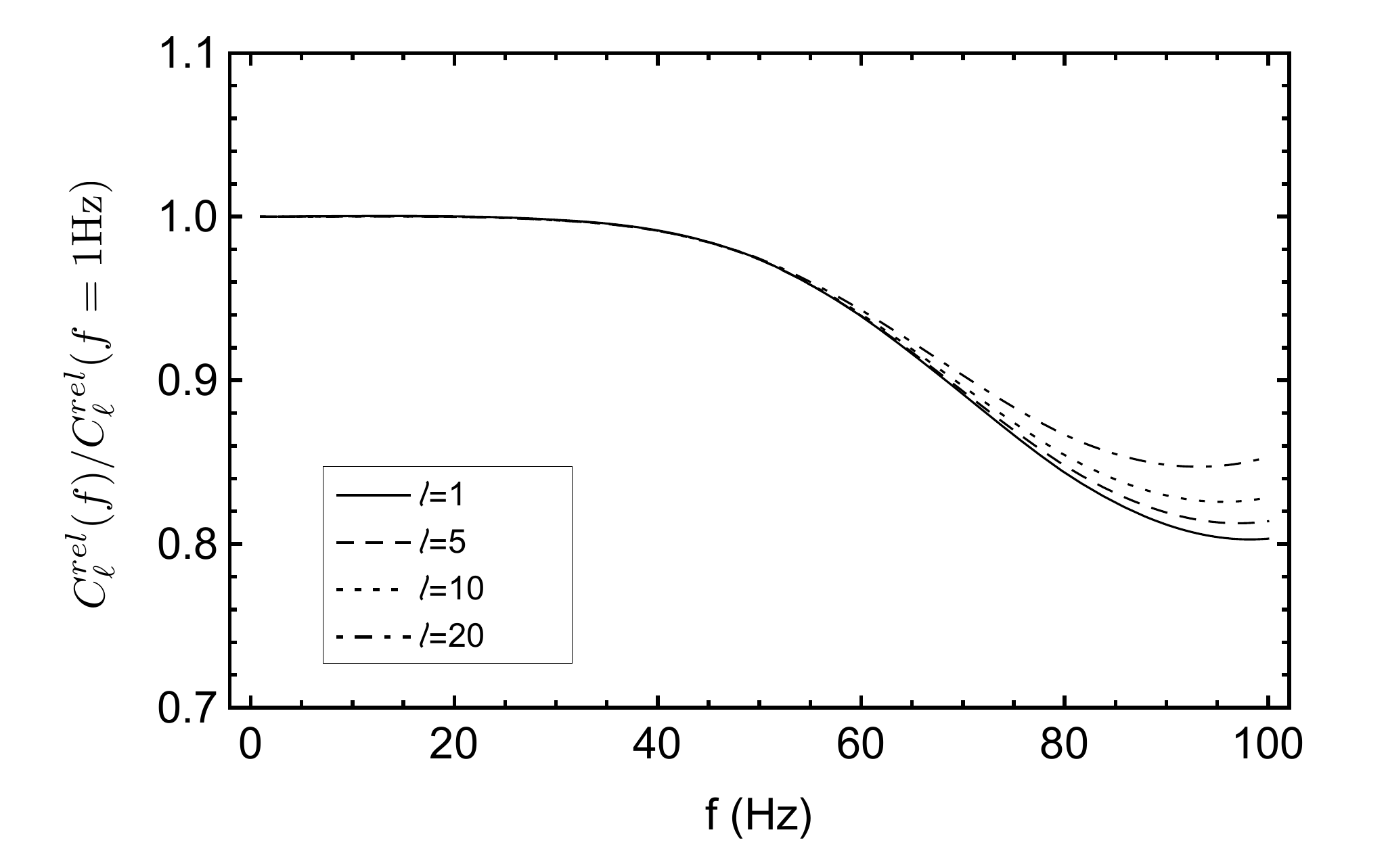}
 \caption{Angular power spectrum as a function of frequency for different multipoles. If the frequency-direction factorization assumption~(\ref{factorization}) were valid, the angular dependence would cancel out when computing relative power spectra. The fact that the relative power spectra normalized over the relative spectra at $f=1$ Hz deviates from 1 for $f>50$ Hz indicates that the factorization hypothesis breaks down in the upper part of the LIGO/Virgo frequency spectrum. The reference astrophysical model used for this plot is defined in \S~\ref{reference}.}\label{freq}
  \end{figure}

Since the angular spectrum on the r.h.s. of Eq.~(\ref{above}) does not depend on frequency while the one on the l.h.s. does, a convenient way to check the consistency of the factorization assumption~(\ref{factorization}) is to verify that the frequency dependence cancels on the r.h.s., i.e. that the angular power spectrum for mergers predicted by our model  scales with frequency as 
 \be
 C_{\ell}(f)\propto f^{2\alpha}\,,
 \ee
with $\alpha=2/3$.  If the factorization hypothesis is valid, the relative fluctuation should not depend on frequency. Figure~\ref{freq} shows that the scaling with frequency is not exactly  a power law.  Analogously, for a fixed value of frequency, for larger $\ell$ the assumption is slightly worse than for small $\ell$. At a frequency of 80 Hz the error one makes assuming a power law is of the order of 20\%. 

\section{Astrophysical models}\label{sub}

We are now in the position to fully investigate different astrophysical scenarios. In \S~\ref{sub4.1} we start by describing the computation of the astrophysical kernel and then discuss the normalisation procedure in \S~\ref{sec:SNR}. We then describe the reference model (\S~\ref{reference}) that was used so far in this article. In \S~\ref{modified} we present a series of modified models to investigate the parameters that have an effect on the angular power spectrum of anisotropies. 

\subsection{Model framework}\label{sub4.1}

The astrophysical sources of GW - such as merging binary BHs and NSs, spinning NSs and supernovae - reside in galaxies and therefore reflect the processes of galactic evolution and star formation. In this work we concentrate on the background from merging binary BHs.

In order to compute the astrophysical kernels $\mathcal{A}$ and $\mathcal{B}$ defined in Eqs.~(\ref{AA}) and (\ref{BB}) we need to compute the GW luminosity $\mathcal{L}_{\Gal}(z, f_{\Gal},  M_{\Gal})$ for each galaxy as a function of its halo mass $M_{\Gal}$ and redshift $z$. We can then sum over the entire galactic population, where the number densities are provided by the halo mass function $\dd n/\dd M_{\Gal} (M_{\Gal},z)$. Our computation follows the formalism we developed in Refs.~\cite{Dvorkin:2016okx,Dvorkin:2016wac,2018MNRAS.479..121D}, which we now briefly describe.

The first step is to calculate the star formation rate (SFR) $\psi(M_{\Gal},t)$, given in units of $M_{\odot}/{\rm yr}$  and the stellar-to-halo mass ratio of a given galaxy using a modified version of the abundance-matching relations of Ref.~\cite{2013ApJ...770...57B}. We use a Salpeter-like initial mass function (IMF)~\cite{1955ApJ...121..161S} to describe the number of stars per unit total stellar mass formed, 
\be\label{salpeter-imf}
\phi=\dd N/\dd M_*\dd M_{\rm tot,*}\propto M_*^{-p}\,, 
\ee
where $M_*$ is the mass of the star at birth.

Having described the total mass of stars formed in each galaxy as a function of time, we also need to model the evolution of massive stars and the nature and mass of their remnants. We assume that the latter depends only on the mass of the progenitor star $M_*$, and on its metallicity $Z$ and is encoded in the function $m=g_s(M_*,Z)$, to be specified for each model. Typically, massive stars ($M_*\gtrsim8M_\odot$) explode as supernovae or collapse to form BH on a timescale of a few Myr. If we assume such short stellar lifetimes, the stellar metallicity tracks the metallicity of the interstellar medium (ISM) given by $Z=Z(M_{\Gal},z)$. We adopt the observational relation of Ref.~\cite{2016MNRAS.456.2140M} for the ISM metallicity as a function of galaxy stellar mass and redshift. We also introduce a cut-off at high BH masses $M_{\rm co}$ which can arise due to pair-instability supernovae, as explained below.

Under these assumptions, the instantaneous BH formation rate at a given cosmic time $t$ (or, equivalently, redshift $z$)  for a galaxy with halo mass $M_{\Gal}$, in units of events per unit BH mass $m$, is given by
\be\label{defR1}
{\cal R}_1(m,t)=\psi[M_{\Gal},t] \phi(M_*)\times \dd M_*/\dd m
\ee 
where $M_*(m)$ and $\dd M_*/\dd m$ are deduced from the relation $m=g_s(M_*,Z)$ (we  assume negligible stellar lifetimes). We then assume that only a fraction $\beta$ of these BHs resides in binary systems that merge within the age of the Universe, so that the rate of formation of the latter is 
\be\label{defR2}
{\cal R}_2(m,t)=\beta {\cal R}_1(m,t)\,.
\ee
As we will show later, this overall factor $\beta$ is used to normalize our model with respect to the number of events observed by aLIGO/aVirgo. Following Ref.~\cite{2018MNRAS.479..121D}, the birth rate of binaries with component masses $(m,m' \leq m)$ is 
\be\label{defRbin}
{\cal R}_{\rm bin}(m,m')= {\cal R}_2(m){\cal R}_2(m')P(m,m')
\ee 
where the distribution function of binary masses $P(m, m')$ is normalized so that $\int {\cal R}_2(m){\cal R}_2(m')P(m,m')\dd m \dd m'=0.5\int {\cal R}_2(m) \dd m$. 

The merger rate depends on the time to coalescence of the binaries which can be expressed as a function of the distribution of the orbital parameters $P(a_{\rm f},e_{\rm f})$ at the time of formation. Since BH binaries are expected to circularize due to gravitational wave radiation reaction before reaching the LIGO/Virgo band (see e.g. Ref.~\cite{2018PhRvD..98h3028L}), we assume circular orbits in what follows, so that the only distribution left to determine is $f(a_{\rm f})$.  Hence, the birth rate of BH binaries (per unit mass squared per unit time and per unit $a_{\rm f}$) is
\be\label{defRf}
\mathcal{R}_f[m,m', a_{\rm f}, t]={\cal R}_{\rm bin}(m,m')f(a_{\rm f})
\ee
from which we deduce that the merger rate at time $t$ is 
\be\label{defRm}
\mathcal{R}_{\m}[m, m', a_{\rm f}, t]= \mathcal{R}_f[m, m', a_{\rm f}, t-\tau_{\rm m}(m, m', a_f)]
\ee
with $\tau_{\rm m}(m, m', a_{\rm f})$ the merger time of the system $(m,m', a_{\rm f})$. 

The GW luminosity of the galaxy is then 
\be
 \mathcal{L}_{\Gal}=\int \dd m\, \dd m'\, \dd a_{\rm f}\, \frac{\dd E}{\dd f}\times \mathcal{R}_{\m}[m, m', a_{\rm f}, t]\,.
\ee
We then need to sum over the entire galactic population as in Eq.~(\ref{AA}), where the integral over the galactic properties $\theta_{\Gal}$ reduces to an integration over $M_{\Gal}$ weighted by the halo mass function given in Ref.~\cite{2008ApJ...688..709T}. The result is the quantity $\mathcal{A}$ presented on Fig.~\ref{ANuZ} (for the reference model defined in \S~\ref{reference} below).

\subsection{Normalization to the number of detected events}\label{sec:SNR}

The overall normalization parameter $\beta$ is adjusted so as to match the number of detections by aLIGO/aVirgo during the O1+O2 observing runs \cite{2018arXiv181112907T}. We therefore require that all our models result in $10$ detectable events over the span of the O1+O2 observation time. In our estimate of the detection rates we follow Ref.~\cite{2018MNRAS.479..121D}, namely we calculate the signal-to-noise rate (SNR) for each binary BH merger produced in the model:
\begin{equation}
 \rho^2=4\sum_i \int \frac{|h(f)|^2}{S_{n,i}(f)}\dd f
 \label{eq:SNR}
\end{equation}
where the index $i$ refers to either Hanford or Livingston detector, $h(f)$ is the GW strain in the observed frequency domain and $S_{n,i}(f)$ are the O2 noise power spectral densities of the corresponding detectors \cite{LIGOcurves1,LIGOcurves2} and we use the correction factor in Ref.~\cite{1993PhRvD..47.2198F} to account for different source orientations. The strain $h(f)$ is calculated using the PhenomB template \cite{2011PhRvL.106x1101A} and assuming zero spins. We define observed events as those with $\rho>8$. The number of sources detectable during O1+O2 is given by multiplying the detection rate by the total observation time $T_{\rm obs}=169.7$ days. 

The resulting values of $\beta$ for each of the models discussed below is shown in Table \ref{tab:beta}. This already shows that while all the astrophysical models are adjusted to predict $10$ detected events during the aLIGO O1+O2 timespan, they predict that the fraction of BH in binaries can range from 0.6\% to 17\%. It can be anticipated that this differences will imprint the AGWB. Also note that while individual mergers are resolved only at low $z$, the AGWB is affected also by higher redshifts. Therefore, even though all models are calibrated to the same number of resolved sources, the resulting AGWB may vary if the high-redshift population of sources differs among the models.

\begin{table}
\begin{center}
\begin{tabular}{c||c|c|c|c|c|c|c}
 \textbf{Model}    &  \emph{Ref}  & \emph{Limongi} & \emph{imf-low} & \emph{imf-hi} & \emph{dMco} & \emph{uMco} & \emph{aconst} \\
\hline
$\beta$ & $0.01$ & $0.013$ & $0.004$ & $0.032$ & $0.006$ & $0.023$ & $0.17$ \\
\hline
\end{tabular}
\caption{\label{bella} Values of the normalization parameter $\beta$ for the different models described in \S~\ref{reference} and~\ref{modified}. The normalization was determined in order  to obtain $10$ detected events during the aLIGO O1+O2 timespan for all of the models.}
\end{center}
\label{tab:beta}
\end{table}

\subsection{Reference  model}\label{reference}

We start by describing the parameter choices we made for our Reference model, used so far in this article. 
\begin{enumerate}
\item The IMF slope for this model is set to $p=2.35$.
\item The BH formation model was chosen as the `delayed' model in Ref.~\cite{2012ApJ...749...91F}. Specifically, we used the functional form provided by Ref.~\cite{2012ApJ...749...91F} to calculate the function $m=g_s(M_*,Z)$. This choice affects the distribution of BH masses and will be discussed in Sec.~\ref{sec:bhmodel}. We also introduced a cutoff mass of $M_{\rm co}=45M_{\odot}$ in the BH mass distribution. The recent analysis~\cite{2018arXiv181112940T} of the population of BHs detected by aLIGO/aVirgo suggests a cutoff at this value, although further observations are needed to confirm it. We will discuss the possible causes of this cutoff in Sec.~\ref{sec:PISN}.
\item We assume $P(m, m')=$~cnst for the distribution of masses in the binaries.
\item We assume that the distribution of the semi-major axis \emph{at formation} is $f(a_{\rm f})\propto a_{\rm f}^{-1}$ with cut-off at $a_{\rm f,min}=0.014$ AU and $a_{\rm f,max}=4000$ AU. The lower bound was chosen so as to ensure that the lightest BH binaries in our model ($5M_{\odot}-5M_{\odot}$) merge within a Hubble time. 
\end{enumerate}

This reference model predicts that the merger rate observed by LIGO/Virgo can be explained provided $\sim 1\%$ of BHs reside in binaries that merge within the Hubble time. We stress that this estimate relies on the assumption of isolated stellar evolution, i.e. that BH formation is not influenced by binary interactions. This may not be the case of close stellar binaries, where mass exchange processes and in particular the co-evolution of the primary compact object and its stellar companion during the common envelope phase may play a major role in the later stages of the secondary evolution~\cite{2016Natur.534..512B,2012ApJ...759...52D,2016A&A...594A..97B}. These processes and their effects on the stochastic background will be further explored in future work.

\subsection{Modified  models}\label{modified}

In order to explore the astrophysical dependencies of the AGWB anisotropies, we consider several models, each varying from the reference model described above in one key aspect, keeping the others fixed. In addition to varying the corresponding parameter, we also need to change the overall efficiency factor $\beta$, as explained above, so that \emph{all} of the models discussed here results in \emph{the same total number of detectable events}. 

\subsubsection{BH formation model \emph{[\emph{limongi} model]}\label{sec:bhmodel}}

BH masses depend on the properties of their stellar progenitors, in particular the mass prior to core collapse and chemical composition, as well as other parameters such as the rotation velocity, see e.g. Refs.~\cite{2011ApJ...730...70O,2016MNRAS.458.2634M,2016A&A...588A..50M,2017hsn..book..513L}. The formation of \emph{binaries} may further depend on such processes as common envelope evolution, dynamical processes in stellar clusters and evolution of hierarchical triple systems, see e.g. Refs.~\cite{2016Natur.534..512B,2012ApJ...759...52D,2016A&A...594A..97B,2016PhRvD..93h4029R}. 

In this article we assume that binary formation process is encoded in the efficiency parameter $\beta$ and the distribution of merger time delays. Furthermore, we explore only one aspect of this complex problem, namely the evolution of isolated massive stars. Recent studies~\cite{2001ApJ...554..548F,2012ApJ...749...91F,2003ApJ...584..971B} suggest that the explosion is powered by neutrinos stored behind the shock and that the explosion energy depends on neutrino heat transport mechanisms, the nature of the hydrodynamic instabilities that convert neutrino thermal energy into kinetic energy that can power the supernova, and the resulting time delay between shock bounce and explosion. 

The Reference model uses the description by Ref.~\cite{2012ApJ...749...91F} which provides an analytic model for a neutrino-driven explosion and calculate the explosion energy, as well as the remnant mass, using numerical pre-collapse stellar models from Ref.~\cite{2002RvMP...74.1015W}. 

Another set of stellar evolution models is provided in Ref.~\cite{2017hsn..book..513L}. These models differ from the ones in Ref.~\cite{2012ApJ...749...91F} in two aspects. First, Ref.~\cite{2017hsn..book..513L} uses a different set of pre-collapse stellar models which vary from \cite{2002RvMP...74.1015W}  in their treatment of convection, mass-loss rate and angular momentum transport. Second, \cite{2017hsn..book..513L} assumed a constant explosion energy in the calculation of the remnant mass, contrary to \cite{2012ApJ...749...91F}. As was shown in \cite{2018MNRAS.479..121D}, these models predict different mass distributions of detectable BHs. In the following, the model \emph{limongi} uses the model described in  Ref.~\cite{2017hsn..book..513L} without stellar rotation.

The parameter $\beta$ derived for this model is very similar to the one in the Reference model, but the mass distribution of the BHs is different, and will affect the resulting AGWB, as we will show below.

\subsubsection{BH mass cutoff due to PISN \emph{[\emph{dMco} and \emph{uMco} models]}} \label{sec:PISN}

Very massive stars (typically in the range $[130-260] M_{\odot}$) are unstable to electron-positron pair creation which may lead to pair-instability supernova (PISN) that disrupt the entire star. In this case, no BH is formed~\cite{2011ApJ...734..102K}. Although direct observational evidence is lacking, the absence of BHs in the mass range $[60-260] M_{\odot}$ may provide an indirect confirmation of this effect. A cutoff in BH mass may be present at even lower masses due to pulsational PISN, where the instability causes short episodes of mass ejection followed by periods of quiescent evolution~\cite{2007Natur.450..390W,2015ASSL..412..199W}. As a result, the stellar mass is reduced below the limit of the onset of the instability, and it was suggested in Ref.~\cite{2018ApJ...856..173T} that this process may lead to an excess of BHs around $\sim 40M_{\odot}$. Recent analysis of the aLIGO/aVirgo events detected during O1+O2 observational runs~\cite{2018arXiv181112940T} provides a tentative measurement of the BH mass cutoff at $M_{\rm co}=45M_{\odot}$ which can be due to PISN . As we will show in what follows, this mass cutoff has an important effect both on the isotropic and anisotorpic stochastic background, suggesting an alternative way to measure this effect. 

In order to explore the sensitivity of the stochastic background to the PISN-induced mass cutoff we varied to  $M_{\rm co}=40M_{\odot}$ [\emph{dMco} model] and  $M_{\rm co}=50M_{\odot}$ [\emph{uMco} model].


\begin{figure*}[!htb]
\includegraphics[width=\columnwidth]{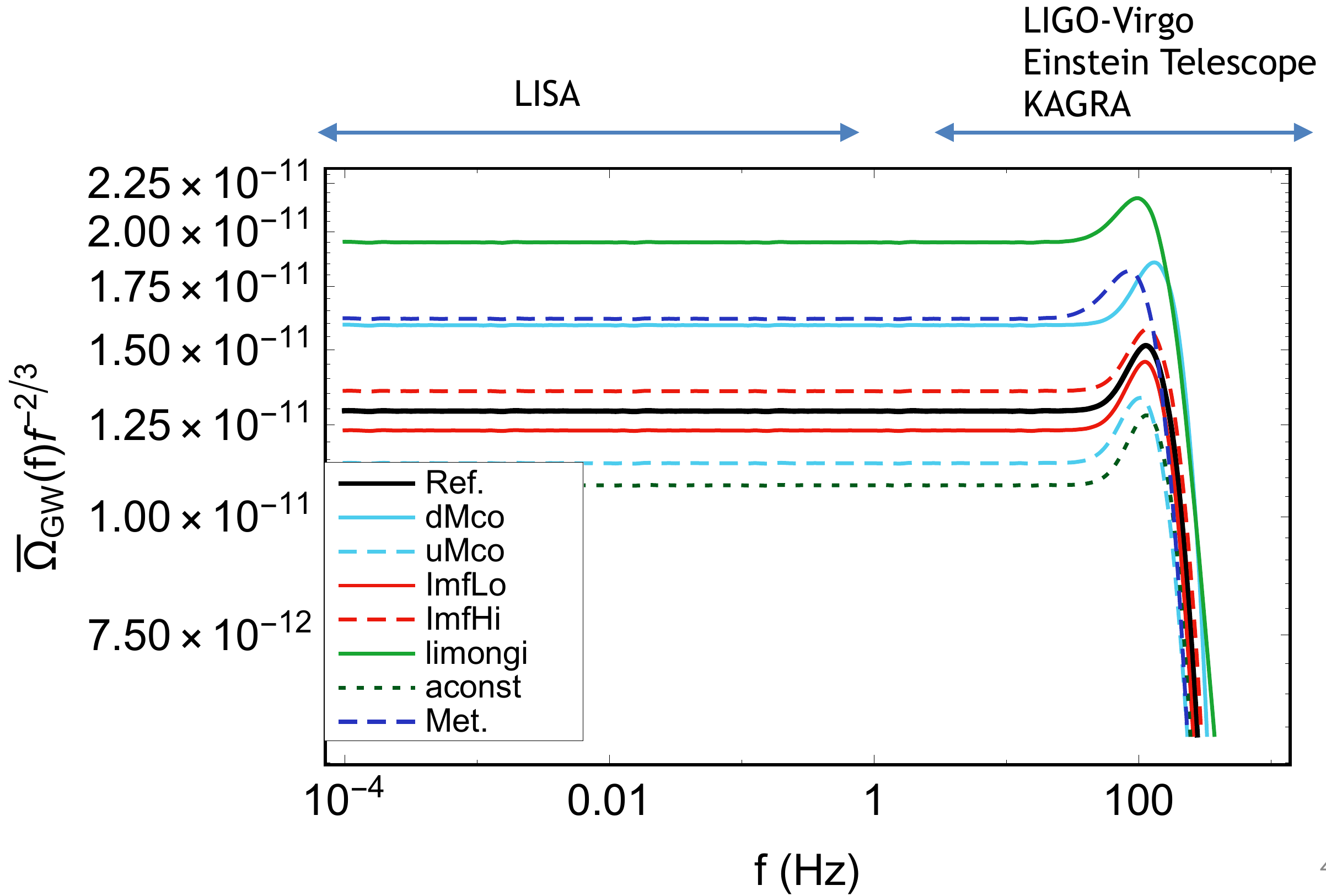}\quad
\includegraphics[width=.92\columnwidth]{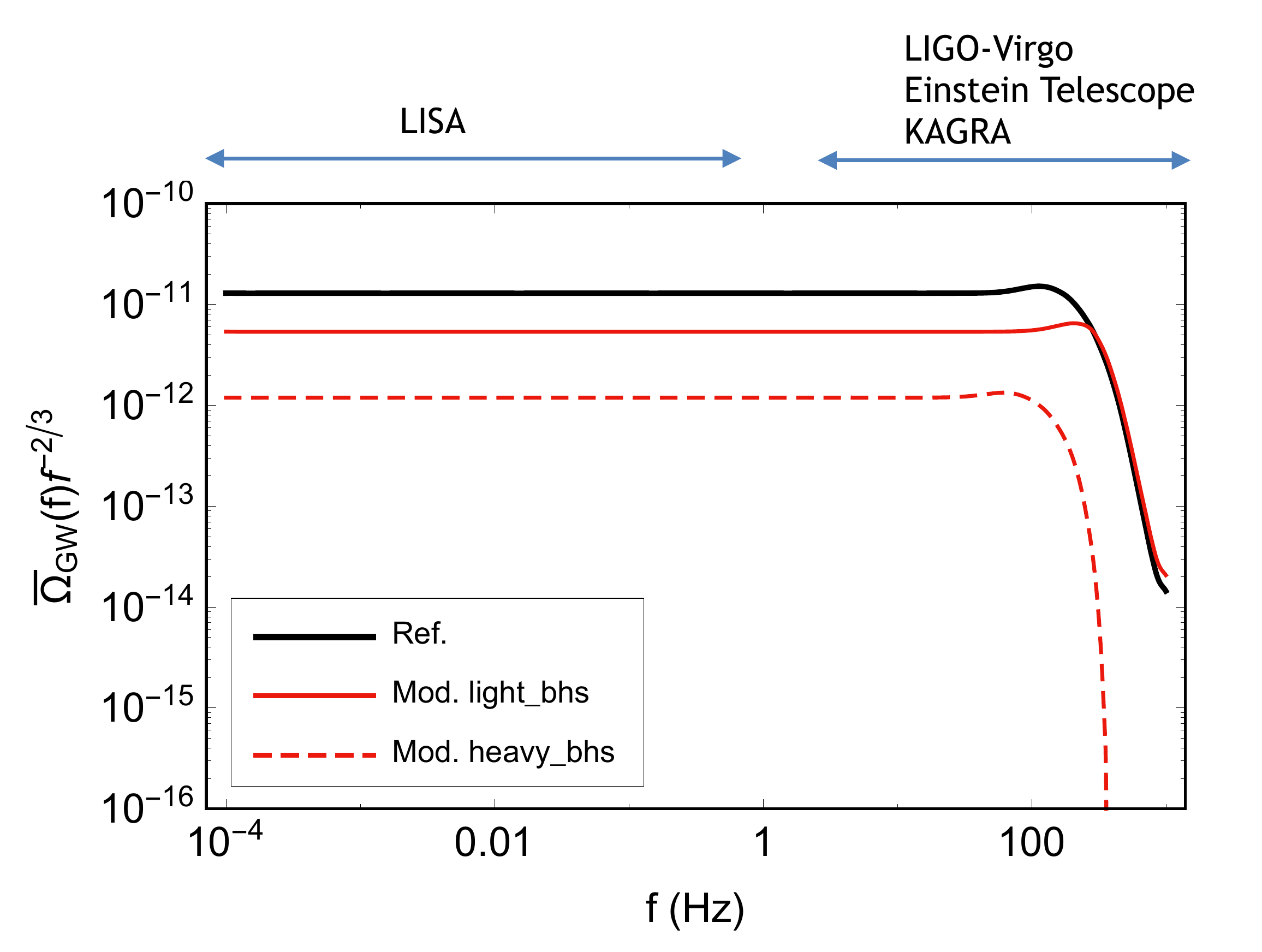}
\caption{Monopole of the energy density of the AGWB multiplied by the power-law $f^{-2/3}$ as a function of frequency, for the 9 astrophysical models described in section~\ref{sub} (Left panel) and comparison of two different BH sub-populations derived from the reference model (Right panel).}\label{AllOmega}
\end{figure*}

\subsubsection{Stellar initial mass function \emph{[\emph{imf-high} and \emph{imf-low} models]}}

The reference model assumes a Salpeter-like IMF with slope $p=2.35$. Interestingly, some studies show that the IMF slope may not be universal (see e.g. the discussion in Ref.~\cite{2014PhR...539...49K}), for example a recent a hint to a more shallow IMF in the Large Magellanic Cloud \cite{2018A&A...618A..73S}. In order to estimate the influence of the IMF we explored two models where the slope was taken to be $p=2.6$ [\emph{imf-high} model] and $p=2.1$ [\emph{imf-low} model]. 

\subsubsection{Distribution of initial separations \emph{[\emph{aconst} model]}}

The reference model assumes the initial separation of the BHs is distributed like $P(a)\propto a^{-1}$. This separation then translates into a distribution of merger delay times, favoring short delay times. We consider the extreme scenario [\emph{aconst} model] of a flat distribution of the initial separations $P(a)\sim const$, which results in longer delay times. 

\subsubsection{Metallicity of progenitor stars}

Metallicity plays an important role in the evolution of massive stars, in particular, high-metallicity stars experience strong winds throughout their lives. As a result, the remnant mass is reduced relative to the low-metallicity case \cite{2010ApJ...715L.138B,2012ApJ...749...91F,2015MNRAS.451.4086S}. In the reference model the metallicity is evolved with the stellar mass following the observational relations of Ref.~\cite{2016MNRAS.456.2140M}. To test the effect of metallicity on the stochastic background, we used a model with constant metallicity of $Z=10^{-3}Z_{\odot}$, which leads to the formation of heavier BHs.

%
%
%

\section{Results}\label{exploration}

We have now defined all the quantities that are required to go through the general computation described in Fig.~\ref{fig_strategy} to compute the monopole (\S~\ref{sub5.1}), the power spectrum (\S~\ref{sub5.2}), and the cross-correlations (\S~\ref{sub5.3}) for the different astrophysical models.

\subsection{Monopole}\label{sub5.1}

Figure~\ref{AllOmega} shows the AGWB monopole as a function of frequency for the various models described in the previous section, for both the LISA and LIGO/Virgo frequency bands. The left panel depicts the results for the 9 individual models discussed above (the AGWB is multiplied by a power-law $f^{-2/3}$ to accentuate the differences in amplitude).  The amplitude varies by a factor of $\sim 2$ between the different models, reflecting the differences in the normalization $\beta$ and source mass distribution.

The right panel of Fig.~\ref{AllOmega} compares the reference model (black) to the contributions of low-mass $<25M_{\odot}$ and high-mass $>25M_{\odot}$ BH populations in red. Both populations are derived from the reference model, where we took into account only the low-mass (high-mass) BHs, respectively, and neither of these models includes 'mixed' (low-mass/high-mass) binaries. As a result, the two red curves do not sum up to the reference model. It can be seen that the AGWB signal is dominated by the high-mass binaries, even though they are outnumbered by the low-mass ones as a result of the power-law stellar mass function. 


\subsection{Angular power spectrum}\label{sub5.2}

\begin{figure*}[!htb]
\includegraphics[width=\columnwidth]{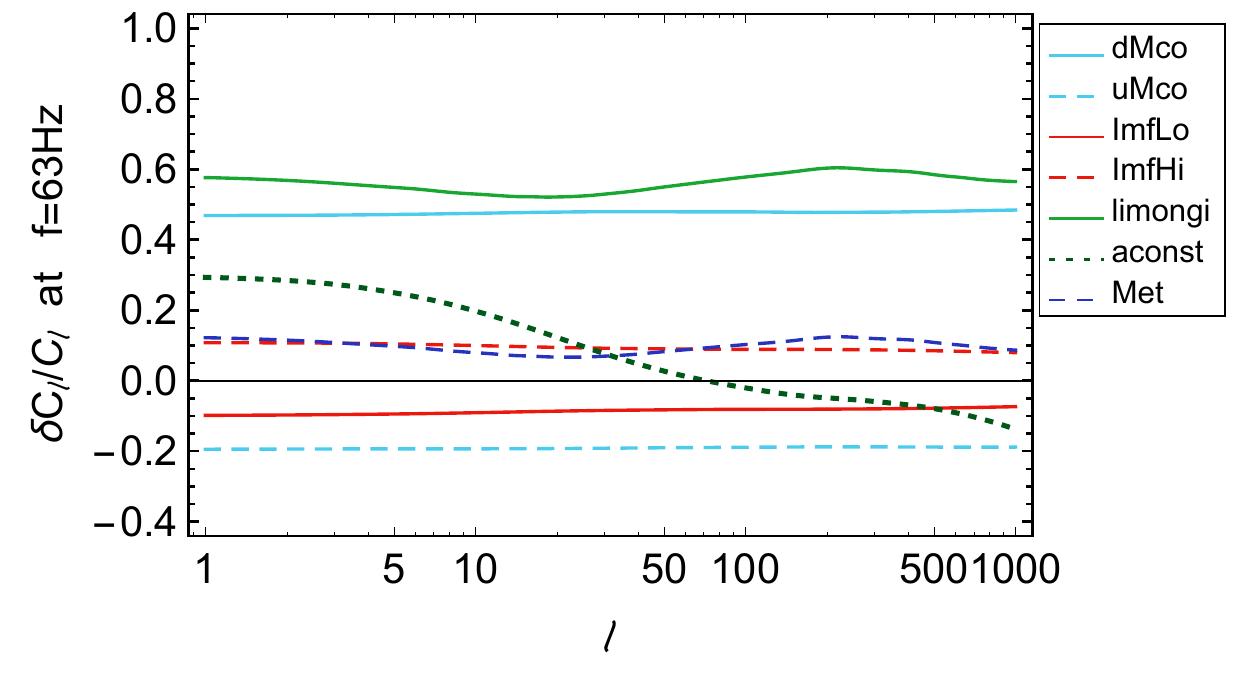}\quad
\includegraphics[width=.94\columnwidth]{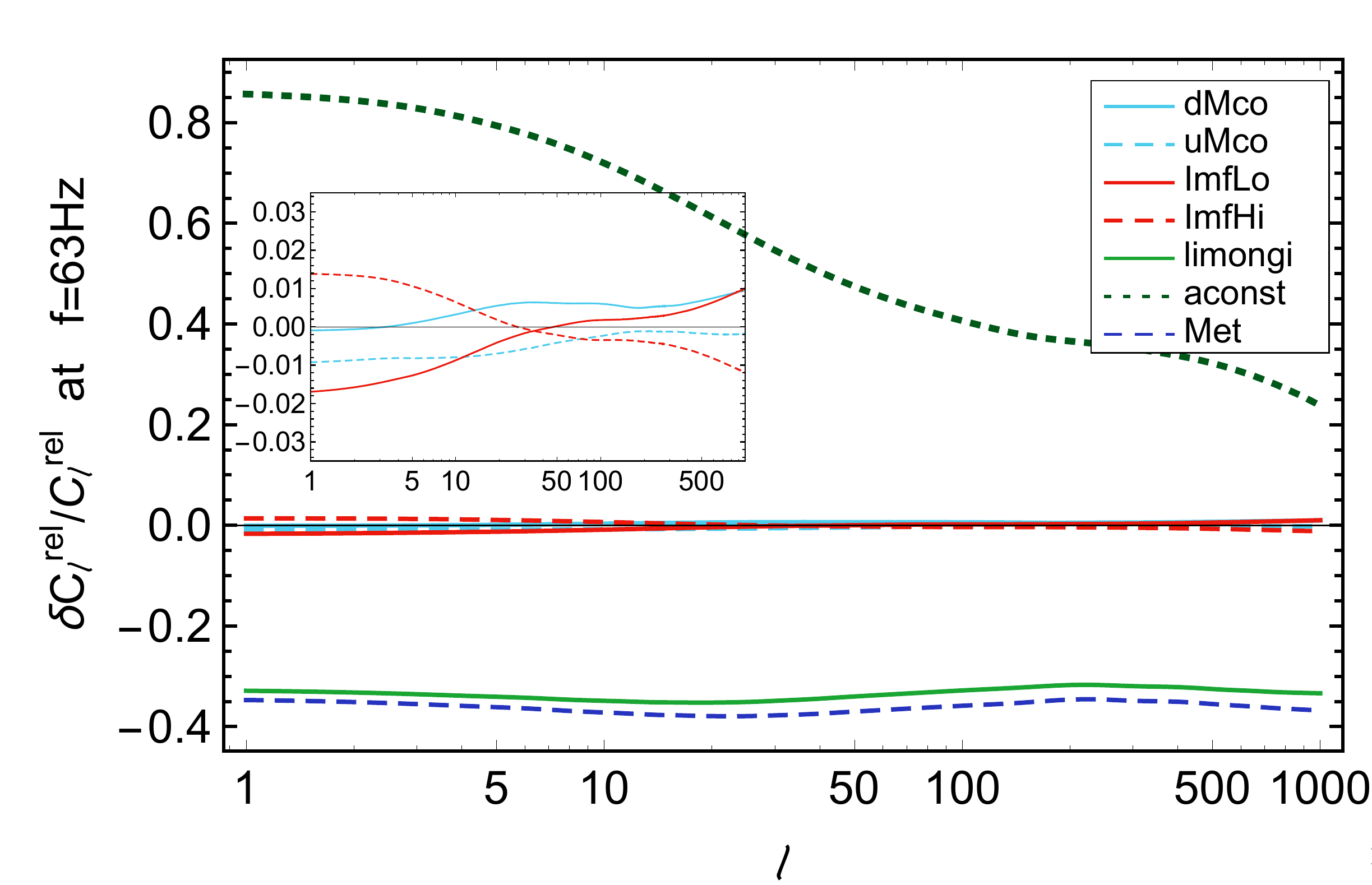}
\caption{Fractional difference between the angular power spectrum of anisotropies in different models and the reference model. On the right panel we show the  fractional difference between \emph{relative} anisotropies, i.e. for each model anisotropies are normalized with respect to the monopole of that model. The frequency is $f=63$ Hz for both panels. On the right panel, the insert is a zoom on small variations.}\label{Clall}
\end{figure*}

\begin{figure*}[!htb]
\includegraphics[width=\columnwidth]{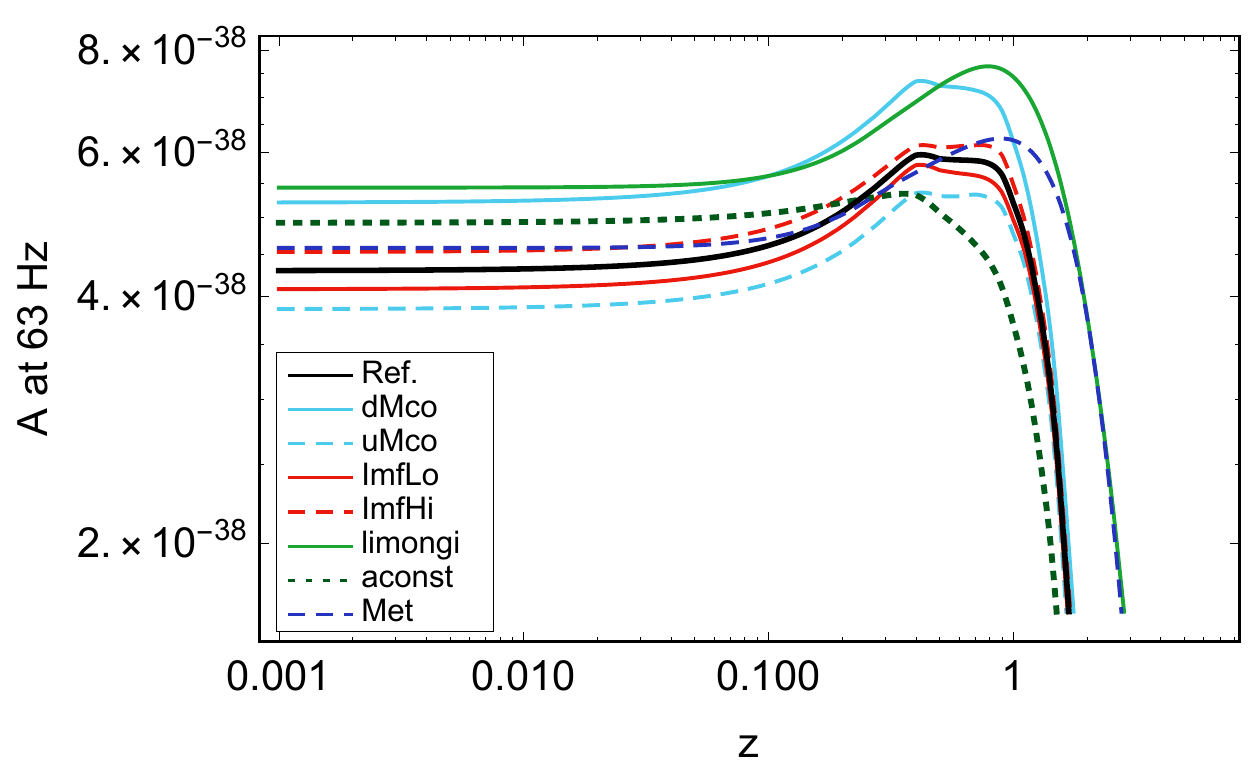}\quad
\includegraphics[width=\columnwidth]{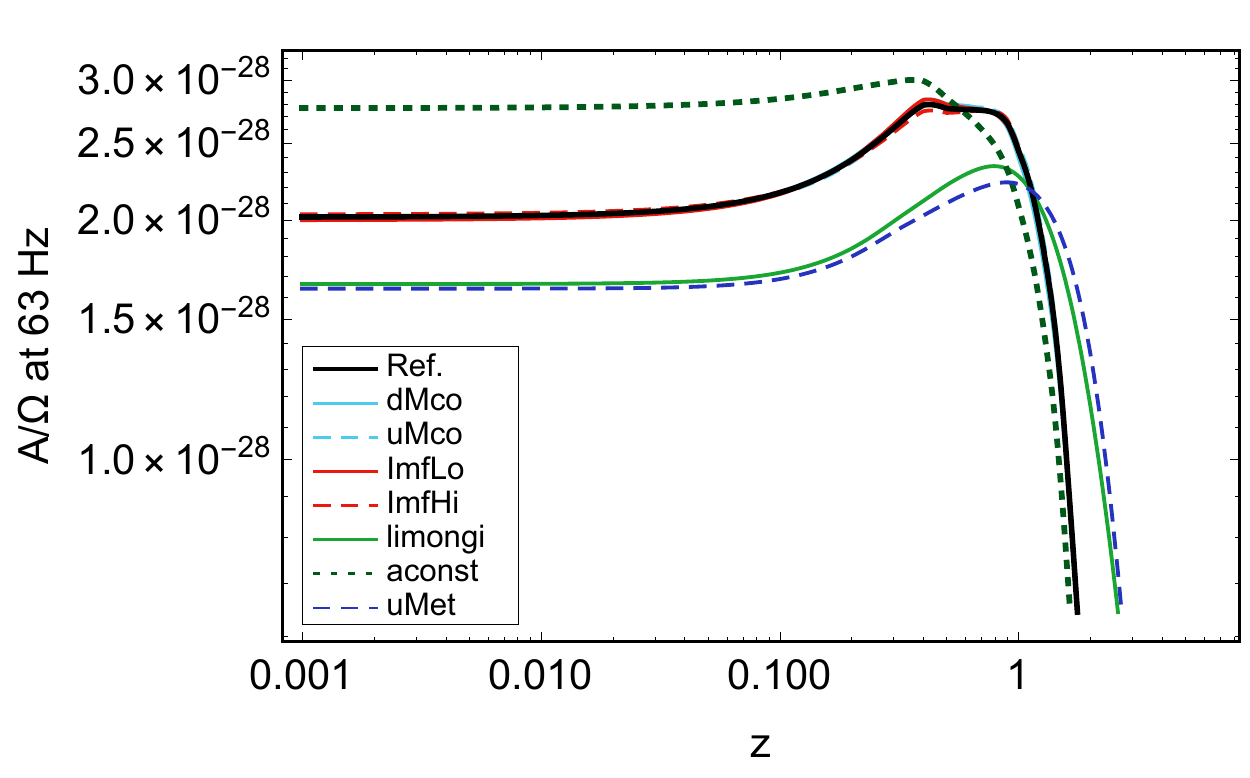}
\caption{Astrophysical kernel $\mathcal{A}$ as a function of redshift for different astrophysical models (Left). The same but normalized over the monopole (Right). The frequency is $f=63$ Hz for both panels.  In both panels the $y$ axis has units erg/cm$^3$.}\label{discussion}
\end{figure*}

\begin{figure*}[!htb]
\includegraphics[width=\columnwidth]{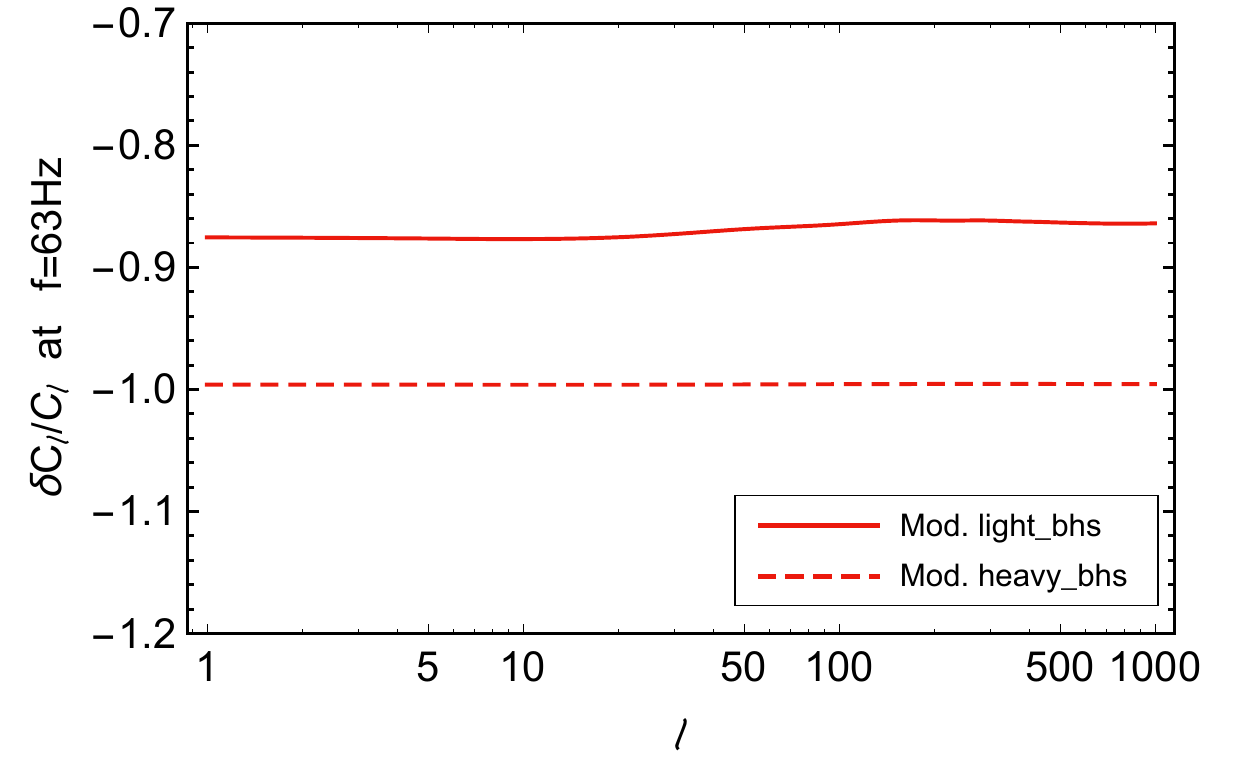}\quad
\includegraphics[width=\columnwidth]{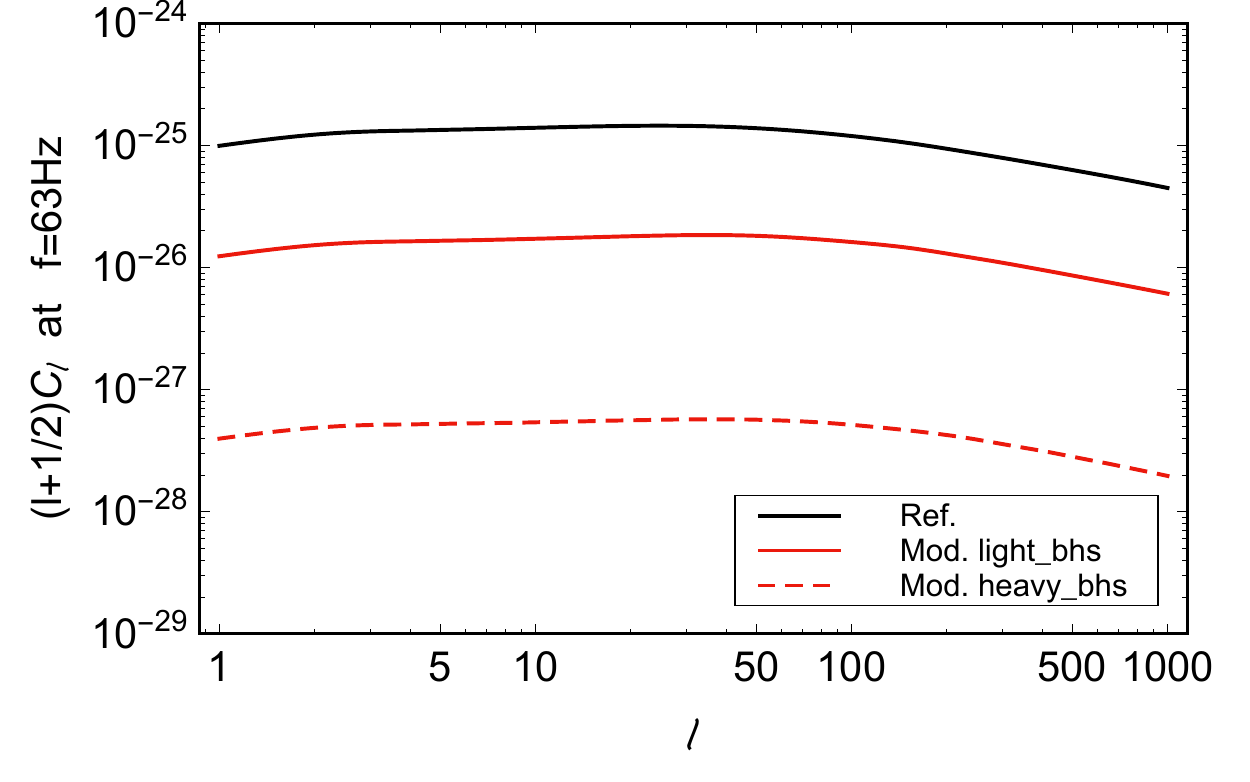}
\caption{Fractional difference between the angular power spectrum of anisotropies in different BH population models and the reference model (Left). Absolute angular power spectrum of the BH models compared to the reference model (Right). We chose a frequency of $f=63$ Hz in the LIGO band.}\label{Clall3}
\end{figure*}

 \begin{figure*}[!htb]
\includegraphics[width=1.02\columnwidth]{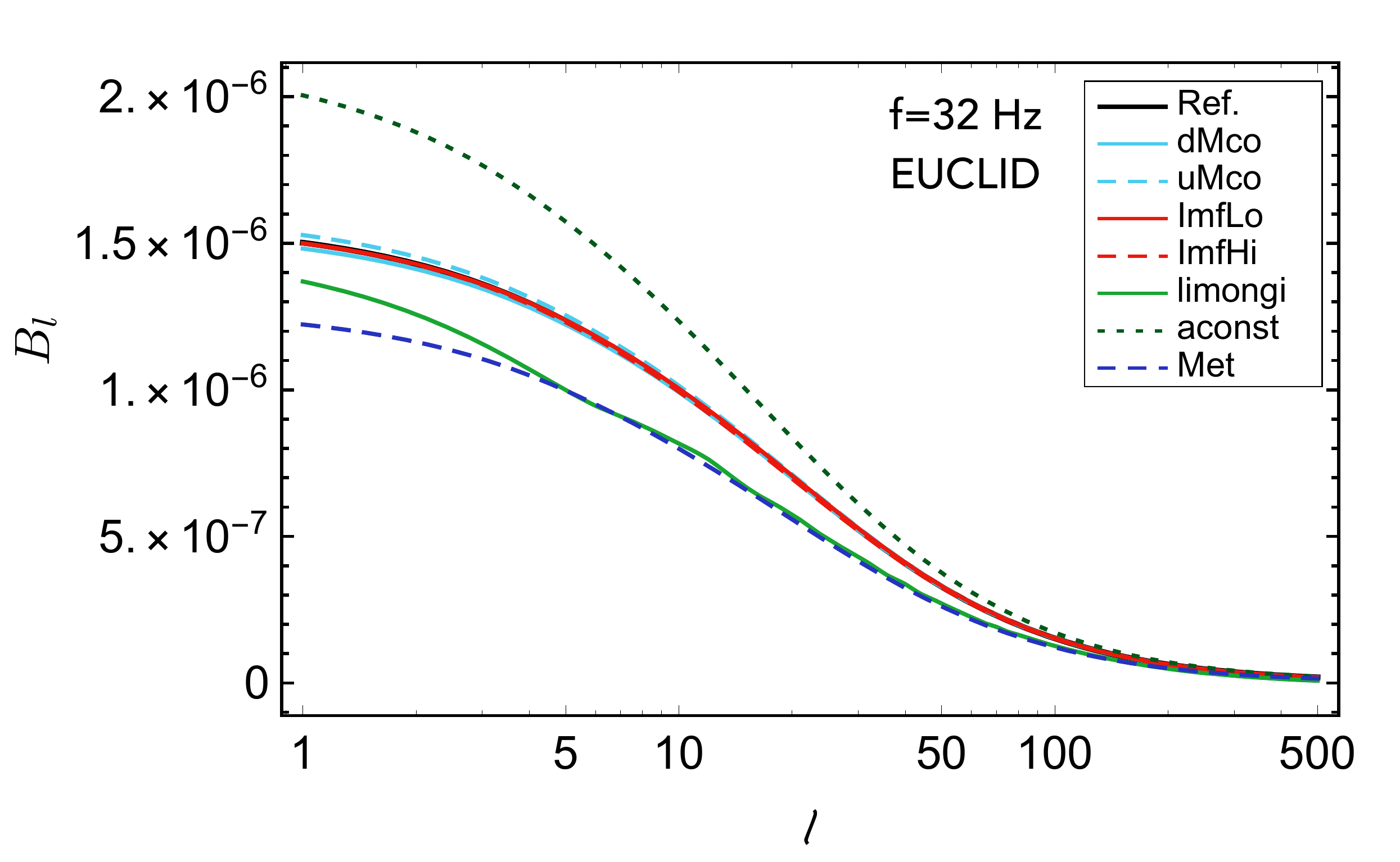}
\includegraphics[width=\columnwidth]{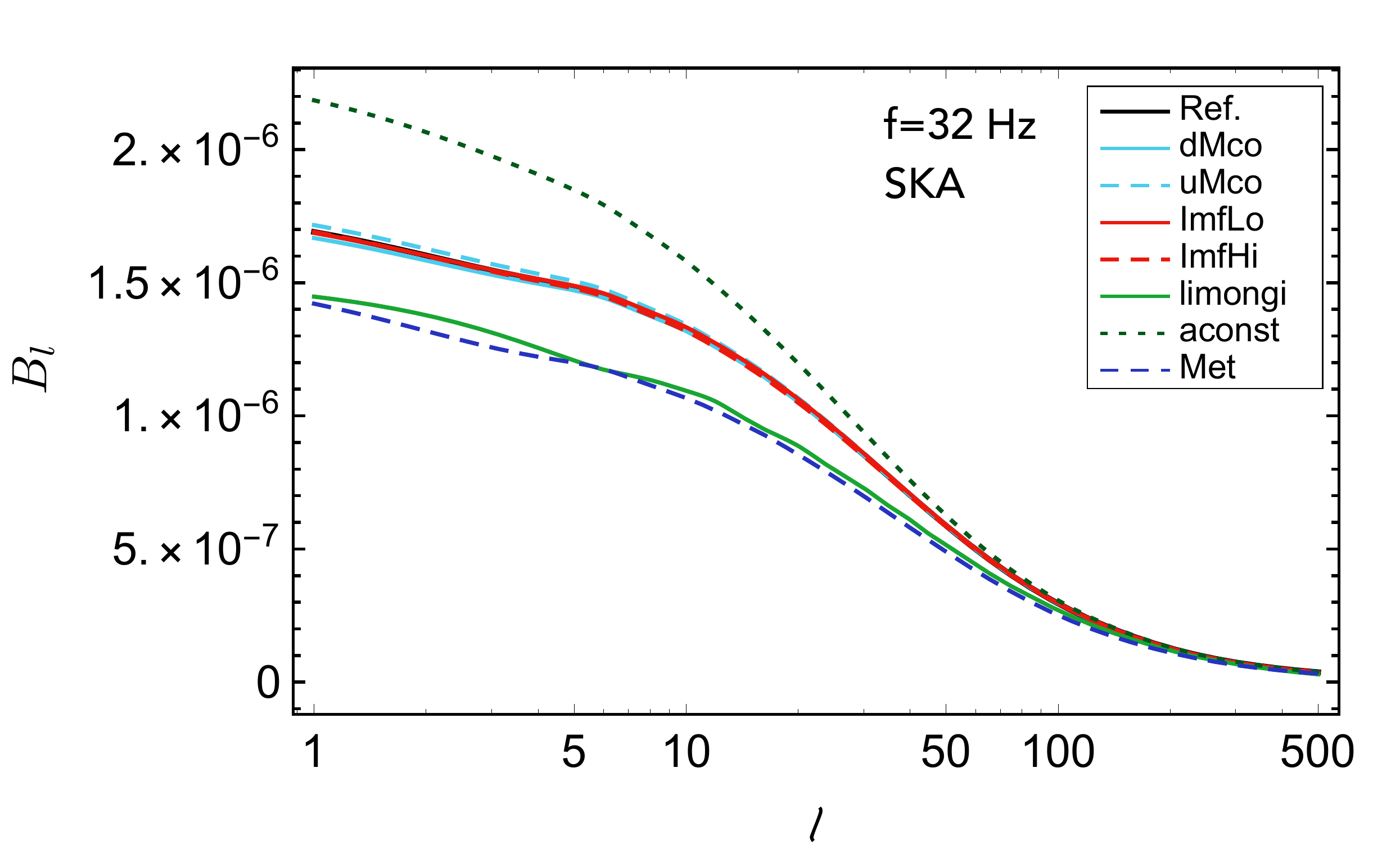}
\caption{Cross-correlation of the AGWB anisotropies with weak lensing convergence for different models and for the SKA~\cite{Andrianomena:2014sya} and Euclid~\cite{2011arXiv1110.3193L} redshift distributions.  The frequency is $f=32$ Hz for both panels. }\label{cross1}
\end{figure*} 

We now focus on the LIGO/Virgo frequency band. In Fig.~\ref{Clall} (left panel) we present the fractional  difference between the power spectra of the various models described  and the angular power spectrum of the reference model, defined as
\begin{equation}
 \frac{\delta C_{\ell}}{C_{\ell}}=\frac{C_{\ell}^{\rm mod}-C_{\ell}^{r\rm ef}}{C_{\ell}^{\rm ref}}
\end{equation}
 where $C_{\ell}^{\rm ref}$ and $C_{\ell}^{\rm mod}$ are the angular power spectra of the reference and modified model, respectively.
 
We can easily conclude that changing the stellar evolution model (and hence the mass distribution of the BH population) and changing the cut-off in mass with respect to the reference model gives a  relative variation of the order of 50\%.  

The right panel of Fig.\,\ref{Clall} shows the same quantity but for the angular power spectra normalized over the monopole (of each model). It can be concluded from these results that changing the distribution of the initial semi-major axis leads to a variation of up to 80\% at low multipoles. We stress that the modified distribution considered here, namely flat in the semi-major axis, has not a strong astrophysical motivation and is taken here for illustrative purposes. The effect of varying the orbital semi-major axis is enhanced when plotting the fractional difference of relative anisotropies (i.e. normalized over the monopole). This can be understood considering that the amplitude of the green dotted curve on the right panel can be (roughly) obtained from the corresponding one in the left panel multiplying it by the ratio between the monopole of the reference model and of the model with new distribution of the semi-major axis. This gives a multiplicative factor of order $1.3$, see Fig.~\ref{AllOmega}, which shifts the curve of fractional differences when going from the left to the right panel. A similar reasoning holds for the other models.

The difference between the models in the left panel of  Fig.~\ref{Clall}  can be explained by noting that the angular power spectrum is sensitive to the astrophysical kernel $\mathcal{A}$, which we plot as a function of redshift in the left panel of Fig.~\ref{discussion}. For a fixed frequency, the monopole of the energy density is sensitive to the integral over redshift of $\mathcal{A}$ while the amplitude of the anisotropies at a given multipole $\ell$  is sensitive to the amplitude of the kernel $\mathcal{A}$ at the redshift $z$ \emph{corresponding} (through Limber) to the multipole $\ell$ considered. Similarly, the amplitude of relative anisotropies (i.e. normalized over the monopole) at a given $\ell$ is sensitive to the amplitude of $\mathcal{A}$ normalized over the monopole, see right panel of Fig.~\ref{discussion}. This also explains why the green dotted curve in Fig.~\ref{Clall} (representing the fractional difference between the anisotropies of the model with modified distribution of initial semi-major axis and the reference one) decreases with multipoles: the slope of $\mathcal{A}$ for this model does not increase between redshift 0.1 and 1, while it does for the reference model.

In Fig. \ref{Clall3} we present the angular power spectra for the models testing different black hole populations and the fractional difference between the power spectrum of the models with only high and only low black hole mass (described in section \ref{sub}) and the angular power spectrum of the reference model.  We observe that the fact that the sum of two angular power spectra (for low and high black hole masses) does not give back the total angular power spectrum is due to the fact that in the division between the sub-population low-high mass binaries we do not consider  binary systems with one high-mass and one low-mass black hole. 

\subsection{Cross-correlations}\label{sub5.3}

 \begin{figure*}[!htb]
\includegraphics[width=\columnwidth]{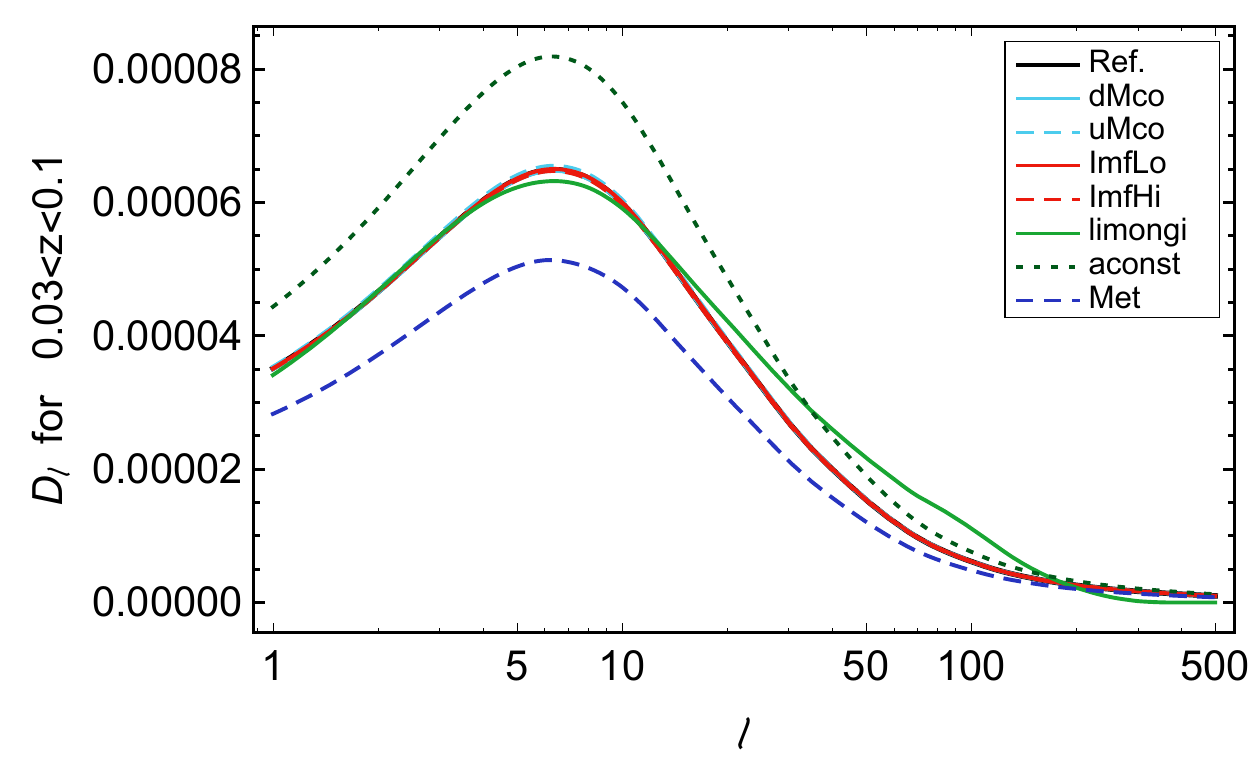}
\includegraphics[width=1.02\columnwidth]{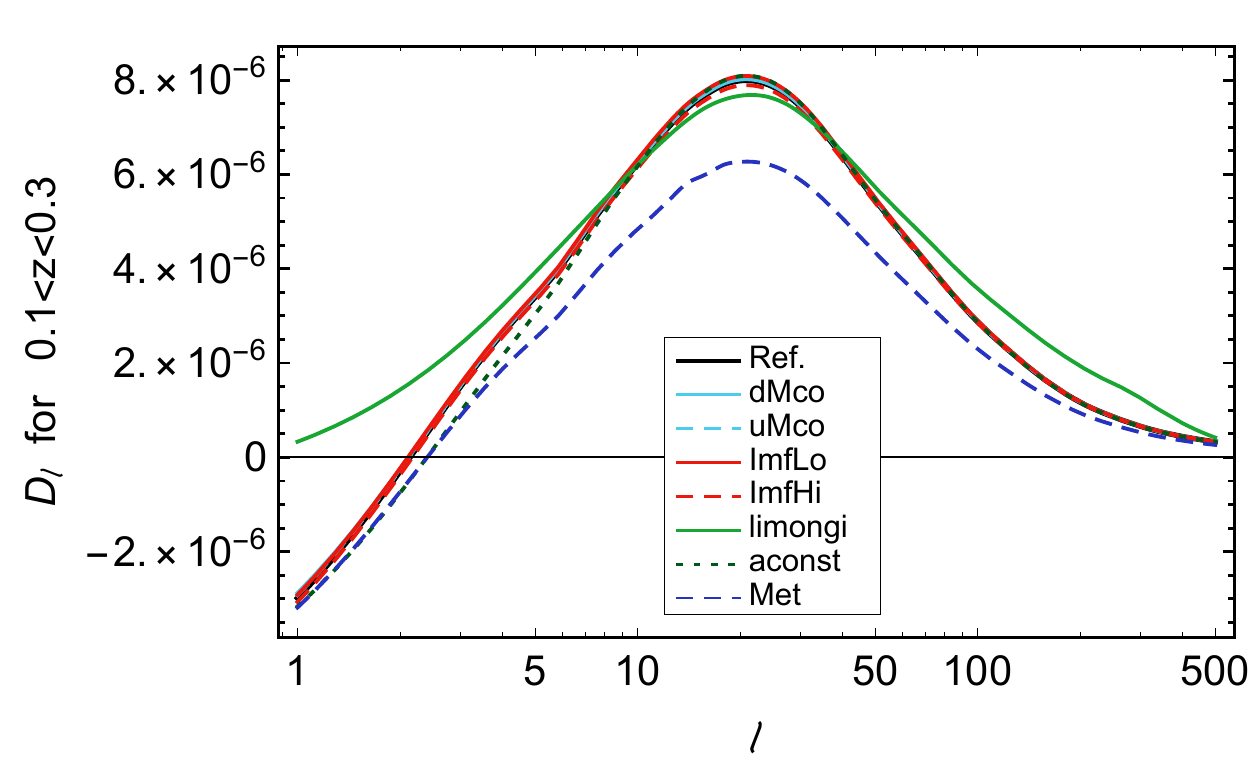}
\caption{Cross-correlation of the AGWB anisotropies with galaxy number counts. We have integrated the spectrum over the range of frequencies in the LIGO band 10 Hz $<f <$100 Hz. }\label{cross2}
\end{figure*} 

Figure~\ref{cross1} shows the cross-correlation between anisotropies of the AGWB and weak lensing convergence, for Euclid and SKA source distributions and for a frequency $f=32$ Hz while Figure~\ref{cross2}  shows the cross-correlation with galaxy number counts for two different redshift bins and integrated over frequency in the LIGO/Virgo band ($10$ Hz $<f<100$ Hz). For both cross-correlations, the effect of changing metallicity and distribution of semi axis gives a typical variation of order 30\%. 

Notice that for each model, the amplitude of the cross-correlation with lensing as a function of multipoles is a biased tracer of the amplitude of the astrophysical kernel $\mathcal{A}$ as a function of redshift, normalized over the monopole (we recall that we are defining the cross-correlations normalizing over the monopole of the AGWB, see eq.  (\ref{Bchi})). On the other side, the cross-correlation with galaxy number counts selects different bins in redshift in the astrophysical kernel $\mathcal{A}$. This can be seen in Fig.~\ref{cross2}: as we shift the window in redshift, the peak of the cross-correlation shifts towards higher multipoles, as expected from the Limber relation between multipoles and redshifts. In a given redshift bin, the amplitude of the cross-correlation is bigger for models with the bigger value of $\mathcal{A}/\bar{\Omega}_{\rm GW}$ in the redshift bins considered.  Hence, interestingly, the cross-correlation with galaxy number counts can help to reconstruct the astrophysical kernel $\mathcal{A}$ as a function of redshifts. 


\section{Contribution of binary neutron stars}\label{NS}

The detection of the binary NS merger by the LIGO/Virgo network \cite{2017PhRvL.119p1101A,2017ApJ...848L..12A} and the estimated rate of mergers in the local Universe of $R=920^{+2220}_{-790}$ Gpc$^{-3}$ yr$^{-1}$ \cite{2018arXiv181112907T} led to the conclusion that these sources may have a comparable contribution to the AGWB relative to binary BHs \cite{Abbott:2017xzg,2019arXiv190302886T}. We may therefore expect that their contribution to the anisotropies of the AGWB will also be important.

While it will be difficult to disentangle the relative contributions of binary BHs and NSs to the overall AGWB, especially in view of the large modeling uncertainty in the binary NS merger rates \cite{2018MNRAS.474.2937C,2019MNRAS.482.2234G}, it is interesting to note that their host galaxies are expected to have different properties. In the isolated BH formation scenario discussed here, BHs masses are heavily influenced by the metallicity of their progenitor stars, as discussed above. Specifically, metal-poor stars retain most of their mass throughout their evolution and collapse to form heavier BHs. As a consequence, these BHs (that also produce stronger GW signal when they merge) form preferentially in high-redshift and/or low-mass galaxies \cite{2016MNRAS.463L..31L,2018MNRAS.474.4997C,2018MNRAS.481.5324M,2019arXiv190300083A}.  In contrast, NSs can also form in metal-rich environments. In view of the different clustering properties of the host galaxy populations, binary BHs and binary NSs can in principle give rise to very different anisotropic components of the AGWB.

In order to estimate the contribution from binary NSs we calculate the formation rate of NSs in our astrophysical model. For simplicity we assume that NSs form  from stars in the mass range $(8,11)M_{\odot}$ with a constant mass of $1.3M_{\odot}$. We follow the formalism described above to calculate the number of detectable sources in the local Universe and normalize the fraction of NSs that reside in binaries that merge within the Hubble time $\beta$ (similarly to the case of BHs) to result in $1$ detection during O1+O2 observing period.

Our results for the anisotropies from binary NS mergers are shown in Fig.\,\ref{BNS}. We present the results in the LIGO-Virgo frequency band. The results in the LISA band look exactly the same.  The value of the background energy density in both the LISA and LIGO-Virgo frequency band is given by $\bar{\Omega}_{GW}(f)\sim 2.15\times 10^{-11} f^{2/3}$.

Fig.\,\ref{BNS} shows that the contribution from binary NSs can be dominant relative to the one from binary BHs. We stress however that this result is model-dependent, and that moreover depends on the (highly uncertain) merger rate of binary NSs. 


\section{Lower frequencies and the LISA band }\label{LISAsec}

The early inspiral phase of merging stellar-mass binary BH that may be observable with  LISA space-borne interferometer is another probe of the astrophysical and cosmological processes discussed in this article. The AGWB from binary BH is expected to be dominant in the LISA band~\cite{2016PhRvD..94j3011D} and below,  and may become a source of confusion noise for some of the other types of sources. Observations with LISA will allow one to study some aspects of resolved and unresolved stellar-mass BH binaries that are difficult to observe with ground-based interferometers.

For example, at the mHz frequencies accessible to LISA, some of the binaries may not be fully circularized, and their residual eccentricities may provide an indication to their formation channel. In particular, binaries formed through dynamical processes in dense stellar cluster can have measurable eccentricities. These can be constrained for the subset of resolved merger, and in addition the distribution of eccentricities of the entire population may also affect the resulting AGWB.

LISA will also allow one to study the AGWB from other types of sources such as close white dwarf binaries (see e.g. Ref.~\cite{2002CQGra..19.1449V}), which may also produce anisotropies \cite{2001PhRvD..64l1501U,2005PhRvD..71b4025K}. Moreover, these anisotropies can potentially be used to distinguish the astrophysical source of stochastic background from the cosmological ones in the early Universe \cite{2018PhRvL.121t1303G}.

\begin{figure}[!htb]
 \includegraphics[width=\columnwidth]{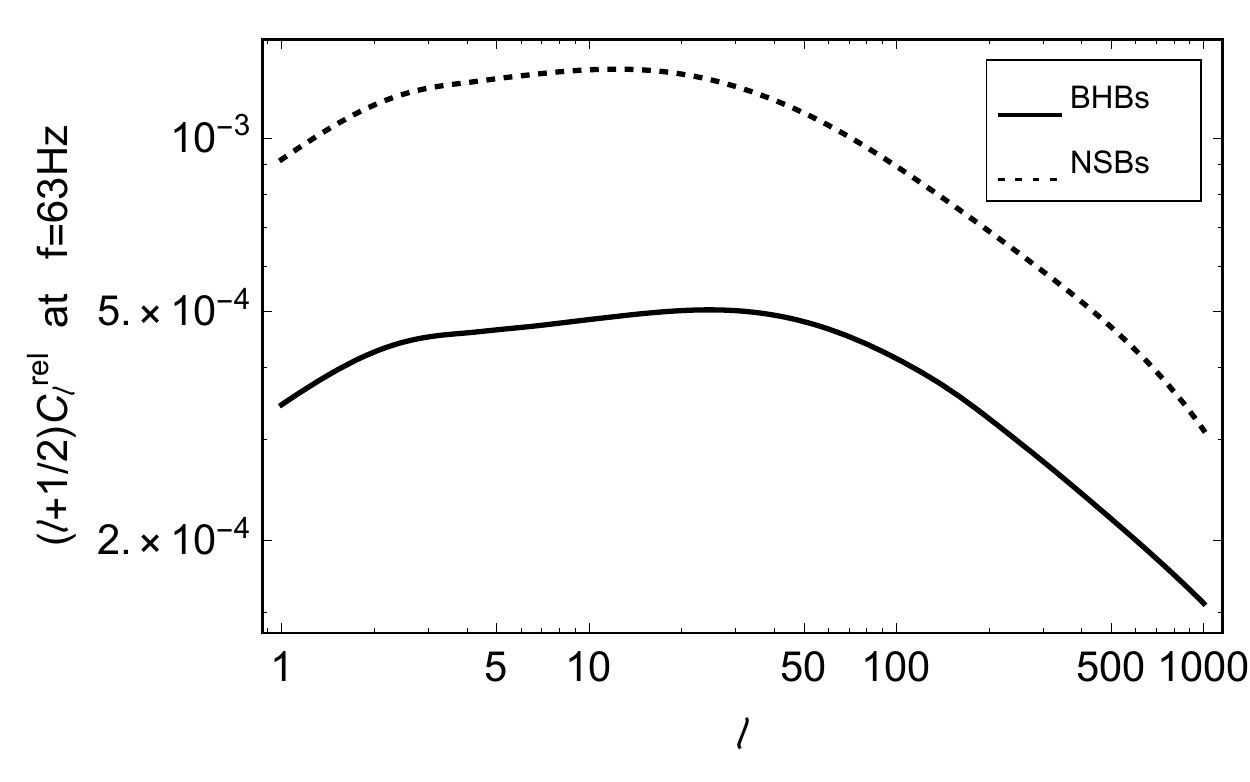} 
\caption{The angular power spectrum of the AGWB anisotropies from merging binary NSs (dotted line) together with the contribution of black hole mergers (for the reference astrophysical model). We have here chosen a frequency of $63$ Hz in the LIGO-Virgo band but the result for relative anisotropies are actually the same in any frequency band (and in particular also in the LISA band). \label{BNS}}
\end{figure}

In Fig.~3 of Ref.~\cite{us} we presented the first prediction of the angular power spectrum of anisotropies in the LISA bands. We note that the relative anisotropies have the same frequency dependence as the monopole, i.e. $C_{\ell}\propto f^{4/3}$, as expected. It follows that the frequency factorization discussed in \S~\ref{fac} is a solid approximation in the LISA frequency band. A plot for relative anisotropies is presented in Fig.~\ref{BNS}, for our reference astrophysical model. The value of the background energy density in both the LISA  frequency band is given by $\bar{\Omega}_{GW}(f)\sim 1.29\times 10^{-11} f^{2/3}$. 

%

\begin{figure*}[!htb]
\includegraphics[width=\columnwidth]{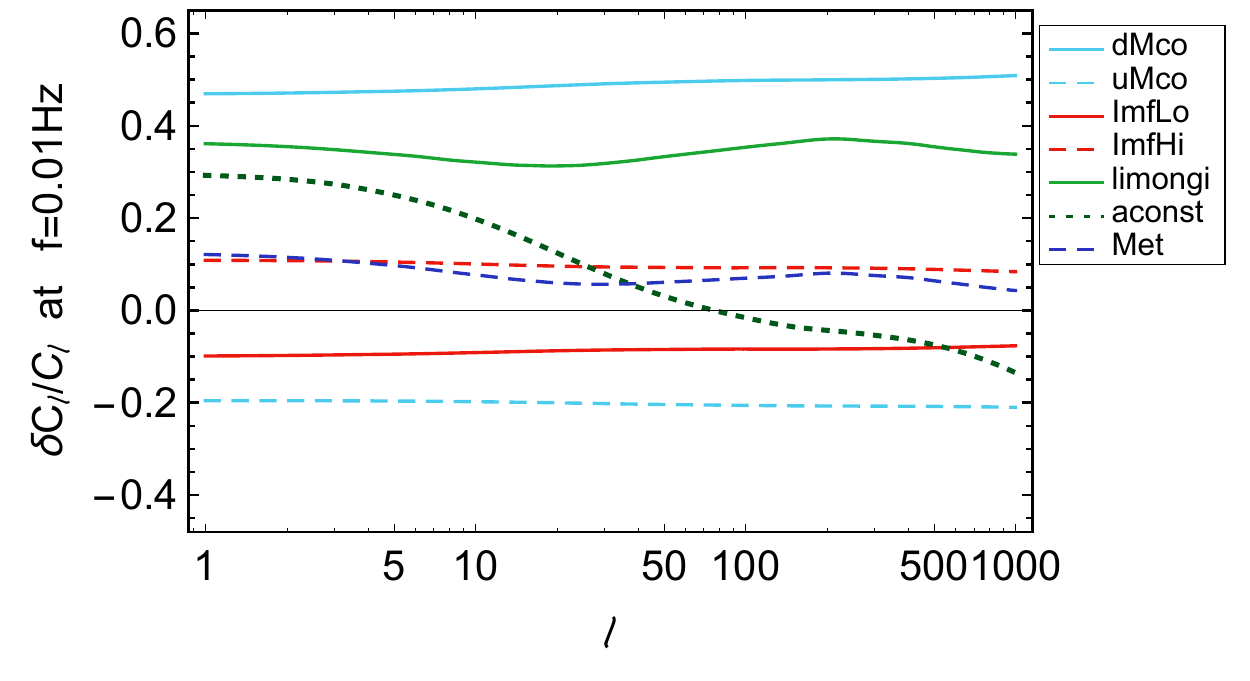}
\includegraphics[width=0.93\columnwidth]{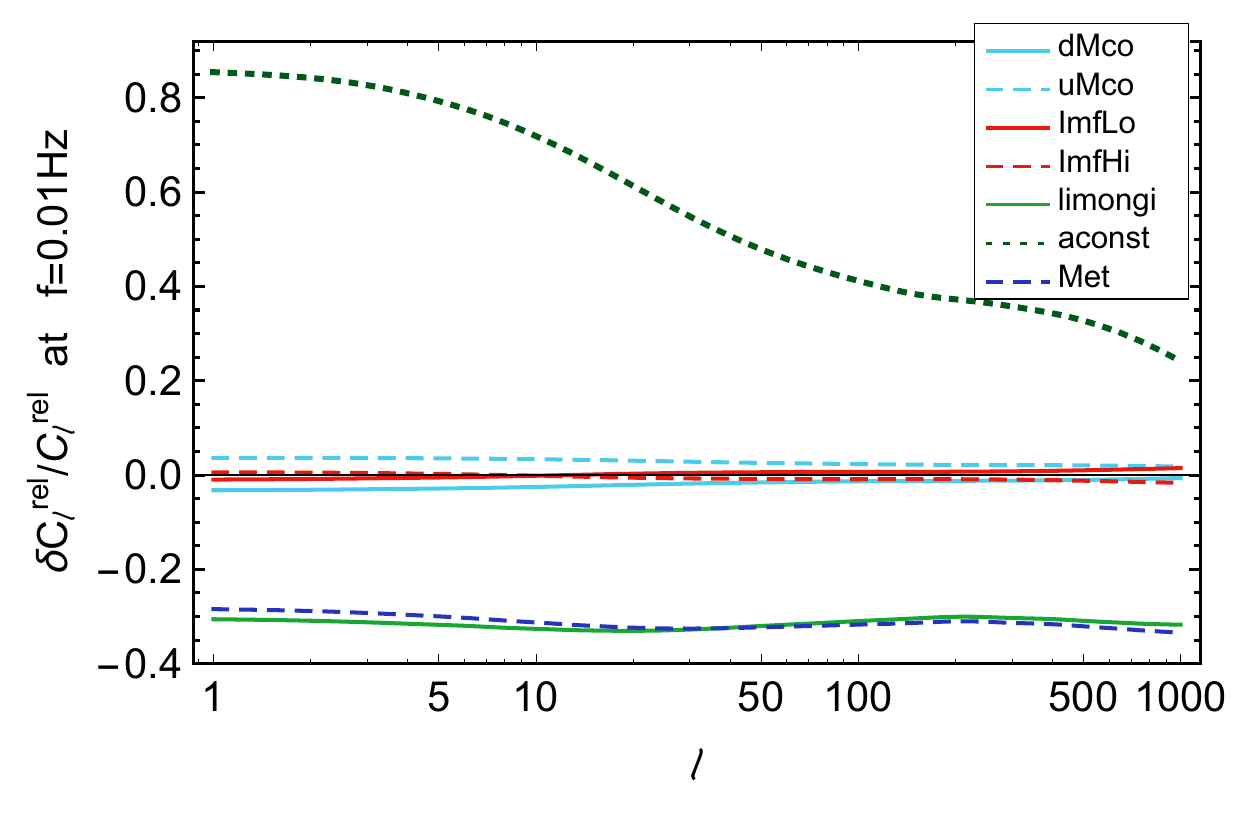}
\caption{Fractional difference between the angular power spectrum of anisotropies in different models and the reference model. The right panel shows the fractional difference between \emph{relative} anisotropies, i.e. for each model anisotropies are normalized with respect to the monopole of that model. We choose a frequency of 0.01 Hz, in the LISA band. }\label{LISA}
\end{figure*}

\begin{figure*}[!htb]
\includegraphics[width=\columnwidth]{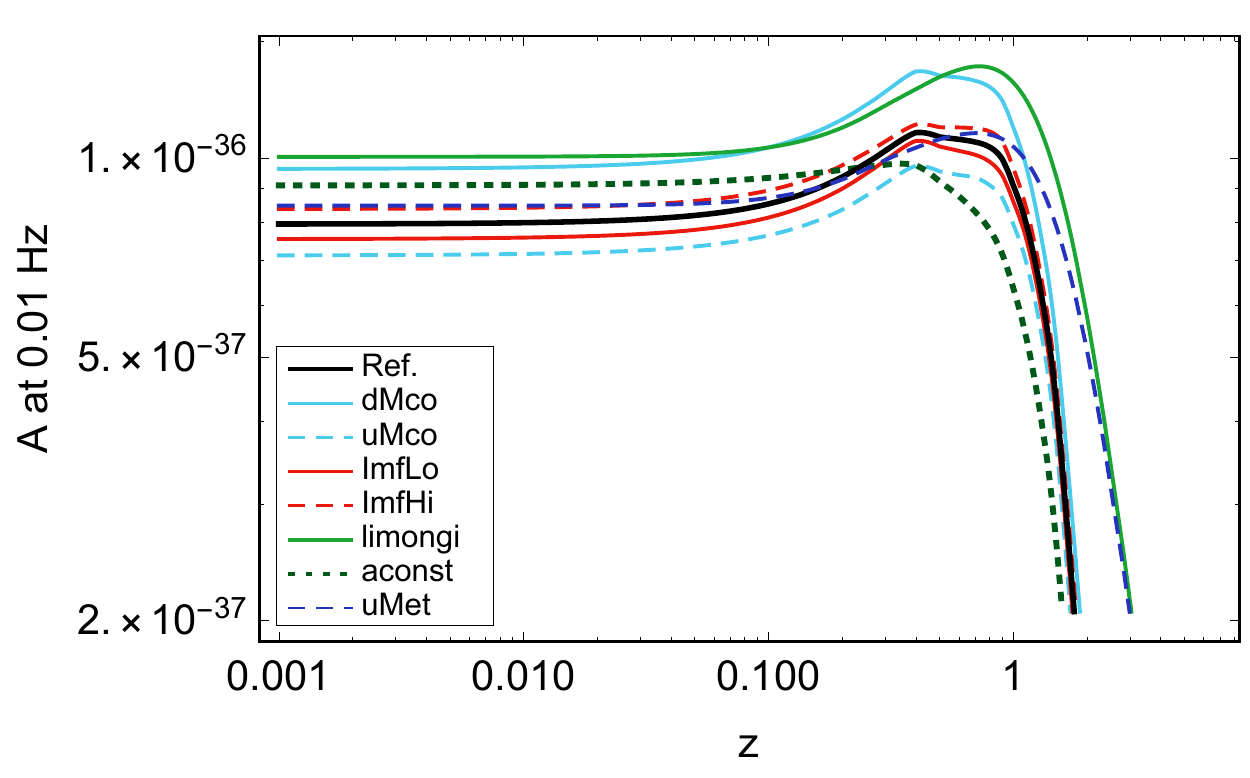}\quad
\includegraphics[width=\columnwidth]{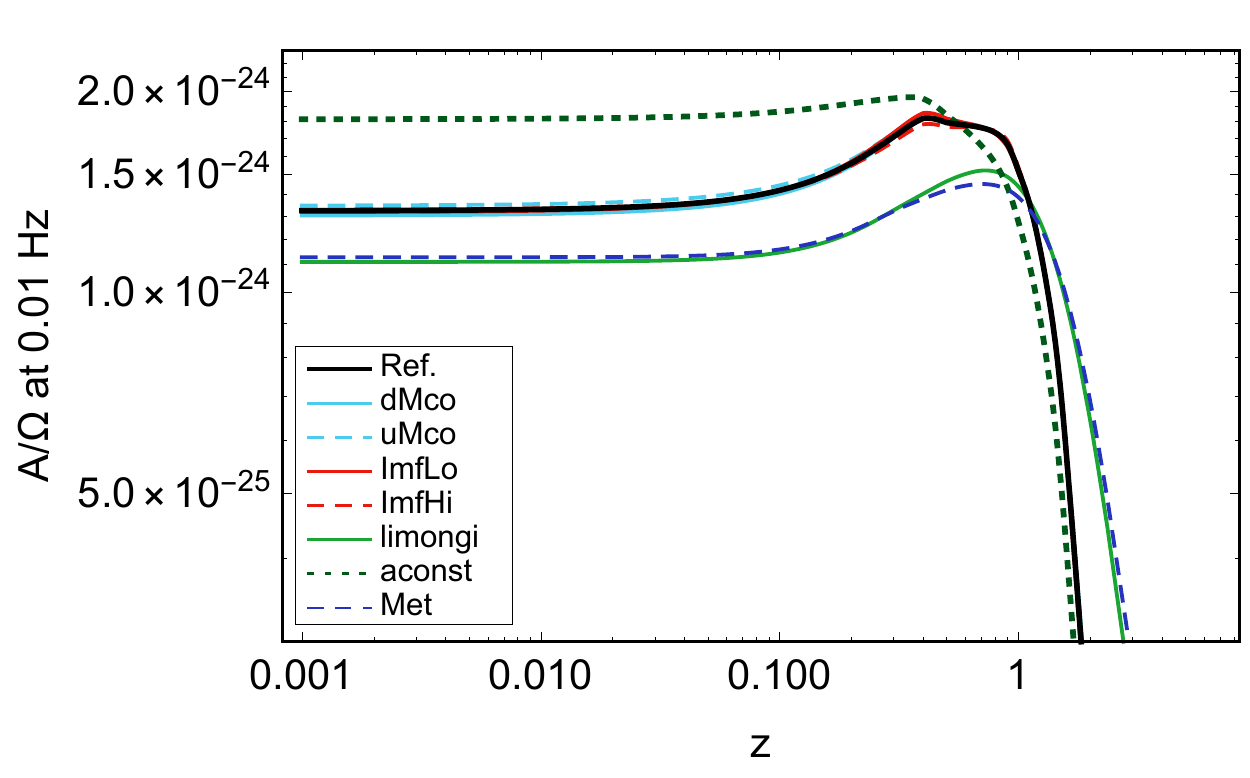}
\caption{Astrophysical kernel $\mathcal{A}$ as a function of redshift, for different astrophysical models (left panel). The same but normalized over the monopole for different astrophysical models (right panel). We chose a frequency $f=0.01$ Hz in the LISA band. In both panels the $y$ axis has units erg/cm$^3$. }\label{discussion2}
\end{figure*}

We compare the anisotropies in the LISA band produced by stellar-mass BH for the same set of astrophysical models in Fig.~\ref{LISA}. As already explained for the LIGO band, for a given model and a given frequency, the amplitude of anisotropies for a given $\ell$ is a tracer of the amplitude of the astrophysical kernel $\mathcal{A}$ for that frequency and at a redshift $z$ related by Limber to the multipole under consideration. Similarly, the amplitude of relative anisotropies as a function of multipoles  is sensitive to the amplitude of $\mathcal{A}$ normalized over the monopole, as a function of redshift. In Fig.~\ref{discussion2} we represent the function $\mathcal{A}$  for the various astrophysical models  and this function normalized over the monopole (left and right panel respectively), as a function of redshift.\footnote{We observe that there is a (small) difference between the results obtained in the LIGO and LISA band; Figs.~\ref{Clall} and \ref{LISA} respectively. This is due to the fact that in the LIGO band  anisotropies have a non-trivial frequency dependence, see \S~\ref{fac}. }

\section{Conclusion}\label{conclusion}

This article explored the astrophysical dependencies of the anisotropies of the AGWB and analyzed the properties of its angular power spectrum. It focused on the contribution to the background coming from BH mergers in both the LIGO-Virgo and LISA frequency bands. In particular, the properties of anisotropies in the LISA band are studied here for the first time. 

After a summary of the framework developed in Ref.~\cite{Cusin:2017mjm, Cusin:2017fwz} to  study AGWB anisotropies and cross-correlations, we analyzed in details the general properties of the angular power spectrum. In particular, using an analytic approach we showed how different redshift bins contribute to the various multipoles of the correlation function.  We also showed that low multipoles are dominated by the contribution of large scales (i.e. small $k$ Fourier modes). It follows that a semi-analytic description of clustering,  like the one used in this work, has to be preferred to a catalogue approach to describe large angular separations  (which are the first one we will hopefully observationally access). We also tested the standard factorization assumption used  in most of present searches (e.g. by LIGO-Virgo), showing that it breaks down in the upper-part of the LIGO spectrum. A general analytic derivation of the shot noise contribution in different frequency bands has also been presented. In particular, we showed that the spatial component of shot noise, due to the discreteness of the galaxy distribution, is present in both LISA and LIGO frequency bands and it dominates the spectrum at high multipoles. Using a toy model, we have also illustrated the possibility to use cross-correlation with galaxy number counts to extract AGWB anisotropies from a  shot-noise dominated map. A future work will be dedicated to studying the possibility of filtering out shot noise using cross-correlations and masking. 

In the second part of the article, we explored the signatures of 9 different astrophysical models, differing from a reference model in one physical ingredient (e.g. distribution function of orbital parameters, cut-off in the BH mass distribution, stellar evolution model for BH generation, IMF, metallicity...). First, from the requirement that the models are all compatible with the LIGO O1+O2 detection rate, we concluded that the fraction of BH in binaries varies from 0.06\% to 17\% across the models. This highlights the actual degeneracy of the models. We then demonstrated that the anisotropies are very sensitive to some astrophysical parameters such as the distribution of the initial semi-major axis of the binary systems, the choice of the metallicity profile and  of the astrophysical model to describe BH generation out of stellar progenitors. More quantitatively, we concluded that changing the distribution of the initial semi axis $a_{\rm f}$ of binary systems from a distribution $\propto 1/a_{\rm f}$ to a flat one induces relative variation in anisotropies (normalized over the monopole) of order 80\% at small multipoles ($\ell<10$). Changing the stellar evolution model or keeping a low constant metallicity leads to a variation of order 40\%. 
These results hold for both LIGO and LISA frequency bands. This result contradicts the claim of Ref.~\cite{Jenkins:2018kxc}, where the authors conclude that the anisotropies do not depend on astrophysical properties of the BH populations. 

The present analysis concludes that the prediction of the angular power spectrum is very sensitive to the astrophysical modeling: astrophysical models differing for the description of stellar evolution and for the choice of the distribution of the orbital semi-major axis  give predictions differing of  $>40$\% in both the LIGO and LISA frequency bands.  For those models, we also compared the results for the cross-correlation with weak lensing convergence (for both SKA and Euclid source distributions) and galaxy number counts, at different redshifts.  At low multipoles, variations are up to order 30\%.  Results for the angular power spectrum in the LISA band are shown to scale with frequency as the monopole. i.e. $C_{\ell}\propto f^{4/3}$. The spectrum of relative anisotropies in the LISA band is therefore frequency independent. 

This result has interesting astrophysical implications.  The AGWB angular power spectrum is an observable we will hopefully have access to observationally in the next few years. From the comparison between future observed intensity maps of the background (for different frequencies) and our theoretical predictions, we will be in the position to set constrains on astrophysical quantities, such as stellar evolution model, metallicity distribution and distribution of orbital parameters, that cannot be accessed  otherwise.  This is the first stage of a long term research program. We analyzed which ingredients of the astrophysical  parametrization give sizable differences in the anisotropies and which do not play a role. It will allow us to refine the astrophysical modeling. We also provide a theoretical understanding of relation between the shape of the angular power spectrum and the shape of the astrophysical kernel.

We emphasize that the monopole and the anisotropies of the AGWB give access to complementary information. While the former is sensitive to the integral over redshift of the astrophysical kernel describing sub-galactic physics (the function $\mathcal{A}$ in this work), anisotropies at different angular separations are sensitive to the amplitude of this kernel at different redshifts. Anisotropies open a new window on the reconstruction of the redshift dependence of the astrophysical kernel: this new observable will help us to test properties of source populations at different redshifts and eventually reconstruct their evolution in time. 

\acknowledgements
It is a pleasure to acknowledge D. Alonso for a careful reading of this work and very interesting discussions on shot noise. We thank G. Lavaux,  F. Lacasa, B. Whiting, I. Michaloliakos,  P. Ferreira, Ruth Durrer and  E. Vangioni  for stimulating discussions. This work was done within the Labex ILP (reference ANR-10-LABX-63), part of the Idex SUPER, and received financial state aid managed by the ANR, as part of the programme Investissements d'avenir under the reference ANR-11-IDEX-0004-02.  GC acknowledges financial support from ERC Grant  No: 693024 and Beecroft Trust.

\appendix
\section{Computation of shot noise}\label{derivation}

We assume a binned galaxy survey. The quantity $\Delta_i$ describes the galaxy overdensity in the bin $i$. Then the quantity $\langle \Delta_i \Delta_j\rangle$ contains the theoretical correlation function and a shot noise contribution 
\be
\langle \Delta_i \Delta_j\rangle=\frac{1}{V_p \bar{n}_G}\delta_{ij}+\xi^{\rm Gal}_{ij}\,,
\ee
where $V_p \bar{n}_G$ is the mean number of galaxies per pixel and $\delta_{ij}$ is the Kronecker symbol. We can take the continuous limit of a binned survey as
\be
\Delta_i\rightarrow \hat{\delta}_{\Gal}(\bx)\,,\qquad \text{and}\quad\delta_{ij}\rightarrow V_p\delta^{(3)}(\bx-\bx')\,,
\ee
where a hat denotes observed quantities and $V_p$ is the pixel volume. Using Eq.~(\ref{sss4}), we can rewrite the \emph{measured} background anisotropies as
\be\label{zut}
\delta\hat{\Omega}_{\rm GW}(\bee)=\int \dd z\, \frac{1}{H}\partial_{z}\left(\frac{\bar{\Omega}_{\rm GW}}{4\pi}\right)\hat{\delta}_{G}(\bee, z)\,,
\ee
or equivalently, using that  $ \partial_{z} \bar{\Omega}_{\rm GW}(z, f)=H\partial_{r} \bar{\Omega}_{\rm GW}(r, f)$ 
\be\label{sssm}
\delta\hat{\Omega}_{\rm GW}(\bee)= \int \dd r\,\partial_{r}\left(\frac{\bar{\Omega}_{\rm GW}}{4\pi}\right)\hat{\delta}_{G}(\bee r)\,.
\ee
We use this result when computing the correlation function of the background anisotropies 
\begin{align}
&\langle \delta\hat{\Omega}_{\rm GW}(\bee) \delta\hat{\Omega}_{\rm GW}(\bee')\rangle=\frac{1}{(4\pi)^2}\int \dd r \frac{\partial \bar \Omega_{\rm GW}}{\partial r}\int \dd r' \frac{\partial \bar \Omega_{\rm GW}}{\partial r'}\nn\\
&\times \left[\langle \delta_{\Gal}(\bx=\bee r)\delta_{\Gal}(\bx'=\bee' r')\rangle +\delta^{(3)}(\bx-\bx') \frac{1}{\bar{n}_{\Gal}}\right]\,. 
\end{align}
The first part in the square brackets is the theoretical result for the correlation function while the second term is the Poisson noise contribution. Writing
\be
 \delta\hat{\Omega}_{\rm GW}(\bee)=\sum_{\ell m}\delta \hat{\Omega}_{\ell m} Y_{\ell m}(\bee)\,,
 \ee
after standard manipulations one finds
\begin{align}
&\langle \delta \hat{\Omega}_{\ell_1 m_1} \delta \hat{\Omega}^*_{\ell_2 m_2}\rangle=\langle \delta \Omega_{\ell_1 m_1} \delta \Omega^*_{\ell_2 m_2}\rangle\\
&+\delta_{m_1m_2}\delta_{\ell_1 \ell_2}\int \frac{\dd r}{(4\pi)^2} \Big|\frac{\partial \bar \Omega_{\rm GW}}{\partial r}\Big|^2 \frac{1}{r^2} \frac{1}{\bar{n}_{\Gal}(r)}\,,\nn
\end{align}
where the first contribution is the theoretical one, the second the shot noise component. Using the standard definition
\be
(2\ell+1)C_{\ell}=\langle \delta \hat{\Omega}_{\ell m} \delta \hat{\Omega}^*_{\ell m} \rangle\,,
\ee
we immediately find for the angular power spectrum
\be\label{qq}
\hat{C}_{\ell}=C_{\ell}+S_n\,,
\ee
where 
\be\label{shots}
S_n=\frac{1}{(4\pi)^2}\int \dd r \Big|\frac{\partial \bar \Omega_{\rm GW}}{\partial r}\Big|^2 \frac{1}{r^2} \frac{1}{\bar{n}_{\Gal}(r)}\,.
\ee
We see that this integral diverges in $r=0$, hence the contribution of Poisson noise depends on the cut-off used to regularize this integral. From an observational point of view, the physical quantity on which the cut-off has to be set is the observed flux: sources with a flux bigger than a given threshold can be resolved and are therefore filtered out. To see where in Eq.~(\ref{shots}) the cut-off in flux appears, we use Eq.~(\ref{link}), i.e.
\be\label{link2}
\partial_{r}\bar{\Omega}_{\rm GW}=\frac{f}{\rho_c} \mathcal{A}(f, r)\,, 
\ee
and we rewrite the astrophysical kernel Eq.~(\ref{AA}) as an integral over emitted luminosity as
\be \label{B2}
\mathcal{A}(f, r)=a^4\int \dd\mathcal{L}_{\Gal}\, \bar{n}_{\Gal}(\mathcal{L}_{\Gal}, r)\mathcal{L}_{\Gal}\,,
\ee
where $\bar{n}_{\Gal}(\mathcal{L}_{\Gal}, r)$ is the average number of galaxies at distance $r$ with luminosity $\mathcal{L}_{\Gal}$. We replace Eq.~(\ref{B2}) in Eq.~(\ref{shots}) and change the variable of integration to redshift. Using that the flux $\Phi$ received per units of frequency from a  source in $z$ is related to the luminosity per units of emitted frequency by $\Phi(f)=1/(4\pi)\mathcal{L}_{\Gal}/(1+z)$, we see that an upper bound on $\Phi$ is translated into a lower bound in redshift and upper bound in luminosity (more precisely, it defines the region of integration in the plane $(z, \mathcal{L}_{\Gal})$. In other terms, we introduce a selection function which is 1 for $ \Phi<\Phi_{\text{cut}}$, and 0 otherwise. Explicitly in Eq.~(\ref{zut}) we introduce the selection function 
\be
W(z, \mathcal{L}_{\Gal})=\left\{
\begin{array}{ccc}
1 &\text{for}& \mathcal{L}_{\Gal}<4\pi\Phi_{\text{cut}}(1+z)\,,\\
0&\text{for}&\mathcal{L}_{\Gal}>4\pi\Phi_{\text{cut}}(1+z)\,.
\end{array}
\right.
\ee
Of course, if one can assume that all galaxies have associated the same luminosity, then the cut-off on flux translates directly into a lower cut-off in redshift (or analogously in $r$). Following the steps described in this section, we find that in Eq.~(\ref{qq}) both the theoretical prediction for the angular power spectrum and the Poisson noise involve an integration with the selection function. The selection function is regularizing the integral defining the shot noise part of the result, see Fig.\ref{PoissonNoise} for an illustration. Choosing a cut-off in flux correspond to filtering out resolvable sources: for a given detector, one can compute the signal to noise ratio as a function of flux and define $\Phi_{\rm cut}$ as the flux corresponding to a signal to noise bigger than a five threshold. For example choosing 5 as such a threshold one can find the cut in flux from 
\be
S/N(\Phi_{\text{cut}})\equiv 5\,.
\ee
The fact that the level of Poisson noise depends on the choice of cut-off in flux, is not surprising and the same result holds for the cosmic infrared background case, see e.g. Ref.~\cite{Ade:2013zsi}. 


\bibliography{myrefs}

\begin{thebibliography}{113}
\expandafter\ifx\csname natexlab\endcsname\relax\def\natexlab#1{#1}\fi
\expandafter\ifx\csname bibnamefont\endcsname\relax
  \def\bibnamefont#1{#1}\fi
\expandafter\ifx\csname bibfnamefont\endcsname\relax
  \def\bibfnamefont#1{#1}\fi
\expandafter\ifx\csname citenamefont\endcsname\relax
  \def\citenamefont#1{#1}\fi
\expandafter\ifx\csname url\endcsname\relax
  \def\url#1{\texttt{#1}}\fi
\expandafter\ifx\csname urlprefix\endcsname\relax\def\urlprefix{URL }\fi
\providecommand{\bibinfo}[2]{#2}
\providecommand{\eprint}[2][]{\url{#2}}

\bibitem[{\citenamefont{Penzias and Wilson}(1965)}]{Penzias:1965wn}
\bibinfo{author}{\bibfnamefont{A.~A.} \bibnamefont{Penzias}} \bibnamefont{and}
  \bibinfo{author}{\bibfnamefont{R.~W.} \bibnamefont{Wilson}},
  \bibinfo{journal}{Astrophys. J.} \textbf{\bibinfo{volume}{142}},
  \bibinfo{pages}{419} (\bibinfo{year}{1965}).

\bibitem[{\citenamefont{{Hauser} and {Dwek}}(2001)}]{2001ARA&A..39..249H}
\bibinfo{author}{\bibfnamefont{M.~G.} \bibnamefont{{Hauser}}} \bibnamefont{and}
  \bibinfo{author}{\bibfnamefont{E.}~\bibnamefont{{Dwek}}},
  \bibinfo{journal}{Annual Review of Astronomy and Astrophysics}
  \textbf{\bibinfo{volume}{39}}, \bibinfo{pages}{249} (\bibinfo{year}{2001}),
  \eprint{astro-ph/0105539}.

\bibitem[{\citenamefont{{Partridge} and {Peebles}}(1967)}]{1967ApJ...148..377P}
\bibinfo{author}{\bibfnamefont{R.~B.} \bibnamefont{{Partridge}}}
  \bibnamefont{and} \bibinfo{author}{\bibfnamefont{P.~J.~E.}
  \bibnamefont{{Peebles}}}, \bibinfo{journal}{Astrophys. J.}
  \textbf{\bibinfo{volume}{148}}, \bibinfo{pages}{377} (\bibinfo{year}{1967}).

\bibitem[{\citenamefont{{Shanks} et~al.}(1991)\citenamefont{{Shanks},
  {Georgantopoulos}, {Stewart}, {Pounds}, {Boyle}, and
  {Griffiths}}}]{1991Natur.353..315S}
\bibinfo{author}{\bibfnamefont{T.}~\bibnamefont{{Shanks}}},
  \bibinfo{author}{\bibfnamefont{I.}~\bibnamefont{{Georgantopoulos}}},
  \bibinfo{author}{\bibfnamefont{G.~C.} \bibnamefont{{Stewart}}},
  \bibinfo{author}{\bibfnamefont{K.~A.} \bibnamefont{{Pounds}}},
  \bibinfo{author}{\bibfnamefont{B.~J.} \bibnamefont{{Boyle}}},
  \bibnamefont{and} \bibinfo{author}{\bibfnamefont{R.~E.}
  \bibnamefont{{Griffiths}}}, \bibinfo{journal}{Nature}
  \textbf{\bibinfo{volume}{353}}, \bibinfo{pages}{315} (\bibinfo{year}{1991}).

\bibitem[{\citenamefont{{Hannestad}}(2006)}]{2006ARNPS..56..137H}
\bibinfo{author}{\bibfnamefont{S.}~\bibnamefont{{Hannestad}}},
  \bibinfo{journal}{Annual Review of Nuclear and Particle Science}
  \textbf{\bibinfo{volume}{56}}, \bibinfo{pages}{137} (\bibinfo{year}{2006}),
  \eprint{hep-ph/0602058}.

\bibitem[{\citenamefont{Abbott et~al.}(2016)}]{TheLIGOScientific:2016wyq}
\bibinfo{author}{\bibfnamefont{B.~P.} \bibnamefont{Abbott}}
  \bibnamefont{et~al.} (\bibinfo{collaboration}{Virgo, LIGO Scientific}),
  \bibinfo{journal}{Phys. Rev. Lett.} \textbf{\bibinfo{volume}{116}},
  \bibinfo{pages}{131102} (\bibinfo{year}{2016}), \eprint{1602.03847}.

\bibitem[{\citenamefont{Regimbau et~al.}(2016)\citenamefont{Regimbau, Evans,
  Christensen, Katsavounidis, Sathyaprakash, and Vitale}}]{Regimbau:2016ike}
\bibinfo{author}{\bibfnamefont{T.}~\bibnamefont{Regimbau}},
  \bibinfo{author}{\bibfnamefont{M.}~\bibnamefont{Evans}},
  \bibinfo{author}{\bibfnamefont{N.}~\bibnamefont{Christensen}},
  \bibinfo{author}{\bibfnamefont{E.}~\bibnamefont{Katsavounidis}},
  \bibinfo{author}{\bibfnamefont{B.}~\bibnamefont{Sathyaprakash}},
  \bibnamefont{and} \bibinfo{author}{\bibfnamefont{S.}~\bibnamefont{Vitale}}
  (\bibinfo{year}{2016}), \eprint{1611.08943}.

\bibitem[{\citenamefont{Mandic et~al.}(2016)\citenamefont{Mandic, Bird, and
  Cholis}}]{Mandic:2016lcn}
\bibinfo{author}{\bibfnamefont{V.}~\bibnamefont{Mandic}},
  \bibinfo{author}{\bibfnamefont{S.}~\bibnamefont{Bird}}, \bibnamefont{and}
  \bibinfo{author}{\bibfnamefont{I.}~\bibnamefont{Cholis}},
  \bibinfo{journal}{Phys. Rev. Lett.} \textbf{\bibinfo{volume}{117}},
  \bibinfo{pages}{201102} (\bibinfo{year}{2016}), \eprint{1608.06699}.

\bibitem[{\citenamefont{Dvorkin
  et~al.}(2016{\natexlab{a}})\citenamefont{Dvorkin, Uzan, Vangioni, and
  Silk}}]{Dvorkin:2016okx}
\bibinfo{author}{\bibfnamefont{I.}~\bibnamefont{Dvorkin}},
  \bibinfo{author}{\bibfnamefont{J.-P.} \bibnamefont{Uzan}},
  \bibinfo{author}{\bibfnamefont{E.}~\bibnamefont{Vangioni}}, \bibnamefont{and}
  \bibinfo{author}{\bibfnamefont{J.}~\bibnamefont{Silk}},
  \bibinfo{journal}{Phys. Rev.} \textbf{\bibinfo{volume}{D 94}},
  \bibinfo{pages}{103011} (\bibinfo{year}{2016}{\natexlab{a}}),
  \eprint{1607.06818}.

\bibitem[{\citenamefont{Nakazato et~al.}(2016)\citenamefont{Nakazato, Niino,
  and Sago}}]{Nakazato:2016nkj}
\bibinfo{author}{\bibfnamefont{K.}~\bibnamefont{Nakazato}},
  \bibinfo{author}{\bibfnamefont{Y.}~\bibnamefont{Niino}}, \bibnamefont{and}
  \bibinfo{author}{\bibfnamefont{N.}~\bibnamefont{Sago}},
  \bibinfo{journal}{Astrophys. J.} \textbf{\bibinfo{volume}{832}},
  \bibinfo{pages}{146} (\bibinfo{year}{2016}), \eprint{1605.02146}.

\bibitem[{\citenamefont{Dvorkin
  et~al.}(2016{\natexlab{b}})\citenamefont{Dvorkin, Vangioni, Silk, Uzan, and
  Olive}}]{Dvorkin:2016wac}
\bibinfo{author}{\bibfnamefont{I.}~\bibnamefont{Dvorkin}},
  \bibinfo{author}{\bibfnamefont{E.}~\bibnamefont{Vangioni}},
  \bibinfo{author}{\bibfnamefont{J.}~\bibnamefont{Silk}},
  \bibinfo{author}{\bibfnamefont{J.-P.} \bibnamefont{Uzan}}, \bibnamefont{and}
  \bibinfo{author}{\bibfnamefont{K.~A.} \bibnamefont{Olive}},
  \bibinfo{journal}{Mon. Not. Roy. Astron. Soc.}
  \textbf{\bibinfo{volume}{461}}, \bibinfo{pages}{3877}
  (\bibinfo{year}{2016}{\natexlab{b}}), \eprint{1604.04288}.

\bibitem[{\citenamefont{Evangelista and Araujo}(2014)}]{Evangelista:2014oba}
\bibinfo{author}{\bibfnamefont{E.~F.~D.} \bibnamefont{Evangelista}}
  \bibnamefont{and} \bibinfo{author}{\bibfnamefont{J.~C.~N.}
  \bibnamefont{Araujo}}, \bibinfo{journal}{Braz. J. Phys.}
  \textbf{\bibinfo{volume}{44}}, \bibinfo{pages}{824} (\bibinfo{year}{2014}),
  \eprint{1504.06605}.

\bibitem[{\citenamefont{Kelley et~al.}(2017)\citenamefont{Kelley, Blecha,
  Hernquist, and Sesana}}]{Kelley:2017lek}
\bibinfo{author}{\bibfnamefont{L.~Z.} \bibnamefont{Kelley}},
  \bibinfo{author}{\bibfnamefont{L.}~\bibnamefont{Blecha}},
  \bibinfo{author}{\bibfnamefont{L.}~\bibnamefont{Hernquist}},
  \bibnamefont{and} \bibinfo{author}{\bibfnamefont{A.}~\bibnamefont{Sesana}}
  (\bibinfo{year}{2017}), \eprint{1702.02180}.

\bibitem[{\citenamefont{Surace et~al.}(2016)\citenamefont{Surace, Kokkotas, and
  Pnigouras}}]{Surace:2015ppq}
\bibinfo{author}{\bibfnamefont{M.}~\bibnamefont{Surace}},
  \bibinfo{author}{\bibfnamefont{K.~D.} \bibnamefont{Kokkotas}},
  \bibnamefont{and}
  \bibinfo{author}{\bibfnamefont{P.}~\bibnamefont{Pnigouras}},
  \bibinfo{journal}{Astron. Astrophys.} \textbf{\bibinfo{volume}{586}},
  \bibinfo{pages}{A86} (\bibinfo{year}{2016}), \eprint{1512.02502}.

\bibitem[{\citenamefont{Talukder et~al.}(2014)\citenamefont{Talukder, Thrane,
  Bose, and Regimbau}}]{Talukder:2014eba}
\bibinfo{author}{\bibfnamefont{D.}~\bibnamefont{Talukder}},
  \bibinfo{author}{\bibfnamefont{E.}~\bibnamefont{Thrane}},
  \bibinfo{author}{\bibfnamefont{S.}~\bibnamefont{Bose}}, \bibnamefont{and}
  \bibinfo{author}{\bibfnamefont{T.}~\bibnamefont{Regimbau}},
  \bibinfo{journal}{Phys. Rev.} \textbf{\bibinfo{volume}{D89}},
  \bibinfo{pages}{123008} (\bibinfo{year}{2014}), \eprint{1404.4025}.

\bibitem[{\citenamefont{Lasky et~al.}(2013)\citenamefont{Lasky, Bennett, and
  Melatos}}]{Lasky:2013jfa}
\bibinfo{author}{\bibfnamefont{P.~D.} \bibnamefont{Lasky}},
  \bibinfo{author}{\bibfnamefont{M.~F.} \bibnamefont{Bennett}},
  \bibnamefont{and} \bibinfo{author}{\bibfnamefont{A.}~\bibnamefont{Melatos}},
  \bibinfo{journal}{Phys. Rev.} \textbf{\bibinfo{volume}{D87}},
  \bibinfo{pages}{063004} (\bibinfo{year}{2013}), \eprint{1302.6033}.

\bibitem[{\citenamefont{Crocker et~al.}(2017)\citenamefont{Crocker, Prestegard,
  Mandic, Regimbau, Olive, and Vangioni}}]{Crocker:2017agi}
\bibinfo{author}{\bibfnamefont{K.}~\bibnamefont{Crocker}},
  \bibinfo{author}{\bibfnamefont{T.}~\bibnamefont{Prestegard}},
  \bibinfo{author}{\bibfnamefont{V.}~\bibnamefont{Mandic}},
  \bibinfo{author}{\bibfnamefont{T.}~\bibnamefont{Regimbau}},
  \bibinfo{author}{\bibfnamefont{K.}~\bibnamefont{Olive}}, \bibnamefont{and}
  \bibinfo{author}{\bibfnamefont{E.}~\bibnamefont{Vangioni}}
  (\bibinfo{year}{2017}), \eprint{1701.02638}.

\bibitem[{\citenamefont{Crocker et~al.}(2015)\citenamefont{Crocker, Mandic,
  Regimbau, Belczynski, Gladysz, Olive, Prestegard, and
  Vangioni}}]{Crocker:2015taa}
\bibinfo{author}{\bibfnamefont{K.}~\bibnamefont{Crocker}},
  \bibinfo{author}{\bibfnamefont{V.}~\bibnamefont{Mandic}},
  \bibinfo{author}{\bibfnamefont{T.}~\bibnamefont{Regimbau}},
  \bibinfo{author}{\bibfnamefont{K.}~\bibnamefont{Belczynski}},
  \bibinfo{author}{\bibfnamefont{W.}~\bibnamefont{Gladysz}},
  \bibinfo{author}{\bibfnamefont{K.}~\bibnamefont{Olive}},
  \bibinfo{author}{\bibfnamefont{T.}~\bibnamefont{Prestegard}},
  \bibnamefont{and} \bibinfo{author}{\bibfnamefont{E.}~\bibnamefont{Vangioni}},
  \bibinfo{journal}{Phys. Rev.} \textbf{\bibinfo{volume}{D92}},
  \bibinfo{pages}{063005} (\bibinfo{year}{2015}), \eprint{1506.02631}.

\bibitem[{\citenamefont{Kowalska et~al.}(2012)\citenamefont{Kowalska, Bulik,
  and Belczynski}}]{Kowalska:2012ba}
\bibinfo{author}{\bibfnamefont{I.}~\bibnamefont{Kowalska}},
  \bibinfo{author}{\bibfnamefont{T.}~\bibnamefont{Bulik}}, \bibnamefont{and}
  \bibinfo{author}{\bibfnamefont{K.}~\bibnamefont{Belczynski}},
  \bibinfo{journal}{Astron. Astrophys.} \textbf{\bibinfo{volume}{541}},
  \bibinfo{pages}{A120} (\bibinfo{year}{2012}), \eprint{1202.3346}.

\bibitem[{\citenamefont{Moore et~al.}(2015)\citenamefont{Moore, Cole, and
  Berry}}]{Moore:2014lga}
\bibinfo{author}{\bibfnamefont{C.~J.} \bibnamefont{Moore}},
  \bibinfo{author}{\bibfnamefont{R.~H.} \bibnamefont{Cole}}, \bibnamefont{and}
  \bibinfo{author}{\bibfnamefont{C.~P.~L.} \bibnamefont{Berry}},
  \bibinfo{journal}{Class. Quant. Grav.} \textbf{\bibinfo{volume}{32}},
  \bibinfo{pages}{015014} (\bibinfo{year}{2015}), \eprint{1408.0740}.

\bibitem[{\citenamefont{{LIGO Scientific
  Collaboration}}(2015)}]{2015CQGra..32g4001L}
\bibinfo{author}{\bibnamefont{{LIGO Scientific Collaboration}}},
  \bibinfo{journal}{Classical and Quantum Gravity}
  \textbf{\bibinfo{volume}{32}}, \bibinfo{eid}{074001} (\bibinfo{year}{2015}),
  \eprint{1411.4547}.

\bibitem[{\citenamefont{{Acernese} et~al.}(2015)\citenamefont{{Acernese},
  {Agathos}, {Agatsuma}, {Aisa}, {Allemandou}, {Allocca}, {Amarni}, {Astone},
  {Balestri}, {Ballardin} et~al.}}]{2015CQGra..32b4001A}
\bibinfo{author}{\bibfnamefont{F.}~\bibnamefont{{Acernese}}},
  \bibinfo{author}{\bibfnamefont{M.}~\bibnamefont{{Agathos}}},
  \bibinfo{author}{\bibfnamefont{K.}~\bibnamefont{{Agatsuma}}},
  \bibinfo{author}{\bibfnamefont{D.}~\bibnamefont{{Aisa}}},
  \bibinfo{author}{\bibfnamefont{N.}~\bibnamefont{{Allemandou}}},
  \bibinfo{author}{\bibfnamefont{A.}~\bibnamefont{{Allocca}}},
  \bibinfo{author}{\bibfnamefont{J.}~\bibnamefont{{Amarni}}},
  \bibinfo{author}{\bibfnamefont{P.}~\bibnamefont{{Astone}}},
  \bibinfo{author}{\bibfnamefont{G.}~\bibnamefont{{Balestri}}},
  \bibinfo{author}{\bibfnamefont{G.}~\bibnamefont{{Ballardin}}},
  \bibnamefont{et~al.}, \bibinfo{journal}{Classical and Quantum Gravity}
  \textbf{\bibinfo{volume}{32}}, \bibinfo{eid}{024001} (\bibinfo{year}{2015}),
  \eprint{1408.3978}.

\bibitem[{\citenamefont{Abbott et~al.}(2017{\natexlab{a}})}]{Evans:2016mbw}
\bibinfo{author}{\bibfnamefont{B.~P.} \bibnamefont{Abbott}}
  \bibnamefont{et~al.} (\bibinfo{collaboration}{LIGO Scientific}),
  \bibinfo{journal}{Class. Quant. Grav.} \textbf{\bibinfo{volume}{34}},
  \bibinfo{pages}{044001} (\bibinfo{year}{2017}{\natexlab{a}}),
  \eprint{1607.08697}.

\bibitem[{\citenamefont{{The LIGO Scientific Collaboration} and {the Virgo
  Collaboration}}(2019{\natexlab{a}})}]{2019arXiv190302886T}
\bibinfo{author}{\bibnamefont{{The LIGO Scientific Collaboration}}}
  \bibnamefont{and} \bibinfo{author}{\bibnamefont{{the Virgo Collaboration}}},
  \bibinfo{journal}{arXiv e-prints} \bibinfo{eid}{arXiv:1903.02886}
  (\bibinfo{year}{2019}{\natexlab{a}}), \eprint{1903.02886}.

\bibitem[{\citenamefont{Maggiore}(2000)}]{Maggiore:1999vm}
\bibinfo{author}{\bibfnamefont{M.}~\bibnamefont{Maggiore}},
  \bibinfo{journal}{Phys. Rept.} \textbf{\bibinfo{volume}{331}},
  \bibinfo{pages}{283} (\bibinfo{year}{2000}), \eprint{gr-qc/9909001}.

\bibitem[{\citenamefont{Allen}(1996)}]{Allen:1996vm}
\bibinfo{author}{\bibfnamefont{B.}~\bibnamefont{Allen}}, in
  \emph{\bibinfo{booktitle}{{Relativistic gravitation and gravitational
  radiation. Proceedings, School of Physics, Les Houches, France, September
  26-October 6, 1995}}} (\bibinfo{year}{1996}), pp. \bibinfo{pages}{373--417},
  \eprint{gr-qc/9604033},
  \urlprefix\url{http://alice.cern.ch/format/showfull?sysnb=0223102}.

\bibitem[{\citenamefont{Smith et~al.}(2006)\citenamefont{Smith, Pierpaoli, and
  Kamionkowski}}]{Smith:2006nka}
\bibinfo{author}{\bibfnamefont{T.~L.} \bibnamefont{Smith}},
  \bibinfo{author}{\bibfnamefont{E.}~\bibnamefont{Pierpaoli}},
  \bibnamefont{and}
  \bibinfo{author}{\bibfnamefont{M.}~\bibnamefont{Kamionkowski}},
  \bibinfo{journal}{Phys. Rev. Lett.} \textbf{\bibinfo{volume}{97}},
  \bibinfo{pages}{021301} (\bibinfo{year}{2006}), \eprint{astro-ph/0603144}.

\bibitem[{\citenamefont{Henrot-Versille
  et~al.}(2015)}]{Henrot-Versille:2014jua}
\bibinfo{author}{\bibfnamefont{S.}~\bibnamefont{Henrot-Versille}}
  \bibnamefont{et~al.}, \bibinfo{journal}{Class. Quant. Grav.}
  \textbf{\bibinfo{volume}{32}}, \bibinfo{pages}{045003}
  (\bibinfo{year}{2015}), \eprint{1408.5299}.

\bibitem[{\citenamefont{{The LIGO Scientific Collaboration} and {the Virgo
  Collaboration}}(2018{\natexlab{a}})}]{2018arXiv181112907T}
\bibinfo{author}{\bibnamefont{{The LIGO Scientific Collaboration}}}
  \bibnamefont{and} \bibinfo{author}{\bibnamefont{{the Virgo Collaboration}}},
  \bibinfo{journal}{arXiv e-prints} \bibinfo{eid}{arXiv:1811.12907}
  (\bibinfo{year}{2018}{\natexlab{a}}), \eprint{1811.12907}.

\bibitem[{\citenamefont{Abbott et~al.}(2018)}]{Abbott:2017xzg}
\bibinfo{author}{\bibfnamefont{B.~P.} \bibnamefont{Abbott}}
  \bibnamefont{et~al.} (\bibinfo{collaboration}{Virgo, LIGO Scientific}),
  \bibinfo{journal}{Phys. Rev. Lett.} \textbf{\bibinfo{volume}{120}},
  \bibinfo{pages}{091101} (\bibinfo{year}{2018}), \eprint{1710.05837}.

\bibitem[{\citenamefont{Shannon et~al.}(2013)}]{Shannon:2013wma}
\bibinfo{author}{\bibfnamefont{R.~M.} \bibnamefont{Shannon}}
  \bibnamefont{et~al.}, \bibinfo{journal}{Science}
  \textbf{\bibinfo{volume}{342}}, \bibinfo{pages}{334} (\bibinfo{year}{2013}),
  \eprint{1310.4569}.

\bibitem[{\citenamefont{Allen and Ottewill}(1997)}]{Allen:1996gp}
\bibinfo{author}{\bibfnamefont{B.}~\bibnamefont{Allen}} \bibnamefont{and}
  \bibinfo{author}{\bibfnamefont{A.~C.} \bibnamefont{Ottewill}},
  \bibinfo{journal}{Phys. Rev.} \textbf{\bibinfo{volume}{D56}},
  \bibinfo{pages}{545} (\bibinfo{year}{1997}), \eprint{gr-qc/9607068}.

\bibitem[{\citenamefont{Cornish}(2001)}]{Cornish:2001hg}
\bibinfo{author}{\bibfnamefont{N.~J.} \bibnamefont{Cornish}},
  \bibinfo{journal}{Class. Quant. Grav.} \textbf{\bibinfo{volume}{18}},
  \bibinfo{pages}{4277} (\bibinfo{year}{2001}), \eprint{astro-ph/0105374}.

\bibitem[{\citenamefont{Mitra et~al.}(2008)\citenamefont{Mitra, Dhurandhar,
  Souradeep, Lazzarini, Mandic, Bose, and Ballmer}}]{Mitra:2007mc}
\bibinfo{author}{\bibfnamefont{S.}~\bibnamefont{Mitra}},
  \bibinfo{author}{\bibfnamefont{S.}~\bibnamefont{Dhurandhar}},
  \bibinfo{author}{\bibfnamefont{T.}~\bibnamefont{Souradeep}},
  \bibinfo{author}{\bibfnamefont{A.}~\bibnamefont{Lazzarini}},
  \bibinfo{author}{\bibfnamefont{V.}~\bibnamefont{Mandic}},
  \bibinfo{author}{\bibfnamefont{S.}~\bibnamefont{Bose}}, \bibnamefont{and}
  \bibinfo{author}{\bibfnamefont{S.}~\bibnamefont{Ballmer}},
  \bibinfo{journal}{Phys. Rev.} \textbf{\bibinfo{volume}{D77}},
  \bibinfo{pages}{042002} (\bibinfo{year}{2008}), \eprint{0708.2728}.

\bibitem[{\citenamefont{Thrane et~al.}(2009)\citenamefont{Thrane, Ballmer,
  Romano, Mitra, Talukder, Bose, and Mandic}}]{Thrane:2009fp}
\bibinfo{author}{\bibfnamefont{E.}~\bibnamefont{Thrane}},
  \bibinfo{author}{\bibfnamefont{S.}~\bibnamefont{Ballmer}},
  \bibinfo{author}{\bibfnamefont{J.~D.} \bibnamefont{Romano}},
  \bibinfo{author}{\bibfnamefont{S.}~\bibnamefont{Mitra}},
  \bibinfo{author}{\bibfnamefont{D.}~\bibnamefont{Talukder}},
  \bibinfo{author}{\bibfnamefont{S.}~\bibnamefont{Bose}}, \bibnamefont{and}
  \bibinfo{author}{\bibfnamefont{V.}~\bibnamefont{Mandic}},
  \bibinfo{journal}{Phys. Rev.} \textbf{\bibinfo{volume}{D80}},
  \bibinfo{pages}{122002} (\bibinfo{year}{2009}), \eprint{0910.0858}.

\bibitem[{\citenamefont{Romano et~al.}(2015)\citenamefont{Romano, Taylor,
  Cornish, Gair, Mingarelli, and van Haasteren}}]{Romano:2015uma}
\bibinfo{author}{\bibfnamefont{J.~D.} \bibnamefont{Romano}},
  \bibinfo{author}{\bibfnamefont{S.~R.} \bibnamefont{Taylor}},
  \bibinfo{author}{\bibfnamefont{N.~J.} \bibnamefont{Cornish}},
  \bibinfo{author}{\bibfnamefont{J.}~\bibnamefont{Gair}},
  \bibinfo{author}{\bibfnamefont{C.~M.~F.} \bibnamefont{Mingarelli}},
  \bibnamefont{and} \bibinfo{author}{\bibfnamefont{R.}~\bibnamefont{van
  Haasteren}}, \bibinfo{journal}{Phys. Rev.} \textbf{\bibinfo{volume}{D92}},
  \bibinfo{pages}{042003} (\bibinfo{year}{2015}), \eprint{1505.07179}.

\bibitem[{\citenamefont{Romano and Cornish}(2016)}]{Romano:2016dpx}
\bibinfo{author}{\bibfnamefont{J.~D.} \bibnamefont{Romano}} \bibnamefont{and}
  \bibinfo{author}{\bibfnamefont{N.~J.} \bibnamefont{Cornish}}
  (\bibinfo{year}{2016}), \eprint{1608.06889}.

\bibitem[{\citenamefont{Renzini and Contaldi}(2018)}]{Renzini:2018vkx}
\bibinfo{author}{\bibfnamefont{A.~I.} \bibnamefont{Renzini}} \bibnamefont{and}
  \bibinfo{author}{\bibfnamefont{C.~R.} \bibnamefont{Contaldi}}
  (\bibinfo{year}{2018}), \eprint{1806.11360}.

\bibitem[{\citenamefont{Abbott
  et~al.}(2017{\natexlab{b}})}]{TheLIGOScientific:2016xzw}
\bibinfo{author}{\bibfnamefont{B.~P.} \bibnamefont{Abbott}}
  \bibnamefont{et~al.} (\bibinfo{collaboration}{Virgo, LIGO Scientific}),
  \bibinfo{journal}{Phys. Rev. Lett.} \textbf{\bibinfo{volume}{118}},
  \bibinfo{pages}{121102} (\bibinfo{year}{2017}{\natexlab{b}}),
  \eprint{1612.02030}.

\bibitem[{\citenamefont{Mingarelli et~al.}(2013)\citenamefont{Mingarelli,
  Sidery, Mandel, and Vecchio}}]{Mingarelli:2013dsa}
\bibinfo{author}{\bibfnamefont{C.~M.~F.} \bibnamefont{Mingarelli}},
  \bibinfo{author}{\bibfnamefont{T.}~\bibnamefont{Sidery}},
  \bibinfo{author}{\bibfnamefont{I.}~\bibnamefont{Mandel}}, \bibnamefont{and}
  \bibinfo{author}{\bibfnamefont{A.}~\bibnamefont{Vecchio}},
  \bibinfo{journal}{Phys. Rev.} \textbf{\bibinfo{volume}{D88}},
  \bibinfo{pages}{062005} (\bibinfo{year}{2013}), \eprint{1306.5394}.

\bibitem[{\citenamefont{Taylor and Gair}(2013)}]{Taylor:2013esa}
\bibinfo{author}{\bibfnamefont{S.~R.} \bibnamefont{Taylor}} \bibnamefont{and}
  \bibinfo{author}{\bibfnamefont{J.~R.} \bibnamefont{Gair}},
  \bibinfo{journal}{Phys. Rev.} \textbf{\bibinfo{volume}{D88}},
  \bibinfo{pages}{084001} (\bibinfo{year}{2013}), \eprint{1306.5395}.

\bibitem[{\citenamefont{Gair et~al.}(2014)\citenamefont{Gair, Romano, Taylor,
  and Mingarelli}}]{Gair:2014rwa}
\bibinfo{author}{\bibfnamefont{J.}~\bibnamefont{Gair}},
  \bibinfo{author}{\bibfnamefont{J.~D.} \bibnamefont{Romano}},
  \bibinfo{author}{\bibfnamefont{S.}~\bibnamefont{Taylor}}, \bibnamefont{and}
  \bibinfo{author}{\bibfnamefont{C.~M.~F.} \bibnamefont{Mingarelli}},
  \bibinfo{journal}{Phys. Rev.} \textbf{\bibinfo{volume}{D90}},
  \bibinfo{pages}{082001} (\bibinfo{year}{2014}), \eprint{1406.4664}.

\bibitem[{\citenamefont{{The LIGO Scientific Collaboration} and {the Virgo
  Collaboration}}(2019{\natexlab{b}})}]{2019arXiv190308844T}
\bibinfo{author}{\bibnamefont{{The LIGO Scientific Collaboration}}}
  \bibnamefont{and} \bibinfo{author}{\bibnamefont{{the Virgo Collaboration}}},
  \bibinfo{journal}{arXiv e-prints} \bibinfo{eid}{arXiv:1903.08844}
  (\bibinfo{year}{2019}{\natexlab{b}}), \eprint{1903.08844}.

\bibitem[{\citenamefont{Cusin et~al.}(2019{\natexlab{a}})\citenamefont{Cusin,
  Durrer, and Ferreira}}]{Cusin:2018avf}
\bibinfo{author}{\bibfnamefont{G.}~\bibnamefont{Cusin}},
  \bibinfo{author}{\bibfnamefont{R.}~\bibnamefont{Durrer}}, \bibnamefont{and}
  \bibinfo{author}{\bibfnamefont{P.~G.} \bibnamefont{Ferreira}},
  \bibinfo{journal}{Phys. Rev.} \textbf{\bibinfo{volume}{D99}},
  \bibinfo{pages}{023534} (\bibinfo{year}{2019}{\natexlab{a}}),
  \eprint{1807.10620}.

\bibitem[{\citenamefont{Cusin et~al.}(2018{\natexlab{a}})\citenamefont{Cusin,
  Dvorkin, Pitrou, and Uzan}}]{Cusin:2018rsq}
\bibinfo{author}{\bibfnamefont{G.}~\bibnamefont{Cusin}},
  \bibinfo{author}{\bibfnamefont{I.}~\bibnamefont{Dvorkin}},
  \bibinfo{author}{\bibfnamefont{C.}~\bibnamefont{Pitrou}}, \bibnamefont{and}
  \bibinfo{author}{\bibfnamefont{J.-P.} \bibnamefont{Uzan}},
  \bibinfo{journal}{Phys. Rev. Lett.} \textbf{\bibinfo{volume}{120}},
  \bibinfo{pages}{231101} (\bibinfo{year}{2018}{\natexlab{a}}),
  \eprint{1803.03236}.

\bibitem[{\citenamefont{Contaldi}(2017)}]{Contaldi:2016koz}
\bibinfo{author}{\bibfnamefont{C.~R.} \bibnamefont{Contaldi}},
  \bibinfo{journal}{Phys. Lett.} \textbf{\bibinfo{volume}{B771}},
  \bibinfo{pages}{9} (\bibinfo{year}{2017}), \eprint{1609.08168}.

\bibitem[{\citenamefont{Cusin et~al.}(2018{\natexlab{b}})\citenamefont{Cusin,
  Pitrou, and Uzan}}]{Cusin:2017mjm}
\bibinfo{author}{\bibfnamefont{G.}~\bibnamefont{Cusin}},
  \bibinfo{author}{\bibfnamefont{C.}~\bibnamefont{Pitrou}}, \bibnamefont{and}
  \bibinfo{author}{\bibfnamefont{J.-P.} \bibnamefont{Uzan}},
  \bibinfo{journal}{Phys. Rev.} \textbf{\bibinfo{volume}{D97}},
  \bibinfo{pages}{123527} (\bibinfo{year}{2018}{\natexlab{b}}),
  \eprint{1711.11345}.

\bibitem[{\citenamefont{Cusin et~al.}(2017)\citenamefont{Cusin, Pitrou, and
  Uzan}}]{Cusin:2017fwz}
\bibinfo{author}{\bibfnamefont{G.}~\bibnamefont{Cusin}},
  \bibinfo{author}{\bibfnamefont{C.}~\bibnamefont{Pitrou}}, \bibnamefont{and}
  \bibinfo{author}{\bibfnamefont{J.-P.} \bibnamefont{Uzan}},
  \bibinfo{journal}{Phys. Rev.} \textbf{\bibinfo{volume}{D96}},
  \bibinfo{pages}{103019} (\bibinfo{year}{2017}), \eprint{1704.06184}.

\bibitem[{\citenamefont{Regimbau}(2011)}]{Regimbau:2011rp}
\bibinfo{author}{\bibfnamefont{T.}~\bibnamefont{Regimbau}},
  \bibinfo{journal}{Res. Astron. Astrophys.} \textbf{\bibinfo{volume}{11}},
  \bibinfo{pages}{369} (\bibinfo{year}{2011}), \eprint{1101.2762}.

\bibitem[{\citenamefont{Jenkins
  et~al.}(2018{\natexlab{a}})\citenamefont{Jenkins, Sakellariadou, Regimbau,
  and Slezak}}]{Jenkins:2018uac}
\bibinfo{author}{\bibfnamefont{A.~C.} \bibnamefont{Jenkins}},
  \bibinfo{author}{\bibfnamefont{M.}~\bibnamefont{Sakellariadou}},
  \bibinfo{author}{\bibfnamefont{T.}~\bibnamefont{Regimbau}}, \bibnamefont{and}
  \bibinfo{author}{\bibfnamefont{E.}~\bibnamefont{Slezak}},
  \bibinfo{journal}{Phys. Rev.} \textbf{\bibinfo{volume}{D98}},
  \bibinfo{pages}{063501} (\bibinfo{year}{2018}{\natexlab{a}}),
  \eprint{1806.01718}.

\bibitem[{\citenamefont{Jenkins
  et~al.}(2018{\natexlab{b}})\citenamefont{Jenkins, O'Shaughnessy,
  Sakellariadou, and Wysocki}}]{Jenkins:2018kxc}
\bibinfo{author}{\bibfnamefont{A.~C.} \bibnamefont{Jenkins}},
  \bibinfo{author}{\bibfnamefont{R.}~\bibnamefont{O'Shaughnessy}},
  \bibinfo{author}{\bibfnamefont{M.}~\bibnamefont{Sakellariadou}},
  \bibnamefont{and} \bibinfo{author}{\bibfnamefont{D.}~\bibnamefont{Wysocki}}
  (\bibinfo{year}{2018}{\natexlab{b}}), \eprint{1810.13435}.

\bibitem[{\citenamefont{Cusin et~al.}(2018{\natexlab{c}})\citenamefont{Cusin,
  Dvorkin, Pitrou, and Uzan}}]{Cusin:2018ump}
\bibinfo{author}{\bibfnamefont{G.}~\bibnamefont{Cusin}},
  \bibinfo{author}{\bibfnamefont{I.}~\bibnamefont{Dvorkin}},
  \bibinfo{author}{\bibfnamefont{C.}~\bibnamefont{Pitrou}}, \bibnamefont{and}
  \bibinfo{author}{\bibfnamefont{J.-P.} \bibnamefont{Uzan}}
  (\bibinfo{year}{2018}{\natexlab{c}}), \eprint{1811.03582}.

\bibitem[{\citenamefont{Maggiore}(2007)}]{Maggiore:1900zz}
\bibinfo{author}{\bibfnamefont{M.}~\bibnamefont{Maggiore}},
  \emph{\bibinfo{title}{{Gravitational Waves. Vol. 1. Theory and Experiments}}}
  (\bibinfo{publisher}{Oxford University Press, 574 p}, \bibinfo{year}{2007}).

\bibitem[{\citenamefont{Bonvin and Durrer}(2011)}]{Bonvin:2011bg}
\bibinfo{author}{\bibfnamefont{C.}~\bibnamefont{Bonvin}} \bibnamefont{and}
  \bibinfo{author}{\bibfnamefont{R.}~\bibnamefont{Durrer}},
  \bibinfo{journal}{Phys. Rev.} \textbf{\bibinfo{volume}{D84}},
  \bibinfo{pages}{063505} (\bibinfo{year}{2011}), \eprint{1105.5280}.

\bibitem[{\citenamefont{Andrianomena et~al.}(2014)}]{Andrianomena:2014sya}
\bibinfo{author}{\bibfnamefont{S.}~\bibnamefont{Andrianomena}}
  \bibnamefont{et~al.}, \bibinfo{journal}{JCAP}
  \textbf{\bibinfo{volume}{1406}}, \bibinfo{pages}{023} (\bibinfo{year}{2014}),
  \eprint{1402.4350}.

\bibitem[{\citenamefont{{Laureijs} et~al.}(2011)\citenamefont{{Laureijs},
  {Amiaux}, {Arduini}, {Augu{\`e}res}, {Brinchmann}, {Cole}, {Cropper},
  {Dabin}, {Duvet}, {Ealet} et~al.}}]{2011arXiv1110.3193L}
\bibinfo{author}{\bibfnamefont{R.}~\bibnamefont{{Laureijs}}},
  \bibinfo{author}{\bibfnamefont{J.}~\bibnamefont{{Amiaux}}},
  \bibinfo{author}{\bibfnamefont{S.}~\bibnamefont{{Arduini}}},
  \bibinfo{author}{\bibfnamefont{J.~.} \bibnamefont{{Augu{\`e}res}}},
  \bibinfo{author}{\bibfnamefont{J.}~\bibnamefont{{Brinchmann}}},
  \bibinfo{author}{\bibfnamefont{R.}~\bibnamefont{{Cole}}},
  \bibinfo{author}{\bibfnamefont{M.}~\bibnamefont{{Cropper}}},
  \bibinfo{author}{\bibfnamefont{C.}~\bibnamefont{{Dabin}}},
  \bibinfo{author}{\bibfnamefont{L.}~\bibnamefont{{Duvet}}},
  \bibinfo{author}{\bibfnamefont{A.}~\bibnamefont{{Ealet}}},
  \bibnamefont{et~al.}, \bibinfo{journal}{ArXiv e-prints}
  (\bibinfo{year}{2011}), \eprint{1110.3193}.

\bibitem[{\citenamefont{Harrison et~al.}(2016)\citenamefont{Harrison, Camera,
  Zuntz, and Brown}}]{Harrison:2016stv}
\bibinfo{author}{\bibfnamefont{I.}~\bibnamefont{Harrison}},
  \bibinfo{author}{\bibfnamefont{S.}~\bibnamefont{Camera}},
  \bibinfo{author}{\bibfnamefont{J.}~\bibnamefont{Zuntz}}, \bibnamefont{and}
  \bibinfo{author}{\bibfnamefont{M.~L.} \bibnamefont{Brown}},
  \bibinfo{journal}{Mon. Not. Roy. Astron. Soc.}
  \textbf{\bibinfo{volume}{463}}, \bibinfo{pages}{3674} (\bibinfo{year}{2016}),
  \eprint{1601.03947}.

\bibitem[{\citenamefont{Ade et~al.}(2016)}]{Ade:2015xua}
\bibinfo{author}{\bibfnamefont{P.~A.~R.} \bibnamefont{Ade}}
  \bibnamefont{et~al.} (\bibinfo{collaboration}{Planck}),
  \bibinfo{journal}{Astron. Astrophys.} \textbf{\bibinfo{volume}{594}},
  \bibinfo{pages}{A13} (\bibinfo{year}{2016}), \eprint{1502.01589}.

\bibitem[{CMB()}]{CMBquick}
\bibinfo{howpublished}{\url{http://www2.iap.fr/users/pitrou/cmbquick.htm}}.

\bibitem[{\citenamefont{Smith et~al.}(2003)\citenamefont{Smith, Peacock,
  Jenkins, White, Frenk, Pearce, Thomas, Efstathiou, and
  Couchmann}}]{Smith:2002dz}
\bibinfo{author}{\bibfnamefont{R.~E.} \bibnamefont{Smith}},
  \bibinfo{author}{\bibfnamefont{J.~A.} \bibnamefont{Peacock}},
  \bibinfo{author}{\bibfnamefont{A.}~\bibnamefont{Jenkins}},
  \bibinfo{author}{\bibfnamefont{S.~D.~M.} \bibnamefont{White}},
  \bibinfo{author}{\bibfnamefont{C.~S.} \bibnamefont{Frenk}},
  \bibinfo{author}{\bibfnamefont{F.~R.} \bibnamefont{Pearce}},
  \bibinfo{author}{\bibfnamefont{P.~A.} \bibnamefont{Thomas}},
  \bibinfo{author}{\bibfnamefont{G.}~\bibnamefont{Efstathiou}},
  \bibnamefont{and} \bibinfo{author}{\bibfnamefont{H.~M.~P.}
  \bibnamefont{Couchmann}} (\bibinfo{collaboration}{VIRGO Consortium}),
  \bibinfo{journal}{Mon. Not. Roy. Astron. Soc.}
  \textbf{\bibinfo{volume}{341}}, \bibinfo{pages}{1311} (\bibinfo{year}{2003}),
  \eprint{astro-ph/0207664}.

\bibitem[{\citenamefont{Marin et~al.}(2013)}]{Marin:2013bbb}
\bibinfo{author}{\bibfnamefont{F.~A.} \bibnamefont{Marin}} \bibnamefont{et~al.}
  (\bibinfo{collaboration}{WiggleZ}), \bibinfo{journal}{Mon. Not. Roy. Astron.
  Soc.} \textbf{\bibinfo{volume}{432}}, \bibinfo{pages}{2654}
  (\bibinfo{year}{2013}), \eprint{1303.6644}.

\bibitem[{\citenamefont{Rassat et~al.}(2008)\citenamefont{Rassat, Amara,
  Amendola, Castander, Kitching, Kunz, Refregier, Wang, and
  Weller}}]{Rassat:2008ja}
\bibinfo{author}{\bibfnamefont{A.}~\bibnamefont{Rassat}},
  \bibinfo{author}{\bibfnamefont{A.}~\bibnamefont{Amara}},
  \bibinfo{author}{\bibfnamefont{L.}~\bibnamefont{Amendola}},
  \bibinfo{author}{\bibfnamefont{F.~J.} \bibnamefont{Castander}},
  \bibinfo{author}{\bibfnamefont{T.}~\bibnamefont{Kitching}},
  \bibinfo{author}{\bibfnamefont{M.}~\bibnamefont{Kunz}},
  \bibinfo{author}{\bibfnamefont{A.}~\bibnamefont{Refregier}},
  \bibinfo{author}{\bibfnamefont{Y.}~\bibnamefont{Wang}}, \bibnamefont{and}
  \bibinfo{author}{\bibfnamefont{J.}~\bibnamefont{Weller}}
  (\bibinfo{year}{2008}), \eprint{0810.0003}.

\bibitem[{\citenamefont{{Anderson} et~al.}(2012)\citenamefont{{Anderson},
  {Aubourg}, {Bailey}, {Bizyaev}, {Blanton}, {Bolton}, {Brinkmann},
  {Brownstein}, {Burden}, {Cuesta} et~al.}}]{2012MNRAS.427.3435A}
\bibinfo{author}{\bibfnamefont{L.}~\bibnamefont{{Anderson}}},
  \bibinfo{author}{\bibfnamefont{E.}~\bibnamefont{{Aubourg}}},
  \bibinfo{author}{\bibfnamefont{S.}~\bibnamefont{{Bailey}}},
  \bibinfo{author}{\bibfnamefont{D.}~\bibnamefont{{Bizyaev}}},
  \bibinfo{author}{\bibfnamefont{M.}~\bibnamefont{{Blanton}}},
  \bibinfo{author}{\bibfnamefont{A.~S.} \bibnamefont{{Bolton}}},
  \bibinfo{author}{\bibfnamefont{J.}~\bibnamefont{{Brinkmann}}},
  \bibinfo{author}{\bibfnamefont{J.~R.} \bibnamefont{{Brownstein}}},
  \bibinfo{author}{\bibfnamefont{A.}~\bibnamefont{{Burden}}},
  \bibinfo{author}{\bibfnamefont{A.~J.} \bibnamefont{{Cuesta}}},
  \bibnamefont{et~al.}, \bibinfo{journal}{MNRAS}
  \textbf{\bibinfo{volume}{427}}, \bibinfo{pages}{3435} (\bibinfo{year}{2012}),
  \eprint{1203.6594}.

\bibitem[{\citenamefont{LoVerde and Afshordi}(2008)}]{LoVerde:2008re}
\bibinfo{author}{\bibfnamefont{M.}~\bibnamefont{LoVerde}} \bibnamefont{and}
  \bibinfo{author}{\bibfnamefont{N.}~\bibnamefont{Afshordi}},
  \bibinfo{journal}{Phys. Rev.} \textbf{\bibinfo{volume}{D78}},
  \bibinfo{pages}{123506} (\bibinfo{year}{2008}), \eprint{0809.5112}.

\bibitem[{\citenamefont{Bernardeau et~al.}(2011)\citenamefont{Bernardeau,
  Pitrou, and Uzan}}]{PitrouFlat}
\bibinfo{author}{\bibfnamefont{F.}~\bibnamefont{Bernardeau}},
  \bibinfo{author}{\bibfnamefont{C.}~\bibnamefont{Pitrou}}, \bibnamefont{and}
  \bibinfo{author}{\bibfnamefont{J.-P.} \bibnamefont{Uzan}},
  \bibinfo{journal}{JCAP} \textbf{\bibinfo{volume}{1102}}, \bibinfo{pages}{015}
  (\bibinfo{year}{2011}), \eprint{1012.2652}.

\bibitem[{\citenamefont{Jenkins and Sakellariadou}(2019)}]{Jenkins:2019uzp}
\bibinfo{author}{\bibfnamefont{A.~C.} \bibnamefont{Jenkins}} \bibnamefont{and}
  \bibinfo{author}{\bibfnamefont{M.}~\bibnamefont{Sakellariadou}}
  (\bibinfo{year}{2019}), \eprint{1902.07719}.

\bibitem[{\citenamefont{Ade et~al.}(2014)}]{Ade:2013zsi}
\bibinfo{author}{\bibfnamefont{P.~A.~R.} \bibnamefont{Ade}}
  \bibnamefont{et~al.} (\bibinfo{collaboration}{Planck}),
  \bibinfo{journal}{Astron. Astrophys.} \textbf{\bibinfo{volume}{571}},
  \bibinfo{pages}{A30} (\bibinfo{year}{2014}), \eprint{1309.0382}.

\bibitem[{\citenamefont{{LIGO Scientific Collaboration} and {Virgo
  Collaboration}}(2017)}]{2017PhRvL.118l1102A}
\bibinfo{author}{\bibnamefont{{LIGO Scientific Collaboration}}}
  \bibnamefont{and} \bibinfo{author}{\bibnamefont{{Virgo Collaboration}}},
  \bibinfo{journal}{Phys. Rev. Letters} \textbf{\bibinfo{volume}{118}},
  \bibinfo{eid}{121102} (\bibinfo{year}{2017}), \eprint{1612.02030}.

\bibitem[{\citenamefont{{Dvorkin} et~al.}(2018)\citenamefont{{Dvorkin}, {Uzan},
  {Vangioni}, and {Silk}}}]{2018MNRAS.479..121D}
\bibinfo{author}{\bibfnamefont{I.}~\bibnamefont{{Dvorkin}}},
  \bibinfo{author}{\bibfnamefont{J.-P.} \bibnamefont{{Uzan}}},
  \bibinfo{author}{\bibfnamefont{E.}~\bibnamefont{{Vangioni}}},
  \bibnamefont{and} \bibinfo{author}{\bibfnamefont{J.}~\bibnamefont{{Silk}}},
  \bibinfo{journal}{Mon. Not. Roy. Astron. Soc.}
  \textbf{\bibinfo{volume}{479}}, \bibinfo{pages}{121} (\bibinfo{year}{2018}),
  \eprint{1709.09197}.

\bibitem[{\citenamefont{{Behroozi} et~al.}(2013)\citenamefont{{Behroozi},
  {Wechsler}, and {Conroy}}}]{2013ApJ...770...57B}
\bibinfo{author}{\bibfnamefont{P.~S.} \bibnamefont{{Behroozi}}},
  \bibinfo{author}{\bibfnamefont{R.~H.} \bibnamefont{{Wechsler}}},
  \bibnamefont{and} \bibinfo{author}{\bibfnamefont{C.}~\bibnamefont{{Conroy}}},
  \bibinfo{journal}{Astrophys. J.} \textbf{\bibinfo{volume}{770}},
  \bibinfo{eid}{57} (\bibinfo{year}{2013}), \eprint{1207.6105}.

\bibitem[{\citenamefont{{Salpeter}}(1955)}]{1955ApJ...121..161S}
\bibinfo{author}{\bibfnamefont{E.~E.} \bibnamefont{{Salpeter}}},
  \bibinfo{journal}{Astrophys. J.} \textbf{\bibinfo{volume}{121}},
  \bibinfo{pages}{161} (\bibinfo{year}{1955}).

\bibitem[{\citenamefont{{Ma} et~al.}(2016)\citenamefont{{Ma}, {Hopkins},
  {Faucher-Gigu{\`e}re}, {Zolman}, {Muratov}, {Kere{\v s}}, and
  {Quataert}}}]{2016MNRAS.456.2140M}
\bibinfo{author}{\bibfnamefont{X.}~\bibnamefont{{Ma}}},
  \bibinfo{author}{\bibfnamefont{P.~F.} \bibnamefont{{Hopkins}}},
  \bibinfo{author}{\bibfnamefont{C.-A.} \bibnamefont{{Faucher-Gigu{\`e}re}}},
  \bibinfo{author}{\bibfnamefont{N.}~\bibnamefont{{Zolman}}},
  \bibinfo{author}{\bibfnamefont{A.~L.} \bibnamefont{{Muratov}}},
  \bibinfo{author}{\bibfnamefont{D.}~\bibnamefont{{Kere{\v s}}}},
  \bibnamefont{and}
  \bibinfo{author}{\bibfnamefont{E.}~\bibnamefont{{Quataert}}},
  \bibinfo{journal}{Mon. Not. Roy. Astron. Soc.}
  \textbf{\bibinfo{volume}{456}}, \bibinfo{pages}{2140} (\bibinfo{year}{2016}),
  \eprint{1504.02097}.

\bibitem[{\citenamefont{{Lower} et~al.}(2018)\citenamefont{{Lower}, {Thrane},
  {Lasky}, and {Smith}}}]{2018PhRvD..98h3028L}
\bibinfo{author}{\bibfnamefont{M.~E.} \bibnamefont{{Lower}}},
  \bibinfo{author}{\bibfnamefont{E.}~\bibnamefont{{Thrane}}},
  \bibinfo{author}{\bibfnamefont{P.~D.} \bibnamefont{{Lasky}}},
  \bibnamefont{and} \bibinfo{author}{\bibfnamefont{R.}~\bibnamefont{{Smith}}},
  \bibinfo{journal}{Phys. Rev. D.} \textbf{\bibinfo{volume}{98}},
  \bibinfo{eid}{083028} (\bibinfo{year}{2018}), \eprint{1806.05350}.

\bibitem[{\citenamefont{{Tinker} et~al.}(2008)\citenamefont{{Tinker},
  {Kravtsov}, {Klypin}, {Abazajian}, {Warren}, {Yepes}, {Gottl{\"o}ber}, and
  {Holz}}}]{2008ApJ...688..709T}
\bibinfo{author}{\bibfnamefont{J.}~\bibnamefont{{Tinker}}},
  \bibinfo{author}{\bibfnamefont{A.~V.} \bibnamefont{{Kravtsov}}},
  \bibinfo{author}{\bibfnamefont{A.}~\bibnamefont{{Klypin}}},
  \bibinfo{author}{\bibfnamefont{K.}~\bibnamefont{{Abazajian}}},
  \bibinfo{author}{\bibfnamefont{M.}~\bibnamefont{{Warren}}},
  \bibinfo{author}{\bibfnamefont{G.}~\bibnamefont{{Yepes}}},
  \bibinfo{author}{\bibfnamefont{S.}~\bibnamefont{{Gottl{\"o}ber}}},
  \bibnamefont{and} \bibinfo{author}{\bibfnamefont{D.~E.}
  \bibnamefont{{Holz}}}, \bibinfo{journal}{Astrophys. J.}
  \textbf{\bibinfo{volume}{688}}, \bibinfo{pages}{709} (\bibinfo{year}{2008}),
  \eprint{0803.2706}.

\bibitem[{\citenamefont{Collaboration}({\natexlab{a}})}]{LIGOcurves1}
\bibinfo{author}{\bibfnamefont{L.~S.} \bibnamefont{Collaboration}},
  \bibinfo{note}{lIGO-G1401390}.

\bibitem[{\citenamefont{Collaboration}({\natexlab{b}})}]{LIGOcurves2}
\bibinfo{author}{\bibfnamefont{L.~S.} \bibnamefont{Collaboration}},
  \bibinfo{note}{lIGO-G1801952}.

\bibitem[{\citenamefont{{Finn} and {Chernoff}}(1993)}]{1993PhRvD..47.2198F}
\bibinfo{author}{\bibfnamefont{L.~S.} \bibnamefont{{Finn}}} \bibnamefont{and}
  \bibinfo{author}{\bibfnamefont{D.~F.} \bibnamefont{{Chernoff}}},
  \bibinfo{journal}{Phys. Rev. D} \textbf{\bibinfo{volume}{47}},
  \bibinfo{pages}{2198} (\bibinfo{year}{1993}), \eprint{gr-qc/9301003}.

\bibitem[{\citenamefont{{Ajith} et~al.}(2011)\citenamefont{{Ajith}, {Hannam},
  {Husa}, {Chen}, {Br{\"u}gmann}, {Dorband}, {M{\"u}ller}, {Ohme}, {Pollney},
  {Reisswig} et~al.}}]{2011PhRvL.106x1101A}
\bibinfo{author}{\bibfnamefont{P.}~\bibnamefont{{Ajith}}},
  \bibinfo{author}{\bibfnamefont{M.}~\bibnamefont{{Hannam}}},
  \bibinfo{author}{\bibfnamefont{S.}~\bibnamefont{{Husa}}},
  \bibinfo{author}{\bibfnamefont{Y.}~\bibnamefont{{Chen}}},
  \bibinfo{author}{\bibfnamefont{B.}~\bibnamefont{{Br{\"u}gmann}}},
  \bibinfo{author}{\bibfnamefont{N.}~\bibnamefont{{Dorband}}},
  \bibinfo{author}{\bibfnamefont{D.}~\bibnamefont{{M{\"u}ller}}},
  \bibinfo{author}{\bibfnamefont{F.}~\bibnamefont{{Ohme}}},
  \bibinfo{author}{\bibfnamefont{D.}~\bibnamefont{{Pollney}}},
  \bibinfo{author}{\bibfnamefont{C.}~\bibnamefont{{Reisswig}}},
  \bibnamefont{et~al.}, \bibinfo{journal}{Physical Review Letters}
  \textbf{\bibinfo{volume}{106}}, \bibinfo{eid}{241101} (\bibinfo{year}{2011}),
  \eprint{0909.2867}.

\bibitem[{\citenamefont{{Fryer} et~al.}(2012)\citenamefont{{Fryer},
  {Belczynski}, {Wiktorowicz}, {Dominik}, {Kalogera}, and
  {Holz}}}]{2012ApJ...749...91F}
\bibinfo{author}{\bibfnamefont{C.~L.} \bibnamefont{{Fryer}}},
  \bibinfo{author}{\bibfnamefont{K.}~\bibnamefont{{Belczynski}}},
  \bibinfo{author}{\bibfnamefont{G.}~\bibnamefont{{Wiktorowicz}}},
  \bibinfo{author}{\bibfnamefont{M.}~\bibnamefont{{Dominik}}},
  \bibinfo{author}{\bibfnamefont{V.}~\bibnamefont{{Kalogera}}},
  \bibnamefont{and} \bibinfo{author}{\bibfnamefont{D.~E.}
  \bibnamefont{{Holz}}}, \bibinfo{journal}{Astrophys. J.}
  \textbf{\bibinfo{volume}{749}}, \bibinfo{eid}{91} (\bibinfo{year}{2012}),
  \eprint{1110.1726}.

\bibitem[{\citenamefont{{The LIGO Scientific Collaboration} and {the Virgo
  Collaboration}}(2018{\natexlab{b}})}]{2018arXiv181112940T}
\bibinfo{author}{\bibnamefont{{The LIGO Scientific Collaboration}}}
  \bibnamefont{and} \bibinfo{author}{\bibnamefont{{the Virgo Collaboration}}},
  \bibinfo{journal}{arXiv e-prints} \bibinfo{eid}{arXiv:1811.12940}
  (\bibinfo{year}{2018}{\natexlab{b}}), \eprint{1811.12940}.

\bibitem[{\citenamefont{{Belczynski}
  et~al.}(2016{\natexlab{a}})\citenamefont{{Belczynski}, {Holz}, {Bulik}, and
  {O'Shaughnessy}}}]{2016Natur.534..512B}
\bibinfo{author}{\bibfnamefont{K.}~\bibnamefont{{Belczynski}}},
  \bibinfo{author}{\bibfnamefont{D.~E.} \bibnamefont{{Holz}}},
  \bibinfo{author}{\bibfnamefont{T.}~\bibnamefont{{Bulik}}}, \bibnamefont{and}
  \bibinfo{author}{\bibfnamefont{R.}~\bibnamefont{{O'Shaughnessy}}},
  \bibinfo{journal}{Nature} \textbf{\bibinfo{volume}{534}},
  \bibinfo{pages}{512} (\bibinfo{year}{2016}{\natexlab{a}}),
  \eprint{1602.04531}.

\bibitem[{\citenamefont{{Dominik} et~al.}(2012)\citenamefont{{Dominik},
  {Belczynski}, {Fryer}, {Holz}, {Berti}, {Bulik}, {Mandel}, and
  {O'Shaughnessy}}}]{2012ApJ...759...52D}
\bibinfo{author}{\bibfnamefont{M.}~\bibnamefont{{Dominik}}},
  \bibinfo{author}{\bibfnamefont{K.}~\bibnamefont{{Belczynski}}},
  \bibinfo{author}{\bibfnamefont{C.}~\bibnamefont{{Fryer}}},
  \bibinfo{author}{\bibfnamefont{D.~E.} \bibnamefont{{Holz}}},
  \bibinfo{author}{\bibfnamefont{E.}~\bibnamefont{{Berti}}},
  \bibinfo{author}{\bibfnamefont{T.}~\bibnamefont{{Bulik}}},
  \bibinfo{author}{\bibfnamefont{I.}~\bibnamefont{{Mandel}}}, \bibnamefont{and}
  \bibinfo{author}{\bibfnamefont{R.}~\bibnamefont{{O'Shaughnessy}}},
  \bibinfo{journal}{Astrophys. J.} \textbf{\bibinfo{volume}{759}},
  \bibinfo{eid}{52} (\bibinfo{year}{2012}), \eprint{1202.4901}.

\bibitem[{\citenamefont{{Belczynski}
  et~al.}(2016{\natexlab{b}})\citenamefont{{Belczynski}, {Heger}, {Gladysz},
  {Ruiter}, {Woosley}, {Wiktorowicz}, {Chen}, {Bulik}, {O'Shaughnessy}, {Holz}
  et~al.}}]{2016A&A...594A..97B}
\bibinfo{author}{\bibfnamefont{K.}~\bibnamefont{{Belczynski}}},
  \bibinfo{author}{\bibfnamefont{A.}~\bibnamefont{{Heger}}},
  \bibinfo{author}{\bibfnamefont{W.}~\bibnamefont{{Gladysz}}},
  \bibinfo{author}{\bibfnamefont{A.~J.} \bibnamefont{{Ruiter}}},
  \bibinfo{author}{\bibfnamefont{S.}~\bibnamefont{{Woosley}}},
  \bibinfo{author}{\bibfnamefont{G.}~\bibnamefont{{Wiktorowicz}}},
  \bibinfo{author}{\bibfnamefont{H.-Y.} \bibnamefont{{Chen}}},
  \bibinfo{author}{\bibfnamefont{T.}~\bibnamefont{{Bulik}}},
  \bibinfo{author}{\bibfnamefont{R.}~\bibnamefont{{O'Shaughnessy}}},
  \bibinfo{author}{\bibfnamefont{D.~E.} \bibnamefont{{Holz}}},
  \bibnamefont{et~al.}, \bibinfo{journal}{Astron. Astrophys.}
  \textbf{\bibinfo{volume}{594}}, \bibinfo{eid}{A97}
  (\bibinfo{year}{2016}{\natexlab{b}}), \eprint{1607.03116}.

\bibitem[{\citenamefont{{O'Connor} and {Ott}}(2011)}]{2011ApJ...730...70O}
\bibinfo{author}{\bibfnamefont{E.}~\bibnamefont{{O'Connor}}} \bibnamefont{and}
  \bibinfo{author}{\bibfnamefont{C.~D.} \bibnamefont{{Ott}}},
  \bibinfo{journal}{Astrophys. J} \textbf{\bibinfo{volume}{730}},
  \bibinfo{eid}{70} (\bibinfo{year}{2011}), \eprint{1010.5550}.

\bibitem[{\citenamefont{{Mandel} and {de Mink}}(2016)}]{2016MNRAS.458.2634M}
\bibinfo{author}{\bibfnamefont{I.}~\bibnamefont{{Mandel}}} \bibnamefont{and}
  \bibinfo{author}{\bibfnamefont{S.~E.} \bibnamefont{{de Mink}}},
  \bibinfo{journal}{Mon. Not. Roy. Astron. Soc.}
  \textbf{\bibinfo{volume}{458}}, \bibinfo{pages}{2634} (\bibinfo{year}{2016}),
  \eprint{1601.00007}.

\bibitem[{\citenamefont{{Marchant} et~al.}(2016)\citenamefont{{Marchant},
  {Langer}, {Podsiadlowski}, {Tauris}, and {Moriya}}}]{2016A&A...588A..50M}
\bibinfo{author}{\bibfnamefont{P.}~\bibnamefont{{Marchant}}},
  \bibinfo{author}{\bibfnamefont{N.}~\bibnamefont{{Langer}}},
  \bibinfo{author}{\bibfnamefont{P.}~\bibnamefont{{Podsiadlowski}}},
  \bibinfo{author}{\bibfnamefont{T.~M.} \bibnamefont{{Tauris}}},
  \bibnamefont{and} \bibinfo{author}{\bibfnamefont{T.~J.}
  \bibnamefont{{Moriya}}}, \bibinfo{journal}{Astron. Astrophys.}
  \textbf{\bibinfo{volume}{588}}, \bibinfo{eid}{A50} (\bibinfo{year}{2016}),
  \eprint{1601.03718}.

\bibitem[{\citenamefont{{Limongi}}(2017)}]{2017hsn..book..513L}
\bibinfo{author}{\bibfnamefont{M.}~\bibnamefont{{Limongi}}},
  \emph{\bibinfo{title}{{Supernovae from Massive Stars}}}
  (\bibinfo{year}{2017}), p. \bibinfo{pages}{513}.

\bibitem[{\citenamefont{{Rodriguez} et~al.}(2016)\citenamefont{{Rodriguez},
  {Chatterjee}, and {Rasio}}}]{2016PhRvD..93h4029R}
\bibinfo{author}{\bibfnamefont{C.~L.} \bibnamefont{{Rodriguez}}},
  \bibinfo{author}{\bibfnamefont{S.}~\bibnamefont{{Chatterjee}}},
  \bibnamefont{and} \bibinfo{author}{\bibfnamefont{F.~A.}
  \bibnamefont{{Rasio}}}, \bibinfo{journal}{Phys. Rev. D}
  \textbf{\bibinfo{volume}{93}}, \bibinfo{eid}{084029} (\bibinfo{year}{2016}),
  \eprint{1602.02444}.

\bibitem[{\citenamefont{{Fryer} and {Kalogera}}(2001)}]{2001ApJ...554..548F}
\bibinfo{author}{\bibfnamefont{C.~L.} \bibnamefont{{Fryer}}} \bibnamefont{and}
  \bibinfo{author}{\bibfnamefont{V.}~\bibnamefont{{Kalogera}}},
  \bibinfo{journal}{Astrophys. J.} \textbf{\bibinfo{volume}{554}},
  \bibinfo{pages}{548} (\bibinfo{year}{2001}), \eprint{astro-ph/9911312}.

\bibitem[{\citenamefont{{Blondin} et~al.}(2003)\citenamefont{{Blondin},
  {Mezzacappa}, and {DeMarino}}}]{2003ApJ...584..971B}
\bibinfo{author}{\bibfnamefont{J.~M.} \bibnamefont{{Blondin}}},
  \bibinfo{author}{\bibfnamefont{A.}~\bibnamefont{{Mezzacappa}}},
  \bibnamefont{and}
  \bibinfo{author}{\bibfnamefont{C.}~\bibnamefont{{DeMarino}}},
  \bibinfo{journal}{Astrophys. J.} \textbf{\bibinfo{volume}{584}},
  \bibinfo{pages}{971} (\bibinfo{year}{2003}), \eprint{astro-ph/0210634}.

\bibitem[{\citenamefont{{Woosley} et~al.}(2002)\citenamefont{{Woosley},
  {Heger}, and {Weaver}}}]{2002RvMP...74.1015W}
\bibinfo{author}{\bibfnamefont{S.~E.} \bibnamefont{{Woosley}}},
  \bibinfo{author}{\bibfnamefont{A.}~\bibnamefont{{Heger}}}, \bibnamefont{and}
  \bibinfo{author}{\bibfnamefont{T.~A.} \bibnamefont{{Weaver}}},
  \bibinfo{journal}{Reviews of Modern Physics} \textbf{\bibinfo{volume}{74}},
  \bibinfo{pages}{1015} (\bibinfo{year}{2002}).

\bibitem[{\citenamefont{{Kasen} et~al.}(2011)\citenamefont{{Kasen}, {Woosley},
  and {Heger}}}]{2011ApJ...734..102K}
\bibinfo{author}{\bibfnamefont{D.}~\bibnamefont{{Kasen}}},
  \bibinfo{author}{\bibfnamefont{S.~E.} \bibnamefont{{Woosley}}},
  \bibnamefont{and} \bibinfo{author}{\bibfnamefont{A.}~\bibnamefont{{Heger}}},
  \bibinfo{journal}{Astrophys. J.} \textbf{\bibinfo{volume}{734}},
  \bibinfo{eid}{102} (\bibinfo{year}{2011}), \eprint{1101.3336}.

\bibitem[{\citenamefont{{Woosley} et~al.}(2007)\citenamefont{{Woosley},
  {Blinnikov}, and {Heger}}}]{2007Natur.450..390W}
\bibinfo{author}{\bibfnamefont{S.~E.} \bibnamefont{{Woosley}}},
  \bibinfo{author}{\bibfnamefont{S.}~\bibnamefont{{Blinnikov}}},
  \bibnamefont{and} \bibinfo{author}{\bibfnamefont{A.}~\bibnamefont{{Heger}}},
  \bibinfo{journal}{Nature} \textbf{\bibinfo{volume}{450}},
  \bibinfo{pages}{390} (\bibinfo{year}{2007}), \eprint{0710.3314}.

\bibitem[{\citenamefont{{Woosley} and {Heger}}(2015)}]{2015ASSL..412..199W}
\bibinfo{author}{\bibfnamefont{S.~E.} \bibnamefont{{Woosley}}}
  \bibnamefont{and} \bibinfo{author}{\bibfnamefont{A.}~\bibnamefont{{Heger}}},
  in \emph{\bibinfo{booktitle}{Very Massive Stars in the Local Universe}},
  edited by \bibinfo{editor}{\bibfnamefont{J.~S.} \bibnamefont{{Vink}}}
  (\bibinfo{year}{2015}), vol. \bibinfo{volume}{412} of
  \emph{\bibinfo{series}{Astrophysics and Space Science Library}}, p.
  \bibinfo{pages}{199}, \eprint{1406.5657}.

\bibitem[{\citenamefont{{Talbot} and {Thrane}}(2018)}]{2018ApJ...856..173T}
\bibinfo{author}{\bibfnamefont{C.}~\bibnamefont{{Talbot}}} \bibnamefont{and}
  \bibinfo{author}{\bibfnamefont{E.}~\bibnamefont{{Thrane}}},
  \bibinfo{journal}{Astrophys. J.} \textbf{\bibinfo{volume}{856}},
  \bibinfo{eid}{173} (\bibinfo{year}{2018}), \eprint{1801.02699}.

\bibitem[{\citenamefont{{Krumholz}}(2014)}]{2014PhR...539...49K}
\bibinfo{author}{\bibfnamefont{M.~R.} \bibnamefont{{Krumholz}}},
  \bibinfo{journal}{Physics Reports} \textbf{\bibinfo{volume}{539}},
  \bibinfo{pages}{49} (\bibinfo{year}{2014}), \eprint{1402.0867}.

\bibitem[{\citenamefont{{Schneider} et~al.}(2018)\citenamefont{{Schneider},
  {Ram{\'\i}rez-Agudelo}, {Tramper}, {Bestenlehner}, {Castro}, {Sana}, {Evans},
  {Sab{\'\i}n-Sanjuli{\'a}n}, {Sim{\'o}n-D{\'\i}az}, {Langer}
  et~al.}}]{2018A&A...618A..73S}
\bibinfo{author}{\bibfnamefont{F.~R.~N.} \bibnamefont{{Schneider}}},
  \bibinfo{author}{\bibfnamefont{O.~H.} \bibnamefont{{Ram{\'\i}rez-Agudelo}}},
  \bibinfo{author}{\bibfnamefont{F.}~\bibnamefont{{Tramper}}},
  \bibinfo{author}{\bibfnamefont{J.~M.} \bibnamefont{{Bestenlehner}}},
  \bibinfo{author}{\bibfnamefont{N.}~\bibnamefont{{Castro}}},
  \bibinfo{author}{\bibfnamefont{H.}~\bibnamefont{{Sana}}},
  \bibinfo{author}{\bibfnamefont{C.~J.} \bibnamefont{{Evans}}},
  \bibinfo{author}{\bibfnamefont{C.}~\bibnamefont{{Sab{\'\i}n-Sanjuli{\'a}n}}},
  \bibinfo{author}{\bibfnamefont{S.}~\bibnamefont{{Sim{\'o}n-D{\'\i}az}}},
  \bibinfo{author}{\bibfnamefont{N.}~\bibnamefont{{Langer}}},
  \bibnamefont{et~al.}, \bibinfo{journal}{Astron. Astrophys.}
  \textbf{\bibinfo{volume}{618}}, \bibinfo{eid}{A73} (\bibinfo{year}{2018}),
  \eprint{1807.03821}.

\bibitem[{\citenamefont{{Belczynski} et~al.}(2010)\citenamefont{{Belczynski},
  {Dominik}, {Bulik}, {O'Shaughnessy}, {Fryer}, and
  {Holz}}}]{2010ApJ...715L.138B}
\bibinfo{author}{\bibfnamefont{K.}~\bibnamefont{{Belczynski}}},
  \bibinfo{author}{\bibfnamefont{M.}~\bibnamefont{{Dominik}}},
  \bibinfo{author}{\bibfnamefont{T.}~\bibnamefont{{Bulik}}},
  \bibinfo{author}{\bibfnamefont{R.}~\bibnamefont{{O'Shaughnessy}}},
  \bibinfo{author}{\bibfnamefont{C.}~\bibnamefont{{Fryer}}}, \bibnamefont{and}
  \bibinfo{author}{\bibfnamefont{D.~E.} \bibnamefont{{Holz}}},
  \bibinfo{journal}{Astrophys. J.} \textbf{\bibinfo{volume}{715}},
  \bibinfo{pages}{L138} (\bibinfo{year}{2010}), \eprint{1004.0386}.

\bibitem[{\citenamefont{{Spera} et~al.}(2015)\citenamefont{{Spera}, {Mapelli},
  and {Bressan}}}]{2015MNRAS.451.4086S}
\bibinfo{author}{\bibfnamefont{M.}~\bibnamefont{{Spera}}},
  \bibinfo{author}{\bibfnamefont{M.}~\bibnamefont{{Mapelli}}},
  \bibnamefont{and}
  \bibinfo{author}{\bibfnamefont{A.}~\bibnamefont{{Bressan}}},
  \bibinfo{journal}{Mon.Not.Roy.Astron.Soc.} \textbf{\bibinfo{volume}{451}},
  \bibinfo{pages}{4086} (\bibinfo{year}{2015}), \eprint{1505.05201}.

\bibitem[{\citenamefont{{The LIGO Scientific Collaboration} and {the Virgo
  Collaboration}}(2017)}]{2017PhRvL.119p1101A}
\bibinfo{author}{\bibnamefont{{The LIGO Scientific Collaboration}}}
  \bibnamefont{and} \bibinfo{author}{\bibnamefont{{the Virgo Collaboration}}},
  \bibinfo{journal}{\prl} \textbf{\bibinfo{volume}{119}}, \bibinfo{eid}{161101}
  (\bibinfo{year}{2017}), \eprint{1710.05832}.

\bibitem[{\citenamefont{{Abbott} et~al.}(2017)\citenamefont{{Abbott}, {Abbott},
  {Abbott}, {Acernese}, {Ackley}, {Adams}, {Adams}, {Addesso}, {Adhikari},
  {Adya} et~al.}}]{2017ApJ...848L..12A}
\bibinfo{author}{\bibfnamefont{B.~P.} \bibnamefont{{Abbott}}},
  \bibinfo{author}{\bibfnamefont{R.}~\bibnamefont{{Abbott}}},
  \bibinfo{author}{\bibfnamefont{T.~D.} \bibnamefont{{Abbott}}},
  \bibinfo{author}{\bibfnamefont{F.}~\bibnamefont{{Acernese}}},
  \bibinfo{author}{\bibfnamefont{K.}~\bibnamefont{{Ackley}}},
  \bibinfo{author}{\bibfnamefont{C.}~\bibnamefont{{Adams}}},
  \bibinfo{author}{\bibfnamefont{T.}~\bibnamefont{{Adams}}},
  \bibinfo{author}{\bibfnamefont{P.}~\bibnamefont{{Addesso}}},
  \bibinfo{author}{\bibfnamefont{R.~X.} \bibnamefont{{Adhikari}}},
  \bibinfo{author}{\bibfnamefont{V.~B.} \bibnamefont{{Adya}}},
  \bibnamefont{et~al.}, \bibinfo{journal}{Astrophys. J. Lett.}
  \textbf{\bibinfo{volume}{848}}, \bibinfo{eid}{L12} (\bibinfo{year}{2017}),
  \eprint{1710.05833}.

\bibitem[{\citenamefont{{Chruslinska} et~al.}(2018)\citenamefont{{Chruslinska},
  {Belczynski}, {Klencki}, and {Benacquista}}}]{2018MNRAS.474.2937C}
\bibinfo{author}{\bibfnamefont{M.}~\bibnamefont{{Chruslinska}}},
  \bibinfo{author}{\bibfnamefont{K.}~\bibnamefont{{Belczynski}}},
  \bibinfo{author}{\bibfnamefont{J.}~\bibnamefont{{Klencki}}},
  \bibnamefont{and}
  \bibinfo{author}{\bibfnamefont{M.}~\bibnamefont{{Benacquista}}},
  \bibinfo{journal}{Mon.Not.Roy.Astron.Soc.} \textbf{\bibinfo{volume}{474}},
  \bibinfo{pages}{2937} (\bibinfo{year}{2018}), \eprint{1708.07885}.

\bibitem[{\citenamefont{{Giacobbo} and {Mapelli}}(2019)}]{2019MNRAS.482.2234G}
\bibinfo{author}{\bibfnamefont{N.}~\bibnamefont{{Giacobbo}}} \bibnamefont{and}
  \bibinfo{author}{\bibfnamefont{M.}~\bibnamefont{{Mapelli}}},
  \bibinfo{journal}{Mon.Not.Roy.Astron.Soc.} \textbf{\bibinfo{volume}{482}},
  \bibinfo{pages}{2234} (\bibinfo{year}{2019}), \eprint{1805.11100}.

\bibitem[{\citenamefont{{Lamberts} et~al.}(2016)\citenamefont{{Lamberts},
  {Garrison-Kimmel}, {Clausen}, and {Hopkins}}}]{2016MNRAS.463L..31L}
\bibinfo{author}{\bibfnamefont{A.}~\bibnamefont{{Lamberts}}},
  \bibinfo{author}{\bibfnamefont{S.}~\bibnamefont{{Garrison-Kimmel}}},
  \bibinfo{author}{\bibfnamefont{D.~R.} \bibnamefont{{Clausen}}},
  \bibnamefont{and} \bibinfo{author}{\bibfnamefont{P.~F.}
  \bibnamefont{{Hopkins}}}, \bibinfo{journal}{Mon.Not.Roy.Astron.Soc.}
  \textbf{\bibinfo{volume}{463}}, \bibinfo{pages}{L31} (\bibinfo{year}{2016}),
  \eprint{1605.08783}.

\bibitem[{\citenamefont{{Cao} et~al.}(2018)\citenamefont{{Cao}, {Lu}, and
  {Zhao}}}]{2018MNRAS.474.4997C}
\bibinfo{author}{\bibfnamefont{L.}~\bibnamefont{{Cao}}},
  \bibinfo{author}{\bibfnamefont{Y.}~\bibnamefont{{Lu}}}, \bibnamefont{and}
  \bibinfo{author}{\bibfnamefont{Y.}~\bibnamefont{{Zhao}}},
  \bibinfo{journal}{Mon.Not.Roy.Astron.Soc.} \textbf{\bibinfo{volume}{474}},
  \bibinfo{pages}{4997} (\bibinfo{year}{2018}), \eprint{1711.09190}.

\bibitem[{\citenamefont{{Mapelli} et~al.}(2018)\citenamefont{{Mapelli},
  {Giacobbo}, {Toffano}, {Ripamonti}, {Bressan}, {Spera}, and
  {Branchesi}}}]{2018MNRAS.481.5324M}
\bibinfo{author}{\bibfnamefont{M.}~\bibnamefont{{Mapelli}}},
  \bibinfo{author}{\bibfnamefont{N.}~\bibnamefont{{Giacobbo}}},
  \bibinfo{author}{\bibfnamefont{M.}~\bibnamefont{{Toffano}}},
  \bibinfo{author}{\bibfnamefont{E.}~\bibnamefont{{Ripamonti}}},
  \bibinfo{author}{\bibfnamefont{A.}~\bibnamefont{{Bressan}}},
  \bibinfo{author}{\bibfnamefont{M.}~\bibnamefont{{Spera}}}, \bibnamefont{and}
  \bibinfo{author}{\bibfnamefont{M.}~\bibnamefont{{Branchesi}}},
  \bibinfo{journal}{Mon.Not.Roy.Astron.Soc.} \textbf{\bibinfo{volume}{481}},
  \bibinfo{pages}{5324} (\bibinfo{year}{2018}), \eprint{1809.03521}.

\bibitem[{\citenamefont{{Artale} et~al.}(2019)\citenamefont{{Artale},
  {Mapelli}, {Giacobbo}, {Sabha}, {Spera}, {Santoliquido}, and
  {Bressan}}}]{2019arXiv190300083A}
\bibinfo{author}{\bibfnamefont{M.~C.} \bibnamefont{{Artale}}},
  \bibinfo{author}{\bibfnamefont{M.}~\bibnamefont{{Mapelli}}},
  \bibinfo{author}{\bibfnamefont{N.}~\bibnamefont{{Giacobbo}}},
  \bibinfo{author}{\bibfnamefont{N.~B.} \bibnamefont{{Sabha}}},
  \bibinfo{author}{\bibfnamefont{M.}~\bibnamefont{{Spera}}},
  \bibinfo{author}{\bibfnamefont{F.}~\bibnamefont{{Santoliquido}}},
  \bibnamefont{and}
  \bibinfo{author}{\bibfnamefont{A.}~\bibnamefont{{Bressan}}},
  \bibinfo{journal}{arXiv e-prints} \bibinfo{eid}{arXiv:1903.00083}
  (\bibinfo{year}{2019}), \eprint{1903.00083}.

\bibitem[{\citenamefont{{Dvorkin} et~al.}(2016)\citenamefont{{Dvorkin}, {Uzan},
  {Vangioni}, and {Silk}}}]{2016PhRvD..94j3011D}
\bibinfo{author}{\bibfnamefont{I.}~\bibnamefont{{Dvorkin}}},
  \bibinfo{author}{\bibfnamefont{J.-P.} \bibnamefont{{Uzan}}},
  \bibinfo{author}{\bibfnamefont{E.}~\bibnamefont{{Vangioni}}},
  \bibnamefont{and} \bibinfo{author}{\bibfnamefont{J.}~\bibnamefont{{Silk}}},
  \bibinfo{journal}{Phys. Rev. D.} \textbf{\bibinfo{volume}{94}},
  \bibinfo{eid}{103011} (\bibinfo{year}{2016}), \eprint{1607.06818}.

\bibitem[{\citenamefont{{Vecchio}}(2002)}]{2002CQGra..19.1449V}
\bibinfo{author}{\bibfnamefont{A.}~\bibnamefont{{Vecchio}}},
  \bibinfo{journal}{Classical and Quantum Gravity}
  \textbf{\bibinfo{volume}{19}}, \bibinfo{pages}{1449} (\bibinfo{year}{2002}).

\bibitem[{\citenamefont{{Ungarelli} and {Vecchio}}(2001)}]{2001PhRvD..64l1501U}
\bibinfo{author}{\bibfnamefont{C.}~\bibnamefont{{Ungarelli}}} \bibnamefont{and}
  \bibinfo{author}{\bibfnamefont{A.}~\bibnamefont{{Vecchio}}},
  \bibinfo{journal}{Phys Rev. D.} \textbf{\bibinfo{volume}{64}},
  \bibinfo{pages}{121501} (\bibinfo{year}{2001}), \eprint{astro-ph/0106538}.

\bibitem[{\citenamefont{{Kudoh} and {Taruya}}(2005)}]{2005PhRvD..71b4025K}
\bibinfo{author}{\bibfnamefont{H.}~\bibnamefont{{Kudoh}}} \bibnamefont{and}
  \bibinfo{author}{\bibfnamefont{A.}~\bibnamefont{{Taruya}}},
  \bibinfo{journal}{Phys. Rev. D.} \textbf{\bibinfo{volume}{71}},
  \bibinfo{eid}{024025} (\bibinfo{year}{2005}), \eprint{gr-qc/0411017}.

\bibitem[{\citenamefont{{Geller} et~al.}(2018)\citenamefont{{Geller}, {Hook},
  {Sundrum}, and {Tsai}}}]{2018PhRvL.121t1303G}
\bibinfo{author}{\bibfnamefont{M.}~\bibnamefont{{Geller}}},
  \bibinfo{author}{\bibfnamefont{A.}~\bibnamefont{{Hook}}},
  \bibinfo{author}{\bibfnamefont{R.}~\bibnamefont{{Sundrum}}},
  \bibnamefont{and} \bibinfo{author}{\bibfnamefont{Y.}~\bibnamefont{{Tsai}}},
  \bibinfo{journal}{Phys. Rev. Letters} \textbf{\bibinfo{volume}{121}},
  \bibinfo{eid}{201303} (\bibinfo{year}{2018}), \eprint{1803.10780}.

\bibitem[{\citenamefont{Cusin et~al.}(2019{\natexlab{b}})\citenamefont{Cusin,
  Dvorkin, Pitrou, and Uzan}}]{us}
\bibinfo{author}{\bibfnamefont{G.}~\bibnamefont{Cusin}},
  \bibinfo{author}{\bibfnamefont{I.}~\bibnamefont{Dvorkin}},
  \bibinfo{author}{\bibfnamefont{C.}~\bibnamefont{Pitrou}}, \bibnamefont{and}
  \bibinfo{author}{\bibfnamefont{J.-P.} \bibnamefont{Uzan}},
  \bibinfo{journal}{submitted to PRL}  (\bibinfo{year}{2019}{\natexlab{b}}).

\end{thebibliography}

\end{document}